\documentclass[preprint,12pt]{elsarticle}

\usepackage{amssymb,amsmath}

\usepackage{multirow} 
\usepackage{bm} 

\journal{Nuclear Physics B}

\newcommand{\Slash}[1]{\ooalign{\hfil/\hfil\crcr$#1$}}
\newcommand{\ket}[1]{\left| \, #1 \, \right\rangle}

\begin{document}

\begin{frontmatter}



\title{Heavy quark symmetry in multi-hadron systems}


\author[a]{Yasuhiro~Yamaguchi}
\author[a]{Shunsuke~Ohkoda}
\author[a,b]{Atsushi~Hosaka}
\author[c]{Tetsuo~Hyodo}
\author[d,e]{Shigehiro~Yasui}
\address[a]{Research Center for Nuclear Physics (RCNP), 
Osaka University, Ibaraki, Osaka, 567-0047, Japan}
\address[b]{J-PARC Branch, KEK Theory Center, Institute of Particle and Nuclear Studies,
KEK, Tokai, Ibaraki, 319-1106, Japan}
\address[c]{Yukawa Institute for Theoretical Physics, Kyoto University, Kyoto, 606-8317, Japan}
\address[d]{KEK Theory Center, Institute of Particle and Nuclear
Studies, High Energy Accelerator Research Organization, 1-1, Oho,
Ibaraki, 305-0801, Japan}
\address[e]{Department of Physics, Tokyo Institute of Technology,
Tokyo 152-8551, Japan}

\begin{abstract}
We discuss the properties of hadronic systems containing one heavy quark in the heavy quark limit. The heavy quark symmetry guarantees the mass degeneracy of the states with total spin and parity $(j-1/2)^{P}$ and $(j+1/2)^{P}$ with $j\geq 1/2$, because the heavy-quark spin is decoupled from the total spin $j$ of the light components called brown muck. We apply this idea to heavy multi-hadron systems, and formulate the general framework to analyze their properties. We demonstrate explicitly the spin degeneracy and the decomposition of the wave functions in exotic heavy hadron systems generated by the one boson exchange potential. 
The masses of the brown muck can be extracted from theoretical and experimental hadron spectra, leading to the color non-singlet spectroscopy.
\end{abstract}


\end{frontmatter}



\section{Introduction}\label{sec:intro}

The study of the exotic hadrons 
provides us with unique opportunities to explore 
fundamental properties of the low energy QCD, such as color confinement, the chiral symmetry breaking, etc. Recently, in the heavy flavor (charm and bottom) sectors, experimental evidences for new candidates of exotic hadrons, such as X, Y, Z, have been reported, and these states are extensively investigated in theoretical works~\cite{Swanson:2006st,Brambilla:2010cs}.
Although there have been many theoretical studies based on various pictures such as multiquarks, hybrids of quarks and gluons, multi-hadrons, and so on, 
we have not yet understood the essential features of exotic heavy hadrons.
In the present article, we approach the structure of the hadronic molecules with a heavy quark from a point of view of the heavy quark symmetry (HQS)~\cite{Isgur:1989vq,Isgur:1989ed,Isgur:1991wq,Rosner:1985dx,Eichten:1993ub,Neubert:1993mb,Manohar:2000dt},
namely the symmetry of the heavy-quark spin, as the fundamental property of heavy hadrons.

In general, for hadrons with a single heavy quark, the HQS leads to the mass degeneracy of two states with different total spin in the heavy quark limit. This is because the spin of the heavy quark is decoupled from the total spin of the other components made of light quarks and gluons. The latter component is called the brown muck, which is everything but the heavy quark.
It is important to note that the brown muck has the conserved total spin $j$, although the brown muck is a non-perturbative object which is dressed by many quarks and gluons like $q^{n}+q^{n}q\bar{q}+q^{n}g+\dots$ with a net quark number $n$.
For $j\neq 0$, the spin degeneracy in the heavy hadrons is realized by the pair states with the total angular momenta, $J=j-1/2$ and $j+1/2$. We call those two states the ``HQS doublet." For $j=0$, there is only one state with $J=1/2$. We call this state the ``HQS singlet."

The HQS is
 seen in the mass spectrum of the charm and bottom hadrons.
For example, the mass splitting between $\bar{D}$ $(J=0)$ and $\bar{D}^{\ast}$ $(J=1)$ mesons is 140 MeV, and that between $B$ and $B^{\ast}$ meson is 45 MeV~\cite{Beringer:1900zz}. Those mass splittings are smaller than ones between $\pi$ and $\rho$ ($\sim 600$ MeV) and that between $K$ and $K^{\ast}$ ($\sim 400$ MeV). Therefore, $\bar{D}$ and $\bar{D}^{\ast}$ ($B$ and $B^{\ast}$) mesons are approximately regarded as the HQS doublet states.
In those cases, the  brown muck is a light quark $q$, which is dressed by quark-antiquark pairs and gluons,  with spin and parity $1/2^+$ in total.

Similar mass degeneracy of HQS doublets is also seen in the baryonic sector. The mass splitting between $\Sigma_{\rm c}$ $(J=1/2)$ and $\Sigma_{\rm c}^{\ast}$ $(J=3/2)$ ($\Sigma_{\rm b}$ and $\Sigma_{\rm b}^{\ast}$) baryons is 65 MeV (20 MeV), which is smaller than 192 MeV between $\Sigma$ and $\Sigma^{*}$.
$\Lambda_{\rm c}$ and $\Lambda_{\rm b}$ with $J=1/2$ in the ground state are regarded as the HQS singlet states, because there is no nearby $J=3/2$ partner.
Recently, two excited bottom baryons $\Lambda_{\rm b}^{\ast}$ have been observed at LHCb~\cite{Aaij:2012da}. 
Although the quantum numbers are not settled yet, assigning $1/2^{-}$ ($3/2^{-}$) for the state with the lower (higher) mass, we see that the mass splitting between $\Lambda_{\rm c}^{\ast}(1/2^{-})$ and $\Lambda_{\rm c}^{\ast}(3/2^{-})$ is 33 MeV and that between $\Lambda_{\rm b}^{\ast}(1/2^{-})$ and $\Lambda_{\rm b}^{\ast}(3/2^{-})$ is only 8 MeV.
Those mass splittings can be compared with 115 MeV between $\Lambda^{\ast}(1/2^{-})$ and $\Lambda^{\ast}(3/2^{-})$.

The brown muck in a heavy baryon with one heavy quark and two light quarks has the same quantum number as a pair of two quarks $qq$.
The cluster of two quarks is called diquark in the constituent picture of the quark model.
Because two quarks can have many possible quantum numbers, such as isospin, total angular momentum and parity $ I(J^{P})$, we can investigate a variety of properties of diquarks in the heavy baryons.
The diquark in $\Lambda_{c}$ ($\Lambda_{b}$) has $I(J^{P})=0(0^{+})$.
The diquark in $\Sigma_{\rm c}$ $(J=1/2)$ and $\Sigma_{\rm c}^{\ast}$ $(J=3/2)$ ($\Sigma_{\rm b}$ and $\Sigma_{\rm b}^{\ast}$) has $1(1^{+})$.
The diquark in $\Lambda_{\rm c}^{\ast}(1/2^{-})$ and $\Lambda_{\rm c}^{\ast}(3/2^{-})$ ($\Lambda_{\rm b}^{\ast}(1/2^{-})$ and $\Lambda_{\rm b}^{\ast}(3/2^{-})$) has $0(1^{-})$.
The analysis of the brown muck in heavy hadrons will also be useful to understand the role of diquarks, not only in the  confinement phase, but also in the deconfinement phase, e.g. the quark-gluon plasma \cite{Shuryak:2003ty,Shuryak:2004tx,Lee:2007wr,Oh:2009zj} and the color superconductivity \cite{Alford:2007xm,Fukushima:2010bq}.
The HQS has been also successfully applied to several excited hadrons with charm and bottom flavors~\cite{Bardeen:2003kt}. It is also relevant to understand the properties of exotic hadrons with hidden charm or bottom quarks \cite{Bondar:2011ev,Voloshin:2011qa,Ohkoda:2011vj,Ohkoda:2012rj}.

We have pointed out in Ref.~\cite{Yasui:2013vca} that the HQS is seen  also in 
 multi-hadron systems with a heavy quark. Let us consider a hadronic molecule (or hadron composite), where a heavy hadron is surrounded by light hadrons to form bound and/or scattering states including resonances.
When the HQS holds, 
we can use spin degrees of freedom to classify states;
the total spin of the hadronic molecule is decomposed into the heavy-quark spin and the total spin of the brown muck~\cite{Yasui:2013vca}. 
The latter has three contributions, namely, a sum of (i) the spins of the light quarks and gluons in the heavy hadron, (ii) the spins of the light hadrons surrounding the heavy hadron and (iii) the relative angular momenta between the heavy hadron and the light hadrons. 
This particular form of the brown muck
is called the light spin-complex (or spin-complex in short), because it is a composite object of light quarks, gluons and light hadrons having the total spin as a conserved quantum number.
It is important to note that the quark-gluon degrees of freedom (i) coexist with the hadronic degrees of freedom (ii) and (iii).
 
Let us show examples for the spin-complex.  First we consider a $\bar{P}^{(\ast)}N$ bound state composed of a heavy meson $\bar{P}^{(\ast)}\sim (\bar{Q}q)_{\mathrm{spin}\,0(1)}$ and a nucleon $N$, where $\bar{Q}$ is a heavy antiquark and $q$ is a light component  in $\bar{P}^{(\ast)}$ \cite{Yasui:2013vca}. Note that $q$ is not simply a single light quark but rather a composite of light quarks and gluons with appropriate quantum numbers in total. Then, the spin-complex for the $\bar{P}^{(\ast)}N$ is denoted by $[Nq]$, which is a composite object of the light component $q$, the nucleon $N$ and the relative angular momentum between $\bar{P}^{(\ast)}$ and $N$. 
The $\bar{P}^{(\ast)}$ molecule has also been considered for the dibaryons of $\bar{P}^{(\ast)}NN$, where the spin-complex is identified with $[NNq]$ \cite{Yamaguchi:2013hsa}.
Yet, another example is the $\bar{P}^{(\ast)}$ meson embedded in nuclear matter \cite{Yasui:2013vca}. The spin-complex there is identified with a sum of $q$ in $\bar{P}^{(\ast)}$ and many pairs of the particle (nucleon) $N$ and the hole $N^{-1}$ generated around the Fermi surface.

The spin-complex provides a new picture for a certain class of brown muck, which is useful in particular for the analysis of hadronic molecules.  It is regarded as a colored effective degree of freedom inside hadrons, just as the constituent quarks and diquarks are.  
It should be noted, however, that there 
is a subtle issue for a criterion about how to separate the spin-complex from other components of the brown muck.
The problem is essentially the same with the difficulty to define 
the structure of hadrons, such as compact multi-quarks and/or extended hadronic molecules, in a model-independent manner \cite{Hyodo:2013nka}.
Therefore, the spin-complex is unambiguously defined only when the model space is specified.
Nevertheless, in this paper, we will show that the spin-complex is a powerful tool to classify the structure of hadrons.

Before closing the introduction, we mention the mass spectrum of the brown muck as the color non-singlet objects~\cite{Lipkin:1987sk,Jaffe:2005md}.
The brown muck is a colored object being (anti-)fundamental representation of the color symmetry.
Nevertheless, it is a well-defined object characterized by its spin-parity and light flavor quantum numbers in the heavy quark limit.
We expect that the brown muck exhibits a rich pattern in mass spectrum, because of its internal structure.
For example, the brown muck in the excited hadrons is heavier than that in the ground state hadrons. 
The mass of the brown muck can be defined in the heavy quark limit with the help of the hadron mass formula in the heavy quark effective theory.
In this paper, we will show that the mass of the brown muck can be extracted from the HQS multiplets both in the charm and bottom sectors.
Thus, with the heavy hadron spectrum,
 we can perform the spectroscopy of the color non-singlet object. 
The study of 
 spin-complexes will be useful to interpret the spectrum of the brown muck in hadronic molecules with a heavy hadron.

The article is organized as follows.
In Section~\ref{sec:general}, we briefly review the HQS and introduce the idea of the spin-complex, and give a general discussion for the wave function of the brown muck. In Section~\ref{sec:examples_exotic}, we show that the spin-complex can be used to classify the structure of hadrons, with the example of the exotic baryons with a heavy antiquark in a potential model.
The present discussion includes the part of the results in our previous work in Ref.~\cite{Yasui:2013vca}.
 In Section~\ref{sec:spectrum}, we discuss the mass spectrum of the brown muck extracted from the experimental data as well as from the predictions in a quark model. Summary and perspectives are given in the last section.

\section{General properties of hadrons with a heavy quark}\label{sec:general}

In this section, we introduce the spin-complex as a convenient tool to express the brown muck, starting from the HQS in QCD. We show that a heavy hadron with the total spin $J\geq 1/2$ 
may have two components with different spin-complex of spin $j = J\pm 1/2$.  
They can be mixed for a finite heavy quark mass, but are decoupled in the heavy quark limit. 
The spin-complex basis is then related to the particle basis, from which the wave functions of the pair states in the HQS doublet are analyzed in terms of the hadronic degrees of freedom. Explicit examples of these components will be given for the $\bar{P}^{(*)}N$ system in Section~\ref{sec:examples_exotic}.

The HQS leads to the systematic expansion of the hadron mass in the inverse powers of the heavy quark mass. This expansion enables us to define the mass of the brown muck, and hence that of the spin-complex,  in the heavy quark limit. We present the basic formula which will be used, in Section~\ref{sec:spectrum}, to extract the spectrum of the brown muck from the experimental data and theoretical predictions with a finite heavy quark mass.

\subsection{Heavy quark symmetry in QCD}\label{subsec:QCD}

We consider that the heavy quark mass $m_{\rm Q}$ is much larger than a typical energy scale of low energy QCD. In this case, an effective field theory with the $1/m_{\rm Q}$ expansion is useful to study the hadrons containing a single heavy quark \cite{Neubert:1993mb,Manohar:2000dt}.
To this end, let us start our discussion first with the heavy quark Lagrangian;
\begin{align}
    {\cal L}_{\rm{HQ}} 
    &= \bar{Q} (iD \hspace{-0.6em} / - m_{\rm{Q}}) Q,
\end{align}
where $Q$ is the heavy quark field, the covariant derivative is defined by $D_{\mu} = \partial_{\mu} + ig_{\mathrm{s}}A_{\mu}^{a}t^{a}$ with the gluon field $A_{\mu}^{a}$, the gauge coupling $g_{\mathrm{s}}$, and $t^{a}=\lambda^{a}/2$ with the Gell-Mann matrices $\lambda^{a}$ ($a=1,\cdots, 8$). The term from light quark and gluon sectors is not relevant in the current discussion. Denoting the four-velocity of the heavy quark as $v^{\mu}$ ($v^2=1$), we decompose the heavy quark field into the positive energy component $Q_{v}(x)$ and the negative energy component ${\cal Q}_{v}(x)$ as 
\begin{align}
    Q(x) 
    &= e^{-i m_{\rm Q} v\cdot x} 
    \left( Q_{v}(x) + {\cal Q}_{v}(x) \right),
\end{align}
by the projections
\begin{align}
    Q_{v}(x)
    &= e^{i m_{\rm Q} v\cdot x}\! \frac{1+v\hspace{-0.5em}/}{2} Q(x), \quad
    {\cal Q}_{v}(x) 
    = e^{i m_{\rm Q} v\cdot x} \frac{1-v\hspace{-0.5em}/}{2} Q(x).
\end{align}
Here we remove the momentum $m_{\rm{Q}}v^{\mu}$ in the original field $Q(x)$ and leave only the residual momenta.
We multiply the projection operators $(1\pm v\hspace{-0.5em}/)/2$ to select the positive (negative) energy component.
In the following discussion, we abbreviate the coordinate ``$x$" in the field. By eliminating ${\cal Q}_{v}$, we obtain the effective Lagrangian for $Q_{v}$,
\begin{align}
    {\cal L}_{\rm{HQET}} 
    &= \bar{Q}_{v} v \!\cdot\! i D Q_{v} 
    + \bar{Q}_{v} \frac{(i D_{\perp})^2}{2m_{\rm Q}} Q_{v} 
    - c(\mu) g_{\mathrm{s}} \bar{Q}_{v} 
    \frac{\sigma_{\mu \nu}G^{\mu \nu}}{4m_{\rm Q}} Q_{v} 
    + {\cal O}(1/m_{\rm Q}^{2}) ,
\label{eq:HQET} 
\end{align}
with $D_{\perp}^{\mu} = D^{\mu} - v^{\mu} \, v\!\cdot\! D$, $G^{\mu\nu}=[D^{\mu},D^{\nu}]/ig_{\mathrm{s}}$, and $\sigma^{\mu\nu}=i[\gamma^{\mu}, \gamma^{\nu}]/2$. Here $c(\mu)$ is the Wilson coefficient for the matching with QCD at the energy scale $\mu$. This is the effective Lagrangian in the heavy quark effective theory (HQET) \cite{Isgur:1989vq,Isgur:1989ed,Isgur:1991wq,Rosner:1985dx,Eichten:1993ub,Neubert:1993mb,Manohar:2000dt}. In the heavy quark mass limit $m_{\rm Q}\to\infty$, only the first term of Eq.~\eqref{eq:HQET} remains and the spin-flip terms involved in $\sigma_{\mu \nu}G^{\mu \nu}$ are suppressed by $1/m_{\rm Q}$. This indicates that the spin of the heavy quark is a conserved quantity, which is known as the HQS.
We will see that the $1/m_{\rm Q}$ expansion and the HQS are essential also for heavy hadrons.

\subsection{Spin-complex}

We now consider the consequences of the HQS in hadronic systems. We are interested in hadrons either with a single heavy quark $(Q)$ or with a single heavy antiquark $(\bar{Q})$, and with arbitrary baryon number $B$. In QCD, such state may be expressed by a superposition of various components with light quarks ($q$) and gluons ($g$) as
\begin{align}
    \ket{{\rm H}_{\rm Q}}
    &=\ket{q^{n}Q}
    \oplus \ket{q^{n} q\bar{q}Q}
    \oplus \ket{q^{n} gQ}\oplus\dots 
    \label{eq:HQQCD}, \\
    \ket{{\rm H}_{\bar{Q}}}
    &=\ket{q^{m}\bar{Q}} 
    \oplus\ket{q^{m} q\bar{q}\bar{Q}}
    \oplus\ket{q^{m} g\bar{Q}}\oplus\dots 
    \label{eq:HQbarQCD},
\end{align}
where $n=3B-1$ and $m=3B+1$ (negative $n$ represents the number of antiquarks, $q^{-n}\equiv \bar{q}^{\,n}$). We then decompose the total spin of this hadron $\vec{J}$ into the spin $\vec{S}$ of the heavy (anti-)quark and the spin $\vec{j}$ of the rest which contains only light degrees of freedom;
\begin{align}
    \vec{J}
    &=\vec{S}+\vec{j} .
    \label{eq:decomposition}
\end{align}
As shown in the previous subsection, the heavy-quark spin $\vec{S}$ is conserved in the heavy quark limit.
Since the total $\vec{J}$ is conserved, the total spin $\vec{j}$ of the light system is also conserved in the heavy quark limit.

For a single hadron system $\ket{\rm H}$, the object which carries the total spin $\vec{j}$ is called ``brown muck'', which is everything but the heavy quark~\cite{Neubert:1993mb}.
In the present discussion, we introduce a notation
\begin{align}
[\alpha]_{f,j^{\cal P}} = q^{n} + q^{n}q\bar{q} + q^{n}g + \dots,
\label{eq:notation_bm}
\end{align}
to express the state of the brown muck $\alpha$ with total $n$ quarks, whose quantum numbers are given by the structure of light flavor $f$ and the total spin-parity $j^{\cal P}$.
For example, when the light flavor ${\rm{SU}}(N_{\rm f})$ symmetry is a good symmetry, $f$ denotes the representation of the ${\rm{SU}}(N_{\rm f})$ symmetry.
In Eq.~(\ref{eq:notation_bm}), the weight of each component depends on $\alpha$.
We note that, as indicated in Eqs.~\eqref{eq:HQQCD} and \eqref{eq:HQbarQCD}, the brown muck is a highly non-perturbative object made of light quarks and gluons.

The brown muck belongs to the color (anti-)triplet, so the strong interaction is at work between the heavy (anti-)quark and the brown muck. Nevertheless, the total spin $\vec{j}$ of the brown muck is well defined through Eq.~\eqref{eq:decomposition} and conserved in the heavy quark limit. In other words, all the interactions which flip $\vec{j}$ (and hence flip the spin of the heavy quark $\vec{S}$) are suppressed in the heavy quark limit, while the interaction which does not flip the spin, such as color electric force, is still active. In this way, the conservation of $\vec{j}$ of the brown muck is realized. In addition, the light-flavor quantum numbers (isospin and strangeness) of the brown muck are identical to those of the heavy hadron, because the heavy quark does not carry them. Thus, the brown muck is a well-defined object in the heavy quark limit, characterized by its spin-parity and light flavors. We emphasize that this viewpoint is useful, not only for theoretical researches, but also for experimental researches in realistic situations with finite heavy quark mass, as we will discuss later.

For normal hadrons like $\bar{Q}q$ mesons and $Qqq$ baryons, the brown mucks are composed of quarks and gluons.
However, this situation may change for exotic hadrons. 
It has been pointed out that, in the heavy sector, there can be hadronic states as ensembles of multiple color singlet objects. For instance, hadronic molecules which are loosely bound systems of hadrons can be generated by inter-hadron forces, even in the exotic sector~\cite{Ohkoda:2011vj,Ohkoda:2012rj,Yasui:2009bz,Yamaguchi:2011xb,Yamaguchi:2011qw,Yamaguchi:2013ty}. In addition, there are investigations on nuclei with heavy hadrons~\cite{Mizutani:2006vq,Yasui:2012rw,Yasui:2013xr} which consists of a heavy hadron and several nucleons. In these cases, 
``everything but the heavy quark'' 
is not simply made of quarks and gluons, but a mixture of quark-gluon components and hadronic degrees of freedom. In Ref.~\cite{Yasui:2013vca}, we proposed to call it ``light spin-complex'' (or ``spin-complex'' in short) to express the composite system of quarks, gluons and hadrons. 
Before discussing the structure of the spin-complex, however, we will first discuss the general property of the brown muck in the next subsection.

\subsection{Brown-muck component basis}

Let us describe a heavy hadron of spin-parity $J^P$ in terms of a heavy (anti-)quark $Q$ ($\bar{Q}$) and a brown muck $[\alpha]_{f,j^{\cal P}}$ as
\begin{align}
    \ket{[\alpha]_{f,j^{\cal P}}Q}_{J^{P}}, \quad
    \ket{[\alpha]_{f,j^{\cal P}}\bar{Q}}_{J^{P}}.
\end{align}
The contents of $[\alpha]_{f,j^{\cal P}}$ was given in Eq.~(\ref{eq:notation_bm}).
 For instance, the brown muck of the ground state $\Lambda_{\rm c}$ with $J^{P}=1/2^{+}$ may contain the scalar diquark $[ud]_{I=0,0^{+}}$ which represents the system of $u$ and $d$ quarks combined into spin zero. Also, the brown muck of a $\bar{D}N$ bound state with $I=0$ can have the spin-complex $[Nq]_{I=0,0^{+}}^{(1,{\rm S})}$ which stands for the system made of a nucleon $N$ and $q$, with total spin combined into triplet and relative angular momentum being S-wave as denoted by $(1,{\rm S})$ in the superscript.
 
In general, several different components $[\alpha]$ can contribute to each brown muck, because $f$ and $j^{\cal P}$ are the only well-defined quantum numbers.
For practical analysis the brown muck can be expanded by a suitable basis, 
for instance, by the diquark basis for heavy baryons and by the spin-complex basis for hadronic molecules.
We introduce a term ``brown-muck component (BMC)" to indicate the component concretely given by particular objects like diquarks and/or spin-complexes.
Various examples are displayed in Section~\ref{sec:examples_exotic} and also in \ref{sec:examples_nonexotic}.

Note that the brown muck is not necessarily a compact cluster. 
The concept of the brown muck rather includes also extended objects like in hadronic molecules.
It should be noted that the ``relative angular momentum'' between the heavy quark and the brown muck is included in the total spin $j$. The parity of the spin-complex is therefore uniquely determined as ${\cal P}=P$ for heavy quark $Q$ and ${\cal P}=-P$ for heavy antiquark $\bar{Q}$. 
In the following,  when $f$ and $\cal P$ are not relevant to the discussion, we sometimes omit $f$ and $\cal P$ for simplicity.

Now we discuss  wave functions more in detail.
As a simple case, we first consider the case of a single BMC.
A heavy hadron $\ket{\rm H}_{J}$ with spin $J\geq 1/2$ can have in general two components $[\alpha]_{J-1/2}Q$ and $[\beta]_{J+1/2}Q$ 
 containing the brown muck $\alpha$ and $\beta$ with total spin $j=J\mp1/2$, respectively, as 
\begin{align}
    \ket{\rm H}_{J}
    =
    \begin{pmatrix}
    C_{\alpha}\ket{[\alpha]_{J-1/2}Q}_{J} 
    \\
    C_{\beta}\ket{[\beta]_{J+1/2}Q}_{J} 
    \end{pmatrix} ,
    \label{def_H}
\end{align}
with coefficients $C_{\alpha}$ and $C_{\beta}$ which satisfy the normalization condition $|C_{\alpha}|^{2}+|C_{\beta}|^{2}=1$.
We suppress the irrelevant parity and flavor indices.
Let us consider the heavy quark limit. In this case, $j=J\mp1/2$ is a good quantum number as shown in the previous subsection, so the two components in ${\rm H}$ should be realized  as independent degrees of freedom: 
\begin{align}
    \ket{J-1/2}_{J}
    &=
    \begin{pmatrix}
    \ket{[\alpha]_{J-1/2}Q}_{J} 
    \\
    0
    \end{pmatrix} 
    \label{eq:Pm} ,\\
    \ket{J+1/2}_{J}
    &=
    \begin{pmatrix}
    0
    \\
    \ket{[\beta]_{J+1/2}Q}_{J} 
    \end{pmatrix}
    \label{eq:Pp},
\end{align}
where we introduce new notations $ \ket{J\mp1/2}_{J}$ to indicate the state containing the brown muck with the total spin $J\mp1/2$ (this should not be confused with the symbol H in Eq.~(\ref{def_H})).
The transition from $\ket{J-1/2}_{J}$ sector to $\ket{J+1/2}_{J}$ sector is suppressed by $1/m_{\rm Q}$ and the Hamiltonian of this system is diagonalized 
by the two basis states (\ref{eq:Pm}) and (\ref{eq:Pp})
in the heavy quark limit. 
The separation of two independent states for $J\geq 1/2$ is the first consequence of the HQS.

Next we consider the other heavy hadron with spin $J+1$. Following the same discussion above,$\ket{J+1/2}_{J+1}$ and $\ket{J+3/2}_{J+1}$ are separated in the heavy quark limit, and we can diagonalize the Hamiltonian.
The $\ket{J+1/2}_{J+1}$ and $\ket{J+3/2}_{J+1}$ component are written in the brown mucks $\beta$ and $\gamma$ with total spin $J+1/2$ and $J+3/2$, respectively, as\footnote{We consider the same model space with Eqs.~\eqref{eq:Pm} and \eqref{eq:Pp}.}
\begin{align}
    \ket{J+1/2}_{J+1}
    &=
    \begin{pmatrix}
    \ket{[\beta]_{J+1/2}Q}_{J+1} \\
    0
 \end{pmatrix} , \\
     \ket{J+3/2}_{J+1}
    &=
    \begin{pmatrix}
     0 \\
     \ket{[\gamma]_{J+3/2}Q}_{J+1}
    \end{pmatrix}.
    \label{eq:HpJm1}
\end{align}
Importantly, the brown muck $[\beta]_{J+1/2}$ of $\ket{J+1/2}_{J+1}$ is identical to that in Eq.~\eqref{eq:Pp}. The only difference is the direction of the heavy-quark spin. Because the heavy-quark spin does not affect the structure of the brown muck, we conclude that the $\ket{J+1/2}_{J}$ and $\ket{J+1/2}_{J+1}$ states containing the common $\beta$ are degenerate in the heavy quark limit. This is the second consequence of the HQS. In the same way, $\ket{J+3/2}_{J+1}$ will be an HQS doublet with $\ket{J+3/2}_{J+2}$.
We note that the brown mucks $\alpha$, $\beta$ and $\gamma$ can be all different in general.
We remark also that, in Eq.~\eqref{eq:Pm}, $\ket{J-1/2}_{J}$ will form an HQS doublet with $\ket{J-1/2}_{J-1}$ for $J\geq 1$, while it does not have a counterpart for $J=1/2$. We call the latter state a HQS singlet. The $J=0$ state has only one component $\ket{1/2}_{0}$ which forms a HQS doublet with $\ket{1/2}_{1}$.

In this way, the HQS indicates the existence of a series of HQS doublet states
\begin{align}
    \bigl\{\ket{J+1/2}_{J},\ket{J+1/2}_{J+1}\bigr\}  ,
    \label{eq:HQdoublet}
\end{align}
with $J\geq 0$, and an HQS singlet state 
\begin{align}
    \ket{0}_{1/2} .
    \label{eq:HQsinglet}
\end{align}
In other words, a HQS doublet is formed for $j\geq 1/2$, while only a HQS singlet is for $j=0$.  

So far we have considered only one state $[\alpha]$ for BMC.
Let us consider the case with multiple number of states $[\alpha_{1}]$, $[\alpha_{2}]$, \dots for BMC.
The heavy hadron can be expanded by the BMC basis set given by
$[\alpha_{1}]_{J-1/2}$, $[\alpha_{2}]_{J-1/2}$, $\dots$ for the brown muck $\alpha_{i}$ with $J-1/2$, and $[\beta_{1}]_{J-1/2}$, $[\beta_{2}]_{J-1/2}$, $\dots$ for the brown muck $\beta_{i}$ with $J+1/2$, which are coupled to the heavy-quark spin 1/2 to form the total spin $J$.
This is expressed as
\begin{align}
    \ket{\rm H}_{J}
    =
    \begin{pmatrix}
    C_{\alpha_{1}}\ket{[\alpha_{1}]_{J-1/2}Q}_{J} \\
    C_{\alpha_{2}}\ket{[\alpha_{2}]_{J-1/2}Q}_{J} \\
    \vdots \\
    C_{\beta_{1}}\ket{[\beta_{1}]_{J+1/2}Q}_{J} \\
    C_{\beta_{2}}\ket{[\beta_{2}]_{J+1/2}Q}_{J} \\
    \vdots \\
    \end{pmatrix} ,
    \label{eq:SCbasis}
\end{align}
with coefficients $C_{\alpha_i}$ and $C_{\beta_{i}}$ for weights of each component $\alpha_{i}$ and $\beta_{i}$ under the normalization condition $\sum_{i} ( |C_{\alpha_{i}}|^{2} + |C_{\beta_{i}}|^{2} ) = 1$.
In the heavy quark limit, the interaction Hamiltonian is independent of the heavy-quark spin and hence it is block-diagonalized into $J-1/2$ sector and $J+1/2$ sector of the brown muck.
Therefore, we obtain two independent states as
\begin{align}
    \ket{J-1/2}_{J}
    &=
    \begin{pmatrix}
    \bar{C}_{\alpha_{1}}\ket{[\alpha_{1}]_{J-1/2}Q}_{J} \\
    \bar{C}_{\alpha_{2}}\ket{[\alpha_{2}]_{J-1/2}Q}_{J} \\
    \vdots \\
    0 \\
    0 \\
    \vdots \\
    \end{pmatrix} ,  \quad
    \ket{J+1/2}_{J}
    =
    \begin{pmatrix}
    0 \\
    0 \\
    \vdots \\
    \bar{C}_{\beta_{1}}\ket{[\beta_{1}]_{J+1/2}Q}_{J} \\
    \bar{C}_{\beta_{2}}\ket{[\beta_{2}]_{J+1/2}Q}_{J} \\
    \vdots \\
    \end{pmatrix} ,
    \label{eq:Ppmulti}
\end{align}
where $\bar{C}_{\alpha_{i}}$ ($\bar{C}_{\beta_{i}}$) represents the relative weight and $\sum_{i} |\bar{C}_{\alpha_{i}}|^{2}  = \sum_{i} |\bar{C}_{\beta_{i}}|^{2} = 1$. These weight factors are not determined simply by the HQS, and depend on the dynamics of the light quark sector. The $J+1$ state in the same model space can have a state
\begin{align}
    \ket{J+1/2}_{J+1}
    &=
    \begin{pmatrix}
    \bar{C}_{\beta_{1}}\ket{[\beta_{1}]_{J+1/2}Q}_{J+1} \\
    \bar{C}_{\beta_{2}}\ket{[\beta_{2}]_{J+1/2}Q}_{J+1} \\
    \vdots \\
    0 \\
    0 \\
    \vdots \\
    \end{pmatrix} ,  \quad
    \ket{J+3/2}_{J+1}
    &=
    \begin{pmatrix}
    0 \\
    0 \\
    \vdots \\
    \bar{C}_{\gamma_{1}}\ket{[\gamma_{1}]_{J+3/2}Q}_{J+1} \\
    \bar{C}_{\gamma_{2}}\ket{[\gamma_{2}]_{J+3/2}Q}_{J+1} \\
    \vdots \\
    \end{pmatrix} ,
    \label{eq:PpJm1multi}
\end{align}
with a similar notation.
We therefore conclude that the structure of the HQS multiplets in Eqs.~\eqref{eq:HQdoublet} and \eqref{eq:HQsinglet} should hold when the brown muck is expanded by several components. Moreover, because the HQS doublet is formed by the same brown muck, the coefficients $\bar{C}_{\beta_{i}}$ in Eq.~\eqref{eq:Ppmulti} should be the same as those in Eq.~\eqref{eq:PpJm1multi}. This means that the wave functions of the HQS doublet are highly correlated with each other.

It should be noted that for a given $J$, it is not necessary that both $\ket{J+1/2}_{J}$ and $\ket{J-1/2}_{J}$ form a hadronic state, because they are made of different brown muck, and it is not necessary that both of them are  stable. On the other hand, if a hadron with spin $J$ exists in the $\ket{J+1/2}_{J}$ ($\ket{J-1/2}_{J}$) channel, there must be a partner $\ket{J+1/2}_{J+1}$ ($\ket{J-1/2}_{J-1}$) with the different total spin $J+1$ ($J-1$). The existence of the HQS doublet in the heavy quark limit leads to an important phenomenological consequence; if we observe a hadron with spin $J$, there can be a spin partner with spin $J\pm 1$ with a similar mass, except for the HQS singlet  of $J=1/2$. In general, the observed hadron with spin $J$ can be either $\ket{J+1/2}_{J}$ or $\ket{J-1/2}_{J}$, but the difference of $\ket{J+1/2}_{J}$ and $\ket{J-1/2}_{J}$ is seen in the wave function.
Such difference should be reflected in the production and decay properties.


In the present discussion, the structure of the brown muck is not specified at all, meaning that the spin degeneracy occurs in the heavy quark limit irrespective of the structure of the brown muck. 
Thanks to this generality, the discussions in this section can be applied not only to conventional hadrons but also to exotic hadrons, multi-hadron states, and their mixtures.

\subsection{Relations between brown-muck component basis and particle basis}

In many discussions of hadronic composites, the wave functions are expressed by the particle basis, namely the relative wave functions between hadrons.
In this subsection we  show a transformation of hadronic states written in terms of a brown muck and a heavy quark to those of a physical particle basis.  

Let us start with an expansion of a hadron state $\ket{\rm H}_{J}$ in terms of a physical particle basis. For example, a heavy baryon state with minimal quark configuration $qqQ$ can be expanded as
\begin{align}
    \ket{\rm H}_{J}
    &=
    \ket{(qqQ)_{J}}
    \oplus
    \sum_{S,L}\ket{(\bar{q}Q)_{s_{M}}(qqq)_{s_{B}}({}^{2S+1}L_{J})}
    \nonumber \\
    &\quad
    \oplus
    \sum_{S,L}\ket{(\bar{q}q)_{s_{M}}(qqQ)_{s_{B}}({}^{2S+1}L_{J})}
    \oplus\dotsb \nonumber \\
    &\equiv  \ket{B_{{\rm Q},J}}
    \oplus\sum_{S,L}\ket{M_{\rm Q}B({}^{2S+1}L_{J})} 
\oplus\sum_{S,L}\ket{MB_{\rm Q}({}^{2S+1}L_{J})}
    \oplus\dotsb ,
    \label{eq:Pbasis}
\end{align}
where the first term corresponds to the ``bare'' heavy baryon state with spin $J$. The bare state does not include the meson-cloud effect, which is represented by the two-body channels in the second (third) term; a heavy meson $M_{\rm Q}$ and a light baryon $B$ (a light meson $M$ and a heavy baryon $B_{\rm Q}$) with relative spin and angular momentum $S$ and $L$. The expansion may also include several $M_{\rm Q}B$ and $MB_{\rm Q}$ channels, as well as many-hadron channels.

The physical hadron including all the virtual states can also be decomposed into the spin-complex basis~\eqref{eq:SCbasis}.
Then, we can relate the particle basis~\eqref{eq:Pbasis} with the spin-complex basis by a unitary matrix $U$ as
\begin{align}
    \begin{pmatrix}
    \ket{B_{{\rm Q},J}} \\
    \ket{M_{\rm Q}B({}^{2S+1}L_{J})} \\
    \vdots \\
    \ket{MB_{\rm Q}({}^{2S+1}L_{J})} \\
    \vdots \\
    \end{pmatrix} 
    &=
    U
    \begin{pmatrix}
    \ket{[\alpha_{1}]_{J-1/2}Q}_{J} \\
    \ket{[\alpha_{2}]_{J-1/2}Q}_{J} \\
    \vdots \\
    \ket{[\beta_{1}]_{J+1/2}Q}_{J} \\
    \ket{[\beta_{2}]_{J+1/2}Q}_{J} \\
    \vdots \\
    \end{pmatrix}  .
    \label{eq:transformation}
\end{align}
Each element of the matrix $U$ can be obtained by the rearrangement of the quark structure. Note that the transformation matrix $U$ is determined, only when the model space is explicitly specified.
In Section~\ref{sec:examples_exotic}, we will show the examples of the basis transformations of an exotic hadron with a $\bar{Q}q$ heavy meson for $M_{\rm Q}$ and a nucleon $N$ for $B$.

The consequences of the HQS become clear by this basis transformation.
Let us suppose that the Hamiltonian $H_{J}$ of the system with total spin $J$ is defined in the particle basis. Using the transformation matrix $U$, we then obtain the Hamiltonian in the 
BMC basis $H^{\rm BMC}_{J}$ as
\begin{align}
    H^{\rm BMC}_{J}
    &=
    U^{-1}
    H_{J}
    U . 
\end{align}
In order to realize the separation of $j_{\pm} = J\pm 1/2$ components in Eq.~\eqref{eq:Ppmulti}, the Hamiltonian in the BMC basis should be block-diagonalized in the heavy quark limit
\begin{align}
    H^{\rm BMC}_{J}
    &=
    \left(
    \begin{array}{cc|cc}
     H_{J,\alpha_{1}}^{\mathrm{BMC}(j_{-})} & \dots & 0  & 0 \\
     \vdots & \ddots & 0 & 0 \\
    \hline
     0 & 0 &  H_{J,\beta_{1}}^{\mathrm{BMC}(j_{+})} & \dots \\
     0 & 0 & \vdots & \ddots
    \end{array}
    \right) \nonumber \\
    &\equiv 
    \begin{pmatrix}
    H_{J}^{\mathrm{BMC}(j_{-})} & 0 \\
    0 & H_{J}^{\mathrm{BMC}(j_{+})} 
    \end{pmatrix} .
    \label{eq:HQsym1}
\end{align}
As we will see below, the block-diagonalization of the Hamiltonian is indeed possible, as far as the Hamiltonian $H_{J}$ is constructed in accordance with the HQS. Note that the off-diagonal terms within each $H_{J}^{\mathrm{BMC}(j_{\pm})}$ are not constrained by the HQS, and it is this term that determines the coefficients $\bar{C}_{\beta_{i}}$ in the bound state wave function~\eqref{eq:Ppmulti}.

The relation to $\ket{{\rm H}_{}}_{J+1}$ is also seen in the BMC basis. The Hamiltonian with spin $J+1$ in the heavy quark limit can be block-diagonalized into 
\begin{align}
    H^{\rm BMC}_{J+1}
    &=
    \begin{pmatrix}
    H_{J+1}^{\mathrm{BMC}(j_{+})} & 0 \\
    0 & H_{J+1}^{\mathrm{BMC}(j_{++})} 
    \end{pmatrix} ,
    \label{eq:HQsym1.5}
\end{align}
with $j_{+}=J+1/2$ and $j_{++}=J+3/2$.
The degeneracy of $\ket{{\rm H}_{}}_{J}$ and $\ket{{\rm H}_{}}_{J+1}$ requires 
\begin{align}
    H_{J}^{\mathrm{BMC}(j_{+})}
    &=
    H_{J+1}^{\mathrm{BMC}(j_{+})} .
    \label{eq:HQsym2}
\end{align}
Strictly, it is sufficient for us that eigenvalues in the two Hamiltonian are the same.
To express this, we introduce $\approx$ as
\begin{eqnarray}
H_{J}^{\mathrm{BMC}(j_{+})} \approx H_{J+1}^{\mathrm{BMC}(j_{+})},
\label{eq:HQsym2.5}
\end{eqnarray}
in the sense that the eigenvalues of $H_{J}^{\mathrm{BMC}(j_{+})}$ are as same as those of $H_{J+1}^{\mathrm{BMC}(j_{+})}$.
For instance, the signs of the off-diagonal components are irrelevant for the eigenvalues. We will see that this happens in the case of the exotic hadron in Section~\ref{sec:examples_exotic}.
Thus, the diagonalization in Eqs.~\eqref{eq:HQsym1} and \eqref{eq:HQsym1.5} and the coincidence of the Hamiltonian in Eq.~\eqref{eq:HQsym2.5} are the consequence of the HQS of the Hamiltonian in the BMC basis.

The particle basis is also useful to analyze the wave function of the eigenstates. For instance, the state $\ket{J+1/2}_{J}$ given by the brown muck in Eq.~\eqref{eq:Ppmulti} is expressed by the particle basis as
\begin{align}
    U\ket{J+1/2}_{J}
    =\begin{pmatrix}
    D_{B_{\rm Q}}\ket{B_{{\rm Q},J}} \\
    D_{M_{\rm Q}B}\ket{M_{\rm Q}B({}^{2S+1}L_{J})} \\
    \vdots \\
    D_{MB_{\rm Q}}\ket{MB_{\rm Q}({}^{2S+1}L_{J})} \\
    \vdots \\
    \end{pmatrix} ,
\end{align}
with
\begin{align}
    D_{X}
    =\sum_{\beta}U_{X,\beta}\bar{C}_{\beta} .
\end{align}
The weights $|D_{X}|^{2}$ ($X=B_{\rm Q}$, $M_{\rm Q}B$, $M_{\rm Q}^{\prime}B^{\prime}$, \dots, $MB_{\rm Q}$, $M^{\prime}B_{\rm Q}^{\prime}$, \dots) represents the probability of finding the state $\ket{X}$ in the eigenstate wave function. Because the transformation matrix $U$ is determined only by the symmetry argument, the structure of the eigenstate is specified once the wave function is determined in terms of the brown muck $\ket{J+1/2}_{J}$. 

\subsection{Mass spectrum of brown muck}\label{sec:massformula}

So far we have discussed the symmetry aspects of heavy hadrons.  
In the end of this section, we discuss some dynamical aspects, and evaluate the mass of the brown muck, which is also useful to the study of the spin-complex in the heavy quark limit.  

In order to define the mass of the brown muck, we decompose the mass of the heavy hadron in terms of $1/m_{\rm Q}$.
Based on the effective Lagrangian (\ref{eq:HQET}), we can expand the mass of the hadron H containing a heavy quark Q as~\cite{Manohar:2000dt}
\begin{align}
M_{\mathrm{H}} = m_{\rm Q} + \bar{\Lambda} - \frac{\lambda_{1}}{2m_{\rm Q}} + 4\vec{S} \!\cdot\! \vec{j} \frac{\lambda_{2}(\mu)}{2m_{\rm Q}}
 + \mathcal{O}(1/m_{\rm Q}^2),
\label{eq:mass_formula}
\end{align}
where we define, in the rest frame with $v_{\mathrm{r}}=(1,\vec{0}\,)$,
\begin{align}
   \bar{\Lambda}
   &= \frac{1}{2} \langle \mathrm{H}_{v_{\mathrm{r}}} |
   \mathcal{H}_{0}| \mathrm{H}_{v_{\mathrm{r}}} \rangle ,
   \label{eq:lambda0} \\
   \lambda_{1}
   &= \frac{1}{2} \langle \mathrm{H}_{v_{\mathrm{r}}} | 
   \overline{Q}_{v_{\mathrm{r}}} (iD_{\perp})^2 Q_{v_{\mathrm{r}}} 
   | \mathrm{H}_{v_{\mathrm{r}}} \rangle , 
   \label{eq:chromoelectric0} \\
   8 \vec{S} \!\cdot\! \vec{j} \lambda_{2}(\mu)
   &= \frac{1}{2} c(\mu) \langle \mathrm{H}_{v_{\mathrm{r}}} | 
   \overline{Q}_{v_{\mathrm{r}}} g_{\mathrm{s}} 
   \sigma_{\alpha\beta} G^{\alpha\beta} Q_{v_{\mathrm{r}}} 
   | \mathrm{H}_{v_{\mathrm{r}}} \rangle , \label{eq:chromomagnetic0}
\end{align}
with denoting the hadron state by $| \mathrm{H}_{v_{\mathrm{r}}} \rangle$. The factor $1/2$ is multiplied due to the normalization of the wave function $\langle \mathrm{H}_{v^{\prime}}(k^{\prime})|\mathrm{H}_{v}(k)\rangle=2v^{0}\delta_{vv^{\prime}}(2\pi)^{3}\delta^{3}(\bm{k}-\bm{k}^{\prime})$. Here $\mathcal{H}_{0}$ is the Hamiltonian obtained from the leading (first) term in $\mathcal{L}_{\mathrm{HQET}}$ (\ref{eq:HQET}) and the light degrees of freedom. In Eq.~(\ref{eq:chromomagnetic0}), $\vec{S}$ and $\vec{j}$ are the operators for the spin of the heavy quark Q and the total spin of the  brown muck \cite{Neubert:1993mb,Manohar:2000dt}, respectively. The dependence of $\mu$ on $\lambda_{2}(\mu)$ originates from 
 the Wilson coefficient $c(\mu)$, because the matching with QCD is done at the energy scale $\mu\simeq m_{\rm Q}$~\cite{Manohar:2000dt}. 
We consider $\lambda_{2}(m_{\rm c})$ and $\lambda_{2}(m_{\rm b})$ for charm and bottom, respectively.
 
The expansion~\eqref{eq:mass_formula} is useful to analyze the QCD properties in heavy hadrons. In fact, $\bar{\Lambda}$, $\lambda_{1}$ and $\lambda_{2}(m_{\rm Q})$ are concerned with the scale anomaly in QCD \cite{Bigi:1994ga,Bigi:1997fj}, the chromoelectric gluons \cite{Neubert:1993zc} (see also \cite{Bigi:1997fj}) and the chromomagnetic gluons, respectively.
There are discussions to utilize the heavy hadrons to probe the gluon dynamics in the (multi-)hadron systems and nuclear systems with a heavy hadron \cite{Yasui:2013iga}.

Now we consider the heavy quark limit where the mass of the hadron H is given by $m_{\rm Q}+\bar{\Lambda}$. We notice that there is no spin dependence, as required by the HQS. In addition, we have defined the brown muck as everything except for the heavy quark, and the mixing with other $j$ components vanishes in the heavy quark limit. Thus, we shall identify $\bar{\Lambda}$ as the mass of the brown muck. 
 We will discuss the way to extract $\bar{\Lambda}$ from the experimental spectrum as well as from the prediction of theoretical models of heavy hadrons in Section~\ref{sec:spectrum}.

\section{Multi-hadrons with a heavy antiquark}\label{sec:examples_exotic}

For the formalism given in the previous section, we will give concrete examples of spin-complex.
We consider exotic baryons with a heavy antiquark with the minimal quark configuration $\bar{Q}qqqq$, and discuss the two-body states whose model space is supplied by a heavy meson and a nucleon.
This state is exotic, because it cannot be reduced to the normal baryons with three quarks as a minimal valence component. In Section~\ref{sec:decomposition}, we will analytically decompose the meson-baryon basis into the spin-complex basis, and show that the HQS doublets/singlets appear. In Section~\ref{sec:model}, we will discuss results by the one-pion exchange potential from the heavy meson effective theory in analytical and numerical calculations, including the $1/m_{\rm Q}$ corrections.

\subsection{Wave function of spin-complex}\label{sec:decomposition} 

We discuss the spin structures of exotic baryons with $\bar{Q}qqqq$. In the current discussion, we assume that there exist exotic baryons which are composed of a heavy meson $\bar{P}\sim (\bar{Q}q)_{\mathrm{spin}\,0}$ or $\bar{P}^{*} \sim (\bar{Q}q)_{\mathrm{spin}\,1}$ and a nucleon $N$. Here $\bar{P}^{(\ast)}$ stands for $\bar{P}$ or $\bar{P}^{\ast}$.\footnote{Note that this notation is different from our previous paper \cite{Yasui:2013vca}, where $P^{(\ast)}$ was used to mean $\bar{Q}q$.} Since the $\bar{P}$ and $\bar{P}^{\ast}$ belongs to the same HQS doublet, we consider the coupled-channel problem of $\bar{P}N$ and $\bar{P}^{\ast}N$ and denote their superposition as $\bar{P}^{(\ast)}N$. We do not include three- or many-hadron channels, higher energy excited channels, and possible compact five-quark states. The exotic baryon is then considered to have a hadronic molecule structure, and the corresponding brown muck is identified as a spin-complex of a quark $q$ and a nucleon $N$.

The $\bar{P}^{(\ast)}N$ states are classified by the quantum numbers: total spin $J$, parity $P$ and isospin $I=0,1$. The relevant channels for given quantum numbers $J^P$ up to $J=7/2$ are summarized in Table~\ref{tab:channel}. As emphasized in Refs.~\cite{Yasui:2009bz,Yamaguchi:2011xb,Yamaguchi:2011qw}, the state mixing between $\bar{P}N$ and $\bar{P}^{\ast}N$ is important due to the mass degeneracy of $\bar{P}N$ and $\bar{P}^{\ast}N$ in heavy quark limit. Moreover, the states with angular momenta $L$ and $L\pm2$ can be also mixed
 by the tensor force. In what follows, we discuss the transformation of the particle basis to the spin-complex basis (the BMC basis spanned exclusively by the spin-complex channels). The isospin indices will be suppressed, which does not affect the basis transformation.

\begin{table}
 \caption{\label{table_qnumbers}Relevant coupled channels of $\bar{P}^{(\ast)}N$ systems for a given quantum number $J^P$. }
{\renewcommand{\arraystretch}{1.5}
 \begin{tabular}{c|cccc}
\hline
$J^P$ &  \multicolumn{4}{c}{channels} \\
\hline
$1/2^-$ &$\bar{P}N(^2{\mathrm S}_{1/2})$&$\bar{P}^\ast N(^2{\mathrm S}_{1/2})$&$\bar{P}^\ast N(^4{\mathrm D}_{1/2})$& \\
$3/2^-$ & $\bar{P}N(^2{\mathrm D}_{3/2})$&$\bar{P}^\ast N(^4{\mathrm S}_{3/2})$&$\bar{P}^\ast
	      N(^4{\mathrm D}_{3/2})$&$\bar{P}^\ast N(^2{\mathrm D}_{3/2})$ \\
$5/2^-$&$\bar{P}N(^2{\mathrm D}_{5/2})$&$\bar{P}^\ast N(^2{\mathrm D}_{5/2})$&$\bar{P}^\ast
	      N(^4{\mathrm D}_{5/2})$&$\bar{P}^\ast N(^4{\mathrm G}_{5/2})$ \\
$7/2^-$&$\bar{P}N(^2{\mathrm G}_{7/2})$&$\bar{P}^\ast N(^4{\mathrm D}_{7/2})$&$\bar{P}^\ast
	      N(^2{\mathrm G}_{7/2})$&$\bar{P}^\ast N(^4{\mathrm G}_{7/2})$ \\
\hline
$1/2^+$ &$\bar{P}N(^2{\mathrm P}_{1/2})$&$\bar{P}^\ast N(^2{\mathrm P}_{1/2})$&$\bar{P}^\ast N(^4{\mathrm P}_{1/2})$& \\
$3/2^+$ & $\bar{P}N(^2{\mathrm P}_{3/2})$&$\bar{P}^\ast N(^2{\mathrm P}_{3/2})$&$\bar{P}^\ast
	      N(^4{\mathrm P}_{3/2})$&$\bar{P}^\ast N(^4{\mathrm F}_{3/2})$ \\
$5/2^+$&$\bar{P}N(^2{\mathrm F}_{5/2})$&$\bar{P}^\ast N(^4{\mathrm P}_{5/2})$&$\bar{P}^\ast
	      N(^2{\mathrm F}_{5/2})$&$\bar{P}^\ast N(^4{\mathrm F}_{5/2})$ \\
$7/2^+$&$\bar{P}N(^2{\mathrm F}_{7/2})$&$\bar{P}^\ast N(^2{\mathrm F}_{7/2})$&$\bar{P}^\ast
	      N(^4{\mathrm F}_{7/2})$&$\bar{P}^\ast N(^4{\mathrm H}_{7/2})$ \\
\hline
 \end{tabular}
}
 \label{tab:channel}
 \end{table}

\subsubsection{Negative parity channels}

Let us discuss negative parity states; $J^{P}=1/2^{-}$, $3/2^{-}$, $5/2^{-}$ and $7/2^{-}$. First we demonstrate concretely the transformation of the basis for $1/2^{-}$. From Table \ref{tab:channel}, the channels for $1/2^-$ in the particle basis are
\begin{eqnarray}
 \bar{P}N(^2{\rm S}_{1/2}),\bar{P}^{\ast}N (^2{\rm S}_{1/2}), \bar{P}^{\ast}N(^4{\rm D}_{1/2}).
 \label{eq:particle_basis_1/2^-}
\end{eqnarray}
In view of the HQS, the wave function is decomposed into the product of a heavy antiquark $\bar{Q}$ and a spin-complex which is composed of the light quarks and gluons in $\bar{P}^{\ast}$ and the nucleon $N$. We denote the spin-complex $[\alpha]$ by $[Nq]$, where $q$ stands for the light components in the $\bar{P}^{(\ast)}$ meson with spin $1/2$. In order to specify the structure and the quantum numbers of the spin-complex, we introduce a notation for the state 
\begin{eqnarray}
[Nq]^{(s_l ,L)}_{j^{\cal P}},
\end{eqnarray}
with $s_l$ being the sum of the spins of $N$ and $q$, $L$ the relative angular momentum between $N$ and $q$, and $j^{\cal P}$  the total spin and parity of the spin-complex. Using this notation, we describe the wave function of the $\bar{P}^{(\ast)}N$ system in the spin-complex basis as
\begin{eqnarray}
 \ket{[Nq]^{(s_l ,L)}_{j^{\cal P}}\bar{Q}}_{J^P} .
\end{eqnarray}
The explicit wave functions of the $1/2^{-}$ state in the spin-complex basis are given as
\begin{eqnarray}
\ket{[Nq]^{(0,{\rm S})}_{0^+}\bar{Q}}_{1/2^-},\quad
\ket{[Nq]^{(1,{\rm S})}_{1^+}\bar{Q}}_{1/2^-},\quad
 \ket{[Nq]^{(1,{\rm D})}_{1^+}\bar{Q}}_{1/2^-} .
 \label{eq:spincomplex_basis_1/2^-}
\end{eqnarray}
We find that there are two kinds of components with $j^{\cal P}=0^{+}$ and $1^{+}$ in the $J^{P}=1/2^{-}$ channel. The particle basis in Eq.~(\ref{eq:particle_basis_1/2^-}) is decomposed to spin-complex basis in Eq.~(\ref{eq:spincomplex_basis_1/2^-}) by utilizing the standard spin-recoupling formula. We denote this transformations by a unitary matrix $U_{J^P}$ as
\begin{eqnarray}
\left(
\renewcommand{\arraystretch}{1.5}
\begin{array}{c}
 \ket{ \bar{P}N(^{2}{\mathrm S}_{1/2}) } \\
 \ket{ \bar{P}^{\ast}N(^{2}{\mathrm S}_{1/2}) } \\
 \ket{ \bar{P}^{\ast}N(^{4}{\mathrm D}_{1/2}) }
\end{array}
\right) = U_{1/2^-}
\left(
\begin{array}{c}
 \ket{ [Nq]^{(0,\mathrm S)}_{0^{+}} \, \bar{Q} }_{1/2^-} \\
 \ket{ [Nq]^{(1, {\mathrm S})}_{1^{+}} \bar{Q} }_{1/2^-} \\
 \ket{ [Nq]^{(1,{\mathrm D})}_{1^{+}} \bar{Q} }_{1/2^-}
\end{array}
\right),
\label{eq:transformation_1/2^-}
\end{eqnarray}
where the explicit form is given by
\begin{eqnarray}
U_{1/2^-}=
\left(
\begin{array}{ccc}
 -\frac{1}{2} & \frac{\sqrt{3}}{2} & 0  \\
 \frac{\sqrt{3}}{2} & \frac{1}{2} & 0  \\
 0 & 0 & -1   
\end{array}
\right).
\label{eq:U_1/2-}
\end{eqnarray}

For the other spin states ($J=3/2$, $5/2$, $7/2$), the particle basis and the spin-complex basis are related as follows,
\begin{eqnarray}
\left(
\renewcommand{\arraystretch}{1.5}
\begin{array}{c}
 \ket{ \bar{P}N(^{2}{\mathrm D}_{3/2}) } \\
 \ket{ \bar{P}^{\ast}N(^{4}{\mathrm S}_{3/2}) } \\
 \ket{ \bar{P}^{\ast}N(^{4}{\mathrm D}_{3/2}) } \\
 \ket{ \bar{P}^{\ast}N(^{2}{\mathrm D}_{3/2}) }
\end{array}
\right) = U_{3/2^-}
\left(
\begin{array}{c}
 \ket{ [Nq]^{(1, {\mathrm S})}_{1^{+}} \bar{Q} }_{3/2^-} \\
 \ket{ [Nq]^{(1, {\mathrm D})}_{1^{+}} \bar{Q} }_{3/2^-} \\
 \ket{ [Nq]^{(0, {\mathrm D})}_{2^{+}} \bar{Q} }_{3/2^-} \\
 \ket{ [Nq]^{(1, {\mathrm D})}_{2^{+}} \bar{Q} }_{3/2^-} 
\end{array}
\right),
\label{eq:transformation_3/2^-}
\end{eqnarray}
with
\begin{eqnarray}
U_{3/2^-}=
\left(
\begin{array}{cccc}
 0 & \frac{\sqrt{6}}{4} & \frac{1}{2} & \frac{\sqrt{6}}{4} \\
 1 & 0 & 0 & 0 \\
 0 & \frac{1}{\sqrt{2}} & 0 & -\frac{1}{\sqrt{2}} \\
 0 & \frac{1}{2\sqrt{2}} & - \frac{\sqrt{3}}{2} & \frac{1}{2\sqrt{2}}
\end{array}
\right),
\label{eq:U_3/2-}
\end{eqnarray}
for $3/2^{-}$,
\begin{eqnarray}
\renewcommand{\arraystretch}{1.5}
 \begin{pmatrix}
  \ket{\bar{P}N(^2{\rm D}_{5/2})} \\
  \ket{\bar{P}^{\ast}N(^2{\rm D}_{5/2})} \\
  \ket{\bar{P}^{\ast}N(^4{\rm D}_{5/2})} \\
  \ket{\bar{P}^{\ast}N(^4{\rm G}_{5/2})}
 \end{pmatrix}
= U_{5/2^-} 
 \begin{pmatrix}
  \ket{[Nq]^{(0,{\rm D})}_{2^+}\bar{Q}}_{5/2^-}  \\
  \ket{[Nq]^{(1,{\rm D})}_{2^+}\bar{Q}}_{5/2^-} \\
  \ket{[Nq]^{(1,{\rm D})}_{3^+}\bar{Q}}_{5/2^-} \\
  \ket{[Nq]^{(1,{\rm G})}_{3^+}\bar{Q}}_{5/2^-}
 \end{pmatrix},
\label{eq:transformation_5/2^-}
\end{eqnarray}
with
\begin{eqnarray}
 U_{5/2^-} = 
\left(
\begin{array}{cccc}
 -\frac{1}{2} & \frac{1}{\sqrt{6}} & \frac{\sqrt{21}}{6} & 0 \\
 \frac{\sqrt{3}}{2} & \frac{1}{3 \sqrt{2}} & \frac{\sqrt{7}}{6} & 0 \\
 0 & \frac{\sqrt{7}}{3} & -\frac{\sqrt{2}}{3} & 0 \\
 0 & 0 & 0 & -1
\end{array}
\right),
\end{eqnarray}
for $5/2^{-}$, and
\begin{eqnarray}
\renewcommand{\arraystretch}{1.5}
 \begin{pmatrix}
  \ket{\bar{P}N(^2{\rm G}_{7/2})} \\
  \ket{\bar{P}^{\ast}N(^4{\rm D}_{7/2})} \\
  \ket{\bar{P}^{\ast}N(^2{\rm G}_{7/2})} \\
  \ket{\bar{P}^{\ast}N(^4{\rm G}_{7/2})}
 \end{pmatrix} =
U_{7/2^-}
\begin{pmatrix}
 \ket{[Nq]^{(1,{\rm D})}_{3^+}\bar{Q}}_{7/2^-} \\
 \ket{[Nq]^{(1,{\rm G})}_{3^+}\bar{Q}}_{7/2^-} \\
 \ket{[Nq]^{(0,{\rm G})}_{4^+}\bar{Q}}_{7/2^-} \\
 \ket{[Nq]^{(1,{\rm G})}_{4^+}\bar{Q}}_{7/2^-}
\end{pmatrix},
\label{eq:transformation_7/2^-}
\end{eqnarray}
with
\begin{eqnarray}
 U_{7/2^-} = 
\left(
\begin{array}{cccc}
 0 & \frac{\sqrt{7}}{4} & \frac{1}{2} & \frac{\sqrt{5}}{4} \\
 1 & 0 & 0 & 0 \\
 0 & \frac{\sqrt{21}}{12} & -\frac{\sqrt{3}}{2} & \frac{\sqrt{15}}{12} \\
 0 & \frac{\sqrt{15}}{6} & 0 & -\frac{\sqrt{21}}{6}
\end{array}
\right),
\end{eqnarray}
for $7/2^{-}$. Each $J$ channel contains the channels with $j=J\pm 1/2$. For higher $J$, the decomposition will be given similarly.

\subsubsection{Positive parity channels}

The states with positive parity can be analyzed in the same way. The wave functions are transformed from the particle basis into the spin-complex basis by a unitary matrix $U_{J^P}$ as
\begin{eqnarray}
\renewcommand{\arraystretch}{1.5}
 \begin{pmatrix}
   \ket{\bar{P}N (^2{\rm P}_{1/2})} \\
  \ket{\bar{P}^{\ast}N(^2{\rm P}_{1/2})} \\
\ket{\bar{P}^{\ast}N(^4{\rm P}_{1/2})} 
 \end{pmatrix}
= U_{1/2^+} 
\begin{pmatrix}
 \ket{[Nq]^{(1,{\rm P})}_{0^-}\bar{Q}}_{1/2^+} \\
 \ket{[Nq]^{(0,{\rm P})}_{1^-}\bar{Q}}_{1/2^+} \\
 \ket{[Nq]^{(1,{\rm P})}_{1^-}\bar{Q}}_{1/2^+} 
\end{pmatrix} \, ,
\label{eq:transformation_1/2^+}
\end{eqnarray}
with %
\begin{eqnarray}
 U_{1/2^+} = 
\left(
\begin{array}{ccc}
 \frac{1}{2} & \frac{1}{2} & \frac{1}{\sqrt{2}} \\
 \frac{1}{2 \sqrt{3}} & -\frac{\sqrt{3}}{2} & \frac{1}{\sqrt{6}} \\
 \sqrt{\frac{2}{3}} & 0 & -\frac{1}{\sqrt{3}}
\end{array}
\right) \, ,
\end{eqnarray}
for $1/2^+$, 
\begin{eqnarray}
\renewcommand{\arraystretch}{1.5}
 \begin{pmatrix}
  \ket{\bar{P}N (^2{\rm P}_{3/2})} \\
  \ket{\bar{P}^{\ast}N (^2{\rm P}_{3/2})} \\
  \ket{\bar{P}^{\ast}N (^4{\rm P}_{3/2})} \\
  \ket{\bar{P}^{\ast}N (^4{\rm F}_{3/2})} 
 \end{pmatrix}
= U_{3/2^+}
\begin{pmatrix}
 \ket{[Nq]^{(0,{\rm P})}_{1^-}\bar{Q}}_{3/2^+} \\
 \ket{[Nq]^{(1,{\rm P})}_{1^-}\bar{Q}}_{3/2^+} \\
 \ket{[Nq]^{(1,{\rm P})}_{2^-}\bar{Q}}_{3/2^+} \\
 \ket{[Nq]^{(1,{\rm F})}_{2^-}\bar{Q}}_{3/2^+}
\end{pmatrix},
\label{eq:transformation_3/2^+}
\end{eqnarray}
with %
\begin{eqnarray}
 U_{3/2^+} = 
\left(
\begin{array}{cccc}
 -\frac{1}{2} & \frac{1}{2 \sqrt{2}} & \frac{\sqrt{10}}{4} & 0 \\
 \frac{\sqrt{3}}{2} & \frac{1}{2 \sqrt{6}} & \frac{\sqrt{5}}{2\sqrt{6}} & 0 \\
 0 & \sqrt{\frac{5}{6}} & -\frac{1}{\sqrt{6}} & 0 \\
 0 & 0 & 0 & -1
\end{array}
\right),
\end{eqnarray}
for $3/2^+$,
\begin{eqnarray}
\renewcommand{\arraystretch}{1.5}
 \begin{pmatrix}
  \ket{\bar{P}N(^2{\rm F}_{5/2})} \\
  \ket{\bar{P}^{\ast}N(^4{\rm P}_{5/2})} \\
  \ket{\bar{P}^{\ast}N(^2{\rm F}_{5/2})} \\
  \ket{\bar{P}^{\ast}N(^4{\rm F}_{5/2})}
 \end{pmatrix}
= U_{5/2^+} 
\begin{pmatrix}
 \ket{[Nq]^{(1,{\rm P})}_{2^-}\bar{Q}}_{5/2^+} \\
 \ket{[Nq]^{(1,{\rm F})}_{2^-}\bar{Q}}_{5/2^+} \\
 \ket{[Nq]^{(0,{\rm F})}_{3^-}\bar{Q}}_{5/2^+} \\
 \ket{[Nq]^{(1,{\rm F})}_{3^-}\bar{Q}}_{5/2^+} 
\end{pmatrix},
\label{eq:transformation_5/2^+}
\end{eqnarray}
with
\begin{eqnarray}
 U_{5/2^+} = 
\left(
\begin{array}{cccc}
 0 & \frac{\sqrt{15}}{6} & \frac{1}{2} & \frac{1}{\sqrt{3}} \\
 1 & 0 & 0 & 0 \\
 0 & \frac{\sqrt{5}}{6} & -\frac{\sqrt{3}}{2} & \frac{1}{3} \\
 0 & \frac{2}{3} & 0 & -\frac{\sqrt{5}}{3}
\end{array}
\right),
\end{eqnarray}
for $5/2^{+}$, and
\begin{eqnarray}
\renewcommand{\arraystretch}{1.5}
 \begin{pmatrix}
  \ket{\bar{P}N(^2{\rm F}_{7/2})} \\
  \ket{\bar{P}^{\ast}N(^2{\rm F}_{7/2})} \\
  \ket{\bar{P}^{\ast}N(^4{\rm F}_{7/2})} \\
  \ket{\bar{P}^{\ast}N(^4{\rm H}_{7/2})}
 \end{pmatrix} 
= U_{7/2^+} 
\begin{pmatrix}
 \ket{[Nq]^{(0,{\rm F})}_{3^-}\bar{Q}}_{7/2^+}  \\
 \ket{[Nq]^{(1,{\rm F})}_{3^-}\bar{Q}}_{7/2^+}  \\
 \ket{[Nq]^{(1,{\rm F})}_{4^-}\bar{Q}}_{7/2^+}  \\
 \ket{[Nq]^{(1,{\rm H})}_{4^-}\bar{Q}}_{7/2^+} 
\end{pmatrix},
\label{eq:transformation_7/2^+}
\end{eqnarray}
with
\begin{eqnarray}
 U_{7/2^+} = 
\left(
\begin{array}{cccc}
 -\frac{1}{2} & \frac{\sqrt{3}}{4} & \frac{3}{4} & 0 \\
 \frac{\sqrt{3}}{2} & \frac{1}{4} & \frac{\sqrt{3}}{4} & 0 \\
 0 & \frac{\sqrt{3}}{2} & -\frac{1}{2} & 0 \\
 0 & 0 & 0 & -1
\end{array}
\right),
\end{eqnarray}
for $7/2^{+}$. Generalization to higher $J$ is straightforward.

\subsubsection{Decomposition of wave functions}
\label{sec:fractions}

The spin-complex basis gives a useful information to investigate properties of the wave function in the particle basis. 
As discussed in Section 2, in the heavy quark limit, there are HQS singlets and doublets.  
For the negative parity sector, the HQS singlet state has $J^P = 1/2^-$ 
with the spin-complex of $j^{\cal P} = 0^+$:
\begin{eqnarray}
 \ket{0^{+}}_{1/2^{-}} = \ket{[Nq]^{(0,{\rm S})}_{0^+}\bar{Q}}_{1/2^-}.
\label{eq:spin_singlet_PbarN}
\end{eqnarray}
We also consider the HQS doublet states with $1/2^{-}$ and $3/2^{-}$ containing the spin-complex with $1^{+}$, which are given by superpositions of two components $[Nq]^{(1,{\rm S})}_{1^+}$ and $[Nq]^{(1,{\rm D})}_{1^+}$:
\begin{align}
 \ket{1^{+}}_{1/2^{-}} &= \sin \theta  \ket{[Nq]^{(1,{\rm S})}_{1^+}\bar{Q}}_{1/2^-} 
+ \cos \theta \ket{[Nq]^{(1,{\rm D})}_{1^+}\bar{Q}}_{1/2^-},
\label{eq:spin_doublet_PbarN_1/2} \\
\ket{1^{+}}_{3/2^{-}} &= \sin \theta  \ket{[Nq]^{(1,{\rm S})}_{1^+}\bar{Q}}_{3/2^-} 
+ \cos \theta \ket{[Nq]^{(1,{\rm D})}_{1^+}\bar{Q}}_{3/2^-}. 
\label{eq:spin_doublet_PbarN_3/2}
\end{align}
The mixing angle $\theta$ determines to $\bar{C}_{\alpha_{i}}$ in Eq.~\eqref{eq:Ppmulti}, which depends on the dynamics in the light components. 
Using Eqs.~(\ref{eq:transformation_1/2^-}) and (\ref{eq:U_1/2-}), we obtain
\begin{eqnarray}
 \ket{[Nq]^{(0,{\rm S})}_{0^+}\bar{Q}}_{1/2^-} = -\frac{1}{2} \ket{\bar{P}N(^2{\rm S}_{1/2})} +\frac{\sqrt{3}}{2}
\ket{\bar{P}^{\ast}N(^2{\rm S}_{1/2})},
\label{eq:transformation_1/2^-_0^+}
\end{eqnarray}
and
\begin{align}   
 \ket{[Nq]^{(1,{\rm S})}_{1^+}\bar{Q}}_{1/2^-} &= \frac{\sqrt{3}}{2}\ket{\bar{P}N(^2{\rm S}_{1/2})} 
+\frac{1}{2}\ket{\bar{P}^{\ast}N(^2{\rm S}_{1/2})},
\label{eq:transformation_1/2^-_1^+_1} \\
 \ket{[Nq]^{(1,{\rm D})}_{1^+}\bar{Q}}_{1/2^-} &= - \ket{\bar{P}^{\ast}N(^4{\rm D}_{1/2})},
\label{eq:transformation_1/2^-_1^+_2}
\end{align}
for $1/2^-$ from Eq.~(\ref{eq:transformation_1/2^-}) and 
\begin{align}
 \ket{[Nq]^{(1,{\rm S})}_{1^+}\bar{Q}}_{3/2^-} &= \ket{\bar{P}^{\ast}N(^4{\rm S}_{3/2})},
\label{eq:transformation_3/2^-_1^+_1} \\
 \ket{[Nq]^{(1,{\rm D})}_{1^+}\bar{Q}}_{3/2^-} &= \frac{\sqrt{6}}{4} \ket{\bar{P}N(^2{\rm D}_{3/2})}
 +\frac{\sqrt{2}}{4} \ket{\bar{P}^{\ast}N(^2{\rm D}_{3/2})}
 \nonumber \\
 &\quad +\frac{\sqrt{2}}{2} \ket{\bar{P}^{\ast}N(^4{\rm D}_{3/2})},
 \label{eq:transformation_3/2^-_1^+_2}
\end{align}
for $3/2^-$ from Eq.~(\ref{eq:transformation_3/2^-}). Therefore, in the particle basis, the HQS singlet state with $1/2^{-}$ and the HQS doublet states with $1/2^{-}$ and $3/2^{-}$ are expressed as 
\begin{eqnarray}
\ket{0^{+}}_{1/2^{-}} = -\frac{1}{2} \ket{\bar{P}N(^2{\rm S}_{1/2})} +\frac{\sqrt{3}}{2}
\ket{\bar{P}^{\ast}N(^2{\rm S}_{1/2})},
\label{eq:spin_doublet_PbarN_1/2_0}
\end{eqnarray}
and
\begin{align}
 \ket{1^{+}}_{1/2^{-}} 
 & = \sin \theta \left( \frac{\sqrt{3}}{2}  \ket{\bar{P}N(^2{\rm S}_{1/2})} 
+\frac{1}{2} \ket{\bar{P}^{\ast}N(^2{\rm S}_{1/2})} \right)  \nonumber \\
&\quad - \cos \theta \ket{\bar{P}^{\ast}N (^4{\rm D}_{1/2})},
\label{eq:spin_doublet_PbarN_1/2_1} \\
 \ket{1^{+}}_{3/2^{-}} 
&=\sin \theta \ket{\bar{P}^{\ast}N(^4{\rm S}_{3/2})} 
 + \cos \theta \Biggl( \frac{\sqrt{6}}{4} \ket{\bar{P}N(^2{\rm D}_{3/2})} 
 \nonumber \\
&\quad +\frac{\sqrt{2}}{4} \ket{\bar{P}^{\ast}N(^2{\rm D}_{3/2})}
+\frac{\sqrt{2}}{2} \ket{\bar{P}^{\ast}N(^4{\rm D}_{3/2})} \Biggr).
\label{eq:spin_doublet_PbarN_3/2_1}
\end{align}
It is important that the ratios of the particle contents in the heavy quark limit are uniquely determined from the symmetry argument.

Now, let us consider the HQS singlet state in Eq.~(\ref{eq:spin_singlet_PbarN}). From the coefficients in Eq.~(\ref{eq:spin_doublet_PbarN_1/2_0}), the ratio of the components in the particle basis is 
\begin{eqnarray}
f(\bar{P}N(^2{\rm S}_{1/2})) : f(\bar{P}^{\ast}N(^2{\rm S}_{1/2})) = 1:3.
\label{eq:fraction_1}
\end{eqnarray}
Next, we consider the HQS doublet states in Eqs.~(\ref{eq:spin_doublet_PbarN_1/2}) and (\ref{eq:spin_doublet_PbarN_3/2}). The ratios of the components in the particle basis are 
\begin{eqnarray}
f(\bar{P}N(^2{\rm S}_{1/2})) : f(\bar{P}^{\ast}N(^2{\rm S}_{1/2})) = 3:1,
\label{eq:fraction_2}
\end{eqnarray}
for $J^P=1/2^-$ from Eq.~(\ref{eq:spin_doublet_PbarN_1/2_1}) and 
\begin{eqnarray}
f(\bar{P}N(^2{\rm D}_{3/2})) : f(\bar{P}^{\ast}N(^2{\rm D}_{3/2}) : f(\bar{P}^{\ast}N(^4{\rm D}_{3/2})) = 3 : 1 : 4,
\label{eq:fraction_3}
\end{eqnarray}
for $J^P=3/2^-$ from Eq.~(\ref{eq:spin_doublet_PbarN_3/2_1}). Those fractions are independent of the mixing angle $\theta$, because they are derived directly from the HQS, and hence they are model-independent results. The mixing angle $\theta$ is not determined from the HQS, but should be determined by light-flavor dynamics as discussed in the next subsection. Similar discussions will be applied to the other $J^{P}$ states.

In this way, the spin-complex basis is closely related to the internal structure of heavy hadrons. The relation between the spin-complex basis and the particle basis is also useful to specify the HQS multiplet. Suppose that we observe a hadron state with $J^{P}=1/2^{-}$. This state belongs either to the HQS singlet containing the $j^{\cal P}=0^{+}$ spin-complex, or to the HQS doublet containing the $1^{+}$ spin-complex. Equations~\eqref{eq:fraction_1} and \eqref{eq:fraction_2} tell us that if the state is in the HQS singlet or doublet, the fractions of $\bar{P}N/\bar{P}^{*}N$ is $1/3$ or $3$, respectively. In this way, the HQS constrains the property of the wave functions depending on the HQS multiplet to which the state belongs. This may be reflected, for instance, in the production and the decay patterns of the heavy hadron. 

\subsection{Analysis with one-pion-exchange potential}\label{sec:model}

\subsubsection{One-pion-exchange potential from heavy meson effective theory}

Let us discuss the dynamics of the $\bar{P}^{(\ast)}N$ systems in the heavy quark limit. We consider the one-pion-exchange potential (OPEP) as a long range force. To determine the interaction of the heavy meson and the pion, we employ the heavy meson effective Lagrangians satisfying the HQS and chiral symmetry \cite{Manohar:2000dt,Casalbuoni:1996pg}. In the heavy meson effective theory, the interaction Lagrangian of the $P^{(\ast)} \sim Q\bar{q}$ meson and the pion $\pi$ is given by
\begin{eqnarray}
{\cal L}_{\pi HH} =   ig_\pi \mbox{Tr} \left[
H_b\gamma_\mu\gamma_5 A^\mu_{ba}\bar{H}_a \right],
\label{LpiHH}
\end{eqnarray}
where the heavy meson field $H_{a}$ is given by the heavy pseudoscalar and vector mesons, $P$ and $P^{\ast}$, as 
\begin{align}
H_a   &= \frac{1+\Slash{v}}{2}\left[P^\ast_{a\,\mu}\gamma^\mu-P_a\gamma_5\right],  \\
\bar H_a  &= \gamma_0 H^\dagger_a \gamma_0 .
\end{align}
Here $v^{\mu}$ is the four-velocity of the heavy meson, and the subscripts $a$, $b$ represent the isospin. The axial current $A^{\mu}_{ba}$ by pions are given as
\begin{eqnarray}
A^{\mu} = \frac{1}{2} \left( \xi^{\dag} \partial^{\mu}\xi - \xi \partial^{\mu} \xi^{\dag} \right),
\end{eqnarray}
where $\xi = \exp(i\hat{\pi}/f_{\pi})$ with the pion decay constant $f_{\pi}=132$ MeV and the pion field is defined by
\begin{eqnarray}
\hat{\pi} =
\left(
\begin{array}{cc}
 \frac{\pi^{0}}{\sqrt{2}} & \pi^{+} \\
 \pi^{-} & -\frac{\pi^{0}}{\sqrt{2}}
\end{array}
\right).
\end{eqnarray}
The coupling constant $g_\pi=0.59$ is determined from the decay width of $D^{\ast} \rightarrow D\pi$ observed in experiments \cite{Beringer:1900zz}. From Eq.~\eqref{LpiHH}, we obtain the pion and heavy meson vertices in the static limit $v^{\mu} = (1, \vec{0})$. As a matter of fact, Lagrangian \eqref{LpiHH} is invariant under the spin transformation for the heavy quark, $H_{a} \rightarrow S H_{a}$ with $S \in \mathrm{SU}(2)_{\mathrm{spin}}$. The interaction Lagrangian for the $\bar{P}^{(*)}$ and the pion, which will be used to construct the $\bar{P}^{(*)}N$ potential, is obtained by changing the overall sign due to the G-parity transformation.

The interaction Lagrangian of the pion and the nucleon is given by the pseudoscalar form, 
\begin{eqnarray}
{\cal L}_{\pi NN} = \sqrt{2} ig_{\pi
NN}\bar{N}\gamma_5 \hat{\pi} N \, , \label{LpiNN}
\end{eqnarray}
where $N =(p,n)^T$ is the nucleon field. The coupling constant for the nucleon is given as $g_{\pi NN}^{2}/4\pi=13.6$ from the phenomenological nuclear potential in Ref.~\cite{Machleidt:2001} (see also Ref.~\cite{Machleidt}). Details are found in Refs.~\cite{Yasui:2009bz,Yamaguchi:2011xb,Yamaguchi:2011qw}.

We may consider the short range interaction supplied by the vector mesons $\omega$ and $\rho$. The vector meson exchange potentials can be also constructed from the vertices of $\bar{P}^{(\ast)}$ and the vector mesons with keeping the HQS as demonstrated in our previous papers~\cite{Yasui:2009bz,Yamaguchi:2011xb,Yamaguchi:2011qw}. 
The results of the spin degeneracy are not modified by the inclusion of the vector meson exchange, as far as the HQS is maintained.

With the above vertices, we construct the OPEP between the $\bar{P}^{(\ast)}$ meson and the nucleon $N$ \cite{Yasui:2009bz,Yamaguchi:2011xb,Yamaguchi:2011qw}. The OPEPs for $\bar{P}N$-$\bar{P}^\ast N$ and $\bar{P}^\ast N$-$\bar{P}^\ast N$ are given as
\begin{align}
 V_{\bar{P}N\!-\!\bar{P}^\ast N}(r)&=- \frac{g_\pi g_{\pi NN}}{\sqrt{2}m_N
 f_\pi}\frac{1}{3} \left[\vec{\varepsilon}\,^\dagger\cdot\vec{\sigma}C(r)+S_\varepsilon(\hat{r})
 T(r)\right]\vec{\tau}_{\bar{P}} \!\cdot\! \vec{\tau}_N, \label{PNP*N} \\
 V_{\bar{P}^\ast N\!-\!\bar{P}^\ast N}(r)&= \frac{g_\pi g_{\pi NN}}{\sqrt{2}m_N
 f_\pi} \frac{1}{3}\left[\vec{T}\cdot\vec{\sigma}C(r)+S_T(\hat{r})
 T(r)\right]\vec{\tau}_{\bar{P}} \!\cdot\! \vec{\tau}_N, \label{P*NP*N}
\end{align}
respectively, by a sum of the central and tensor forces. 
In Eqs.~(\ref{PNP*N}) and (\ref{P*NP*N}), $\vec{\varepsilon}$ ($\vec{\varepsilon}^{\,\dag}$) is the polarization vector of the incoming (outgoing) $\bar{P}^\ast$, $\vec{T}$ is the spin-one operator of $\bar{P}^\ast$, and $S_\varepsilon(\hat{r})$ [$S_T(\hat{r})$] is the tensor operator $S_{\cal O}(\hat{r})=3(\vec{\cal O}\cdot\hat{r})(\vec{\sigma}\cdot\hat{r})-\vec{\cal O}\cdot\vec{\sigma}$ with $\hat{r}=\vec{r}/r$ and $r=|\vec{r}\,|$ for $\vec{{\cal O}}=\vec{\varepsilon}$ ($\vec{T}$), where $\vec{r}$ is the relative position vector between $\bar{P}^{(\ast)}$ and $N$. $\vec{\sigma}$ are the Pauli matrices acting on the nucleon spin. $\vec{\tau}_{\bar{P}}$ ($\vec{\tau}_N$) are isospin operators for $\bar{P}^{(\ast)}$ ($N$). The functions $C(r)$ and $T(r)$ for the central and tensor parts are
\begin{align}
 &C(r)=\int\frac{d^3 \vec{q}}{(2\pi)^3}
 \frac{m_{\pi}^2}{\vec{q}^{\,\,2}+m_{\pi}^2}e^{i\vec{q}\cdot\vec{r}}F(\vec{q}\,)\,
 , \label{Cpote}\\
 &S_{\cal O}(\hat{r})T(r)=\int\frac{d^3 \vec{q}}{(2\pi)^3}
 \frac{-\vec{q}^{\,\,2}}{\vec{q}^{\,\,2}+m_{\pi}^2}S_{\cal O}(\hat{q})e^{i\vec{q}\cdot\vec{r}}F(\vec{q}\,)
 , \label{Tpote}
\end{align}
with $\hat{q}=\vec{q}/|\vec{q}\,|$, where the dipole-type form factor $F(\vec{q}\,)=(\Lambda_{N}^2-m_{\pi}^2)(\Lambda_{\bar{P}}^2-m_{\pi}^2)/(\Lambda_{N}^2+|\vec{q}\,|^2)(\Lambda_{\bar{P}}^2+|\vec{q}\,|^2)$ with cutoff parameters $\Lambda_N$ and $\Lambda_{\bar{P}}$ are introduced for the spatial sizes of hadrons as discussed in Refs.~\cite{Yasui:2009bz,Yamaguchi:2011xb,Yamaguchi:2011qw}.

We note that the OPEP for $\bar{P}N$-$\bar{P}N$ does not exist, because the $\bar{P}\bar{P}\pi$ vertex is forbidden by parity conservation. Instead, the $\bar{P}N$-$\bar{P}N$ interaction is effectively supplied from the mixing $\bar{P}N$ and $\bar{P}^{\ast}N$ ($\bar{P}N \rightarrow \bar{P}^{\ast}N \rightarrow \bar{P}N$), as emphasized in Refs.~\cite{Yasui:2009bz,Yamaguchi:2011xb,Yamaguchi:2011qw}.

\subsubsection{Negative parity channels}\label{subsubsec:negative}

Let us discuss concretely the channels with $J^{P}=1/2^{-}$ and $3/2^{-}$, whose particle bases are given by Eqs.~\eqref{eq:transformation_1/2^-} and \eqref{eq:transformation_3/2^-}, respectively.
From Eqs.~(\ref{PNP*N}) and (\ref{P*NP*N}), the Hamiltonians in the particle basis are given by
\begin{eqnarray}
H_{1/2^-} =
\left(
\begin{array}{ccc}
 K_0 & \sqrt{3} \, {C} & -\sqrt{6} \, {T}  \\
\sqrt{3} \, {C} & K_0-2 \, {C} & -\sqrt{2} \, {T} \\
-\sqrt{6} \, {T} & -\sqrt{2} \, {T} & K_2 + ({C} - 2\, {T})
\end{array}
\right), \\
H_{3/2^-} =
\left(
\begin{array}{cccc}
 K_2 & \sqrt{3}\, {T} & -\sqrt{3} \, {T} & \sqrt{3}\,{C} \\
\sqrt{3}\,{T} &K_0 + {C} & 2\,{T} & {T} \\
-\sqrt{3}\,{T} & 2\,{T} & K_2 + {C} & -{T} \\
\sqrt{3}\,{C} & {T} & -{T} & K_2 -2\,{C}
\end{array}
\right),
\end{eqnarray}
for $1/2^{-}$ and $3/2^{-}$, respectively. Here the kinetic terms are
\begin{eqnarray}
K_{L} = - \frac{1}{2\mu}\left( \frac{\partial^2}{\partial r^2 }+ \frac{2}{r} \frac{\partial}{\partial r} - \frac{L(L+1)}{r^2} \right),
\label{eq:kinetic_term}
\end{eqnarray}
for angular momentum $L$ with the reduced mass $\mu=m_{N}$ in the heavy quark limit, and we have defined
\begin{eqnarray}
{C}=\kappa\, C(r), \,
{T} = \kappa\, T(r),
\end{eqnarray}
with $\kappa = (g_{\pi}g_{\pi NN}/\sqrt{2} m_N f_{\pi}) (\vec{\tau}_{\bar{P}} \!\cdot\! \vec{\tau}_N/3)$. We emphasize again that the $\bar{P}N$ and $\bar{P}^{\ast}N$ states can be mixed, and accordingly the states with different angular momenta can also be mixed by the off-diagonal components of the tensor force.
The tensor force induces the strong attractions, as known in the nucleon-nucleon interaction in nuclear physics. Thus, the mixing effects of $\bar{P}N$ and $\bar{P}^{\ast}N$ are essentially important to switch on the strong tensor force in the OPEP.

Now, let us rewrite the Hamiltonians $H_{1/2^-}$ and $H_{3/2^-}$ in the spin-complex basis by using the unitary matrices $U_{1/2^{-}}$ and $U_{3/2^{-}}$ in Eqs.~(\ref{eq:U_1/2-}) and (\ref{eq:U_3/2-}). The results are given as
\begin{align}
 H_{1/2^-}^{\mathrm{SC}} &=
U_{1/2^-}^{-1} H_{1/2^-} U_{1/2^-} \nonumber \\
&=
\left(
\begin{array}{c|cc}
 K_{0} - 3\,{C} & 0 & 0 \\
 \hline
 0 & K_{0} + {C} & -2\sqrt{2} \,{T} \\
 0 & -2\sqrt{2} \,{T} & K_{2} +  ({C} - 2\,{T})
\end{array}
\right) \nonumber \\
&\equiv 
\left(
\begin{array}{c|c}
 H_{1/2^-}^{\mathrm{SC}(0^+)} & 0 \\
 \hline
 0 & H_{1/2^-}^{\mathrm{SC}(1^+)}
\end{array}
\right), \\
 H_{3/2^{-}}^{\mathrm{SC}} &= U_{3/2^-}^{-1} H_{3/2^-} U_{3/2^-} \nonumber \\
 &=
\left(
\begin{array}{cc|cc}
 K_{0} + {C} & 2\sqrt{2}\,{T} & 0 & 0 \\
 2\sqrt{2}\,{T} & K_{2} + ({C} - 2\,{T}) & 0 & 0 \\
\hline
 0 & 0 & K_{2} - 3\,{C} & 0 \\
 0 & 0 & 0 & K_{2} + ({C} + 2\,{T})
\end{array}
\right) \nonumber \\
&\equiv
\left(
\begin{array}{c|c}
 H_{3/2^-}^{\mathrm{SC}(1^+)} & 0 \\
 \hline
 0 & H_{3/2^-}^{\mathrm{SC}(2^+)}
\end{array}
\right). 
\end{align}
As in the previous section, we introduce the notation $H_{J^{P}}^{\mathrm{SC}(j^{\cal P})}$ for the Hamiltonian for the $J^{P}$ state containing the spin-complex with $j^{\cal P}$.
To denote the spin-complex basis, we use the superscript ``SC" instead of ``BMC".
Thus, with the spin-complex basis, we obtain the block-diagonal forms. From the results, we find that the terms with the $1^{+}$ spin-complex in the $1/2^{-}$ and $3/2^{-}$ channels are identical, except for the signs of the off-diagonal components which are irrelevant for the eigenvalues:
\begin{eqnarray}
H_{1/2^-}^{\mathrm{SC}(1^+)} \approx H_{3/2^-}^{\mathrm{SC}(1^+)},
\end{eqnarray}
where $\approx$ stands for the equality of the eigenvalues as introduced in the previous section. This means that the eigenstates with $1/2^{-}$ and $3/2^{-}$ containing the spin-complex with $1^{+}$ form the HQS doublet whose masses are completely degenerate. Similarly, we will obtain $H_{3/2^-}^{\mathrm{SC}(2^+)}\approx H_{5/2^-}^{\mathrm{SC}(2^+)}$ suggesting that the $3/2^{-}$ and $5/2^{-}$ states containing the spin-complex with $2^{+}$ belong to the HQS doublet, as shown below. On the other hand, there is no corresponding component to $H_{1/2^-}^{\mathrm{SC}(0^+)}$. Therefore, the $1/2^{-}$ state containing the spin-complex with $0^{+}$ belongs to the HQS singlet. Those are exactly what we have discussed in the previous section. 
The spin degeneracy is shown generally in the heavy {\it quark} effective theory in Eq.~(\ref{eq:HQET}), where fundamental degrees of freedom are quarks and gluons. 
Interestingly, the spin degeneracy is shown also for hadronic molecules whose eigenstates are induced from the heavy {\it meson} effective theory with the hadronic degrees of freedom.

For completeness, we present the results with $J^{P}=5/2^{-}$ and $7/2^{-}$ in the negative parity sector. In the particle basis, the Hamiltonians are given as 
\begin{eqnarray}
 H_{5/2^-} =
\begin{pmatrix}
 K_2 & \sqrt{3}C & \sqrt{\frac{6}{7}}T & -\frac{6}{\sqrt{7}}T \\
 \sqrt{3}C & K_2 -2C & \sqrt{\frac{2}{7}} T & -2 \sqrt{\frac{3}{7}} T \\
 \sqrt{\frac{6}{7}} T & \sqrt{\frac{2}{7}}T & K_2 +(C+\frac{10}{7}T) & \frac{4}{7}\sqrt{6}T \\
 -\frac{6}{\sqrt{7}}T & -2 \sqrt{\frac{3}{7}}T & \frac{4}{7}\sqrt{6}T & K_4 +(C-\frac{10}{7}T)
\end{pmatrix},
\end{eqnarray}
for $5/2^{-}$ and %
\begin{eqnarray}
 H_{7/2^-} =
\begin{pmatrix}
 K_4 & 3\sqrt{\frac{3}{7}}T & \sqrt{3}C & -\sqrt{\frac{15}{7}}T \\
 3\sqrt{\frac{3}{7}}T & K_2 +(C-\frac{4}{7}T) & \frac{3}{\sqrt{7}}T & \frac{6}{7}\sqrt{5}T \\
 \sqrt{3}C & \frac{3}{\sqrt{7}}T & K_4 -2C & -\sqrt{\frac{5}{7}}T \\
 -\sqrt{\frac{15}{7}}T & \frac{6}{7}\sqrt{5}T & -\sqrt{\frac{5}{7}}T & K_4 +(C+\frac{4}{7}T)
\end{pmatrix},
\end{eqnarray}
for $7/2^{-}$. %
Using Eqs.~\eqref{eq:transformation_5/2^-} and \eqref{eq:transformation_7/2^-}, we obtain the Hamiltonian in the spin-complex basis as
\begin{align}
 H_{5/2^{-}}^{\mathrm{SC}} &= U_{5/2^-}^{-1} H_{5/2^-} U_{5/2^-} \nonumber \\
&= 
\left(
\begin{array}{cc|cc}
 K_2 -3 C & 0 & 0 & 0 \\
 0 & K_2+ (C+2 T) & 0 & 0 \\
\hline
 0 & 0 & K_2 + (C-\frac{4}{7}T) & \frac{12 \sqrt{3}}{7}T \\
 0 & 0 & \frac{12 \sqrt{3}}{7}T & K_4 +(C-\frac{10}{7}T)
\end{array}
\right) \nonumber \\
&\equiv
\left(
\begin{array}{c|c}
 H^{\mathrm{SC}(2^+)}_{5/2^-} & 0 \\
\hline
 0 & H^{\mathrm{SC}(3^+)}_{5/2^-}
\end{array}
\right),
\end{align}
for $5/2^{-}$ and %
\begin{align}
 H_{7/2^{-}}^{\mathrm{SC}} &= U_{7/2^-}^{-1}H_{7/2^-}U_{7/2^-} \nonumber \\
&=
\left(
\begin{array}{cc|cc}
K_2+ (C-\frac{4}{7}T) & \frac{12 \sqrt{3}}{7}T & 0 & 0 \\
 \frac{12 \sqrt{3}}{7}T & K_4 +(C-\frac{10}{7}T) & 0 & 0 \\
\hline
 0 & 0 & K_4 -3 C & 0 \\
 0 & 0 & 0 & K_4 +(C+2 T)
\end{array}
\right) \nonumber \\
&\equiv
\left(
\begin{array}{c|c}
 H^{\mathrm{SC}(3^+)}_{7/2^-} & 0 \\
 \hline
 0 & H^{\mathrm{SC}(4^+)}_{7/2^-}
\end{array}
\right),
\end{align} 
for $7/2^{-}$. %
Therefore, we confirm the equivalence of the eigenvalues
\begin{eqnarray}
 H^{\mathrm{SC}(2^+)}_{3/2^-} \approx H^{\mathrm{SC}(2^+)}_{5/2^-},
\end{eqnarray}
and 
\begin{eqnarray}
 H^{\mathrm{SC}(3^+)}_{5/2^-} \approx H^{\mathrm{SC}(3^+)}_{7/2^-},
\end{eqnarray}
which indicate that the $3/2^{-}$ and $5/2^{-}$ states containing the spin-complex with $2^{+}$ belong to the HQS doublet, and the $5/2^{-}$ and $7/2^{-}$ states containing the spin-complex with $3^{+}$ also belong to the HQS doublet. A partner of the remaining $H^{\mathrm{SC}(4^+)}_{7/2^-}$ is considered to be in the $J^P=9/2^{-}$ channel.

\subsubsection{Positive parity channels}

Similarly, we consider the positive party channels with $J^{P}=1/2^{+}$, $3/2^{+}$, $5/2^{+}$ and $7/2^{+}$.
In the particle basis, the Hamiltonians are
\begin{align}
 H_{1/2^+} &=
\begin{pmatrix}
 K_1 & \sqrt{3}C & -\sqrt{6}T \\
 \sqrt{3}C & K_1 - 2C & -\sqrt{2}T \\
 -\sqrt{6}T & -\sqrt{2}T & K_1 + (C - 2T)
\end{pmatrix},
\end{align}
for $1/2^{+}$, %
\begin{align}
 H_{3/2^+} &=
\begin{pmatrix}
 K_1 & \sqrt{3}C & \sqrt{\frac{3}{5}}T & -3\sqrt{\frac{3}{5}}T \\
 \sqrt{3}C & K_1 - 2C & \frac{1}{\sqrt{5}}T & -\frac{3}{\sqrt{5}}T \\
 \sqrt{\frac{3}{5}}T & \frac{1}{\sqrt{5}}T & K_1 + (C + \frac{8}{5}T) & \frac{6}{5}T \\
 -3\sqrt{\frac{3}{5}}T & -\frac{3}{\sqrt{5}}T & \frac{6}{5}T & K_3 + (C-\frac{8}{5}T)
\end{pmatrix},
\end{align}
for $3/2^{+}$, %
\begin{align}
 H_{5/2^+} &=
\begin{pmatrix}
 K_3 & \frac{3}{5}\sqrt{10}T & \sqrt{3}C & -2\sqrt{\frac{3}{5}}T \\
 \frac{3}{5}\sqrt{10}T & K_1 +(C-\frac{2}{5}T) & \sqrt{\frac{6}{5}}T & \frac{4}{5}\sqrt{6}T \\
 \sqrt{3}C & \sqrt{\frac{6}{5}}T & K_3 -2C & -\frac{2}{\sqrt{5}}T \\
 -2\sqrt{\frac{3}{5}}T & \frac{4}{5}\sqrt{6}T & -\frac{2}{\sqrt{5}}T & K_3 +(C+\frac{2}{5}T)
\end{pmatrix},
\end{align}
for $5/2^{+}$, and %
\begin{align}
 H_{7/2^+} &=
\begin{pmatrix}
 K_3 & \sqrt{3}C & T & -\sqrt{5}T \\
 \sqrt{3}C & K_3 -2C & \frac{1}{\sqrt{3}}T & -\sqrt{\frac{5}{3}}T \\
 T & \frac{1}{\sqrt{3}}T & K_3 +(C+\frac{4}{3}T) & \frac{2}{3}\sqrt{5}T \\
 -\sqrt{5}T & -\sqrt{\frac{5}{3}}T & \frac{2}{3}\sqrt{5}T & K_5 +(C-\frac{4}{3}T)
\end{pmatrix},
\end{align}
for $7/2^{+}$. %
In the spin-complex basis, the Hamiltonians are obtained as 
\begin{align}
 H_{1/2^{+}}^{\mathrm{SC}} &= U^{-1}_{1/2^+} H_{1/2^+} U_{1/2^+} \nonumber \\
&=
\left(
\begin{array}{c|cc}
 K_1 + (C-4 T) & 0 & 0 \\
\hline
 0 & K_1 -3 C & 0 \\
 0 & 0 & K_1 + (C+2 T)
\end{array}
\right) \nonumber \\
&\equiv
\left(
\begin{array}{c|c}
 H^{\mathrm{SC}(0^-)}_{1/2^+} & 0 \\
 \hline
 0 & H^{\mathrm{SC}(1^-)}_{1/2^+}
\end{array}
\right),
\label{eq:HSC1/2+}
\end{align}
for $1/2^{+}$, %
\begin{align}
 H_{3/2^{+}}^{\mathrm{SC}} &= U^{-1}_{3/2^+} H_{3/2^+} U_{3/2^+} \nonumber \\
&=
\left(
\begin{array}{cc|cc}
 K_1 -3 C & 0 & 0 & 0 \\
 0 & K_1 + (C+2 T) & 0 & 0 \\
\hline
 0 & 0 & K_1 + (C-\frac{2}{5}T) & \frac{6 \sqrt{6}}{5}T \\
 0 & 0 & \frac{6 \sqrt{6}}{5}T & K_3 +(C-\frac{8}{5}T)
\end{array}
\right) \nonumber \\
&\equiv
\left(
\begin{array}{c|c}
 H^{\mathrm{SC}(1^-)}_{3/2^+} & 0 \\
 \hline
 0 & H^{\mathrm{SC}(2^-)}_{3/2^+}
\end{array}
\right),
\label{eq:HSC3/2+}
\end{align}
for $3/2^{+}$,
\begin{align}
 H_{5/2^{+}}^{\mathrm{SC}} &= U_{5/2^+}^{-1}H_{5/2^+}U_{5/2^+} \nonumber \\
&=
\left(
\begin{array}{cc|cc}
 K_1 +(C-\frac{2}{5}T) & \frac{6 \sqrt{6} }{5}T & 0 & 0 \\
 \frac{6 \sqrt{6} }{5}T & K_3 +(C-\frac{8}{5}T) & 0 & 0 \\
\hline
 0 & 0 & K_3 -3 C & 0 \\
 0 & 0 & 0 & K_3 +(C+2 T)
\end{array}
\right) \nonumber \\
&\equiv
\left(
\begin{array}{c|c}
 H^{\mathrm{SC}(2^-)}_{5/2^+} & 0 \\
 \hline
 0 & H^{\mathrm{SC}(3^-)}_{5/2^+}
\end{array}
\right),
\end{align}
for $5/2^{+}$, and %
\begin{align}
 H_{7/2^{+}}^{\mathrm{SC}} &= U_{7/2^+}^{-1}H_{7/2^+}U_{7/2^+} \nonumber \\
&=
\left(
\begin{array}{cc|cc}
 K_3 -3 C & 0 & 0 & 0 \\
 0 & K_3 +(C+2 T) & 0 & 0 \\
 \hline
 0 & 0 & K_3 +(C-\frac{2}{3}T) & \frac{4 \sqrt{5} }{3}T \\
 0 & 0 & \frac{4 \sqrt{5}}{3}T & K_5 +(C-\frac{4}{3}T)
\end{array}
\right) \nonumber \\
&\equiv
\left(
\begin{array}{c|c}
 H^{\mathrm{SC}(3^-)}_{7/2^+} & 0 \\
 \hline
 0 & H^{\mathrm{SC}(4^-)}_{7/2^+}
\end{array}
\right),
\end{align}
for $7/2^{+}$. 
Therefore, we find that the Hamiltonian
\begin{eqnarray}
H^{\mathrm{SC}(0^{-})}_{1/2^{+}},
\end{eqnarray}
has no partner in any other diagonal components, and hence the $1/2^{+}$ state containing the spin-complex with $0^{-}$ is classified as the HQS singlet. On the other hand, we find the relations
\begin{align}
H^{\mathrm{SC}(1^{-})}_{1/2^{+}} & \approx H^{\mathrm{SC}(1^{-})}_{3/2^{+}}, \\
H^{\mathrm{SC}(2^{-})}_{3/2^{+}} & \approx H^{\mathrm{SC}(2^{-})}_{5/2^{+}}, \\
H^{\mathrm{SC}(3^{-})}_{5/2^{+}} & \approx H^{\mathrm{SC}(3^{-})}_{7/2^{+}},
\end{align}
indicating that the $1/2^{+}$ and $3/2^{+}$ ($3/2^{+}$ and $5/2^{+}$ or $5/2^{+}$ and $7/2^{+}$) states containing the spin-complex with $1^{-}$ ($2^{-}$ or $3^{-}$) belong to the HQS doublet. It is naturally expected that the state with $J^{P}=7/2^{+}$ and $j^{\cal P}=3^{-}$ has a partner in the $J^{P}=9/2^{+}$ channel.

\subsubsection{Spin degeneracy for general $J^{P}$}

From the analysis above, we find a general relation for $j \ge 1$
\begin{eqnarray}
 H^{\mathrm{SC}(j^{\mathcal{P}})}_{j-1/2^{P}} \approx H^{\mathrm{SC}(j^{\mathcal{P}})}_{j+1/2^{P}},
 \label{eq:identity_hamiltonian}
\end{eqnarray}
with $P=-\mathcal{P}$. This means that the $j\pm 1/2^{P}$ states containing the spin-complex with $j^{\mathcal{P}}$ belong to the HQS doublet. The $1/2^{P}$ state containing the spin-complex with $0^{\mathcal{P}}$ belongs to the HQS singlet. Those results are consistent with what is expected form the HQS.

Comments are in order. First, the off-diagonal components in Hamiltonian $H_{J^P}^{\mathrm{SC}(j^{\cal P})}$ with the spin-complex basis are responsible for the mixing among the components with same $j^{\cal P}$, which determine the mixing angle in Eqs.~\eqref{eq:spin_doublet_PbarN_1/2} and \eqref{eq:spin_doublet_PbarN_3/2}. The strength of the off-diagonal components is not constrained by the HQS, and depends on the chosen potential. In fact, the inclusion of the vector meson exchange potential modifies the strength of the off-diagonal components, as shown in \ref{sec:appendix1}. In addition, although some of the off-diagonal components in $H_{J^P}^{\mathrm{SC}(j^{\cal P})}$ vanish in the present potential, they can be finite with general interactions. Nevertheless, the mass degeneracy of the HQS doublet and the fraction of the components such as Eqs.~\eqref{eq:fraction_1}, \eqref{eq:fraction_2}, and \eqref{eq:fraction_3} are always guaranteed, thanks to the relation~\eqref{eq:identity_hamiltonian}. 

Second, to realize these properties, we should include all possible channels which are related by the HQS. If some channels are missing, Eq.~\eqref{eq:identity_hamiltonian} does not hold.
In fact, if we switch off the mixing between $\bar{P}N$ and $\bar{P}^{*}N$, the mass degeneracy does not occur, because the $\bar{P}$ and $\bar{P}^{*}$ are in the HQS doublet.
In the same way, the use of the single coupling constant $g_{\pi}$ for the $\bar{P}\bar{P}^{*}\pi$ and $\bar{P}^{*}\bar{P}^{*}\pi$ vertices is also necessary, as it is required by the HQS.
We remark that the coupling to different angular momentum states is not always necessary. Indeed, as shown in \ref{sec:examples_nonexotic}, the relation~\eqref{eq:identity_hamiltonian} is still valid with only the S-wave states.
The mixing of angular momentum is necessary rather for switching on the strong tensor force.

Third, in the present analysis, we have considered the spin-complex basis with the $Nq$ configuration. Even when other components such as $\Delta q$ and $N \pi q$ are considered, the Hamiltonian can be block-diagonalized in the spin-complex basis, as far as we respect the HQS to construct the interaction potential. In this way, spin-complex basis is useful to grasp the model-independent property of heavy hadrons in the framework of effective models.

\subsubsection{Numerical results}

In this section, let us show the numerical results of the properties of the $\bar{P}^{(\ast)}N$ system. In addition to the $\pi$ exchange potential, we also examine the $\rho$ and $\omega$ exchange potentials shown in \ref{sec:appendix1}. In the present discussion, we use finite masses $m_{\bar{P}}$ and $m_{\bar{P}^*}= m_{\bar{P}}+\Delta m$ for heavy mesons $\bar{P}$ and $\bar{P}^{\ast}$, respectively, by introducing the mass difference $\Delta m$, in order to control the breaking of the HQS. Accordingly, different reduced masses $\mu$ are used in the $\bar{P}N$ and $\bar{P}^{*}N$ channels and $\Delta m$ is added in the diagonal terms in the $\bar{P}^{*}N$ channels. The mass difference is parametrized as $\Delta m/m_{P^{*}}=(0.617\text{ GeV}/m_{P^{*}})^{2.25}$ so as to fit empirically the experimental data in the strange, charm, and bottom sectors~\cite{Yamaguchi:2011xb}. We denote the mesons in the heavy quark limit $m_{\bar{P}}=m_{\bar{P}^{*}}\to \infty$ as $\bar{P}_{\rm Q}^{(\ast)}$. The cutoff parameter of the form factor is chosen to be $\Lambda_N=830$ MeV ($\Lambda_N=846$ MeV) for the $\pi$ ($\pi\rho\,\omega$) potential, and we use $\Lambda_{\bar{P}_{\rm Q}}=1.12\Lambda_N$ estimated in the heavy quark limit~\cite{Yamaguchi:2011xb}.

\begin{figure}[tb]
 \centering
 \includegraphics[width=8cm,clip]{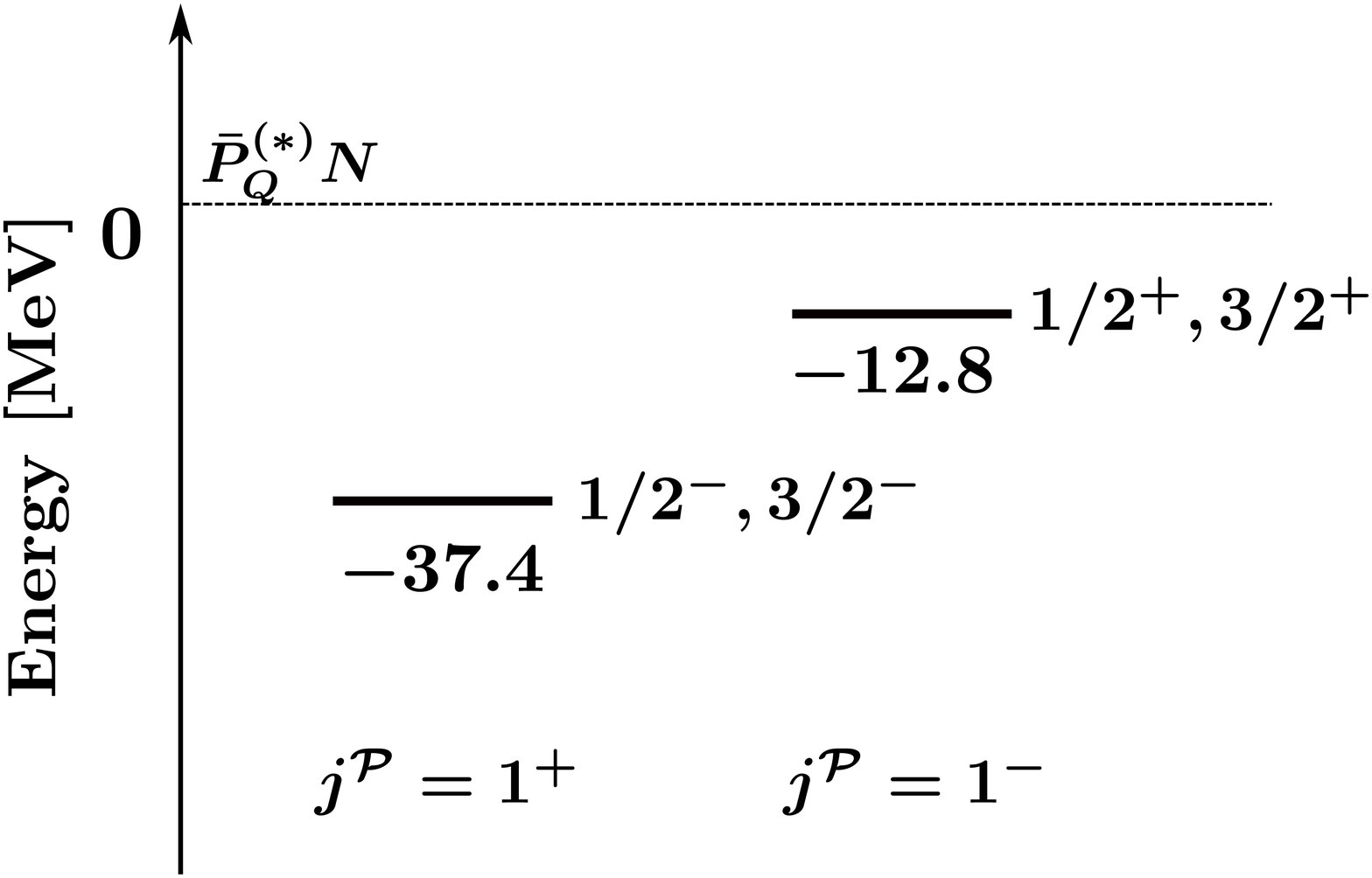}
 \caption{Energy levels of the bound $\bar{P}_{\rm Q}^{(\ast)}N$ states with $I=0$ and $J^P=1/2^{\pm}$ and $3/2^{\pm}$ in the heavy quark limit, when the $\pi\rho\,\omega$ potential is used.}
 \label{energylevel}
\end{figure}

To begin with, we consider the $\bar{P}_{\rm Q}^{(\ast)}N$ state in the heavy quark limit. By solving the coupled-channel Schr\"odinger equations numerically, we obtain bound states with $(I,J^P)=(0,1/2^{\pm})$ and $(0,3/2^{\pm})$ as displayed in Fig.~\ref{energylevel}. We find the mass degeneracy in the $J^P=1/2^-$ and $3/2^-$ channels, and also in the $J^P=1/2^+$ and $3/2^+$ channels. The lowest energy states are of $1/2^-$ and $3/2^-$, which have the same binging energy $34.1$ MeV with the $\pi$ potential and $37.4$ MeV with the $\pi\rho\,\omega$ potential. These states are considered to form an HQS doublet with the spin-complex $j^{\cal P}=1^+$.
This becomes clear when the wave functions are decomposed in the following. Since the $\pi$ exchange potential predominates in the $\bar{P}^{(\ast)}N$ states as discussed in Refs.~\cite{Yasui:2009bz,Yamaguchi:2011xb,Yamaguchi:2011qw}, the results with the $\pi\rho\,\omega$ potential are not very far from those with the $\pi$ potential. The secondly bound states are found in the $1/2^+$ and $3/2^+$ channels with a binding energy $11.8$ MeV for the $\pi$ potential and $12.8$ MeV for the $\pi\rho\,\omega$ potential. Therefore, the degenerate states seem to belong the HQS doublet with $j^{\cal P}=1^-$ from the viewpoint of the analytical argument given in Section~\ref{subsubsec:negative}.

The properties of the HQS doublets, in particular those of the spin-complex, are reflected in the fractions in the particle basis. We obtain the mixing ratios of the wave functions of the bound states in Table~\ref{mixingratio}. For the bound state for $J^P=1/2^-$ with $\pi\rho\,\omega$ potential, the mixing ratios of the S-wave states are 64.0 \% for the $\bar{P}N(^2{\rm S}_{1/2})$ channel and 21.3 \% for the $\bar{P}^\ast N(^2{\rm S}_{1/2})$ channel, and hence the fraction of the wave functions is $f(\bar{P}N(^2{\rm S}_{1/2})):f(\bar{P}^\ast N(^2{\rm S}_{1/2}))=3:1$. This result is consistent with the fraction in Eq.~\eqref{eq:fraction_2} and shows that the state is made of the spin-complex with $j^{\cal P}=1^{+}$. Similarly, we obtain the fraction for the $J^P=3/2^-$ as $f(\bar{P}N(^2{\rm D}_{3/2})):f(\bar{P}^\ast N(^2{\rm D}_{3/2})):f(\bar{P}^\ast N(^4{\rm D}_{3/2}))=3:1:4$, which coincides with the fractions in Eq~\eqref{eq:fraction_3}. Therefore, it is confirmed that the wave functions of the $1/2^-$ and $3/2^-$ states are equivalent in the spin-complex basis and they are made of the $1^{+}$ spin-complex. The mixing ratios of $\bar{P}^{\ast}N(^{4}{\rm D}_{1/2})$ for $1/2^-$ [$\bar{P}^{\ast}N(^{4}{\rm S}_{3/2})$ for $3/2^-$] are different because of the off-diagonal components in the spin-complex basis. This effect is characterized by the mixing angle $\theta$ in Eqs.~\eqref{eq:spin_doublet_PbarN_1/2_1} and \eqref{eq:spin_doublet_PbarN_3/2_1}. It is also the case in the positive parity sector; the mixing ratios of the $1/2^{+}$ and $3/2^{+}$ states follow the fractions derived in the spin-complex basis. Because the absence of the off-diagonal components in $H^{\mathrm{SC}(1^-)}_{1/2^+}$ and $H^{\mathrm{SC}(1^-)}_{3/2^+}$ in Eqs.~\eqref{eq:HSC1/2+} and \eqref{eq:HSC3/2+} or in Eqs.~\eqref{eq:AppHSC1/2+} and \eqref{eq:AppHSC3/2+} in the $\pi$ or $\pi\rho\,\omega$ model, we obtain the identical mixing ratios in two potentials. In the positive parity sector, the model dependence lies only in the diagonal components, which results in the difference of the binding energies.

We evaluate the spatial sizes of $\bar{P}_{\rm Q}^{(\ast)}N$ from the mean-squared-radius, 
namely the expectation value of the distance $r$ between $\bar{P}_{\rm Q}^{(\ast)}$ and $N$,
 which can be expressed in the particle basis 
\begin{align}
   \sqrt{\langle r^{2}\rangle }
   &= \sqrt{ \sum_{i}\langle (\bar{P}^{(*)}N)_{i} |
   \ r^{2}\  | (\bar{P}^{(*)}N)_{i} \rangle},
\end{align}
with $i$ being $^{2S+1}L_{J}$ of $\bar{P}_{\rm Q}^{(\ast)}N$.
Interestingly, we find $\sqrt{\langle r^{2}\rangle }$ of $1/2^-$ and $3/2^-$ are the same values, 1.2 fm for the $\pi$ potential and 1.1 fm for the $\pi\rho\,\omega$ potential, respectively. We also obtain the same sizes of the $1/2^+$ and $3/2^+$ states, 1.6 fm for the $\pi$ and $\pi\rho\,\omega$ potentials.
Those results are consistent with the fact that the spin-complex in the $1/2^-$ ($1/2^+$) state is exactly the same as that in the $3/2^-$ ($3/2^+$) state.
Although the mixing ratios in the particle basis are quite different in the $1/2^{\pm}$ and $3/2^{\pm}$ states, the sum of the all channels gives the same results, thanks to the equivalence of the wave functions in spin-complex basis. In this sense, the spin-complex basis is more appropriate to express the structure of the hadrons in the heavy quark limit.

\begin{table}[tb]
 \caption{Mixing ratios of the $\bar{P}_{\rm Q}^{(*)}N$ channels in the bound states with $I=0$ in the heavy quark limit.}
 \label{mixingratio}
 \begin{center}  
 {\renewcommand{\arraystretch}{1.5}
  \begin{tabular}{c|cccc}
   \hline
   $J^{P}=1/2^-$&$\bar{P}N(^2{\rm S}_{1/2})$ &$\bar{P}^\ast N(^2{\rm S}_{1/2})$ &$\bar{P}^\ast N(^4{\rm D}_{1/2})$
   & --- \\\hline
   $\pi$&63.6 \% &21.2 \%&15.2 \%&--- \\
   $\pi\rho\,\omega$&64.0 \%&21.3 \%&14.7 \%&--- \\ \hline
   $J^{P}=3/2^-$&$\bar{P}N(^2{\rm D}_{3/2})$ &$\bar{P}^\ast N(^4{\rm S}_{3/2})$ &$\bar{P}^\ast N(^4{\rm D}_{3/2})$
   &$\bar{P}^\ast N(^2{\rm D}_{3/2})$  \\\hline
   $\pi$&5.7 \%&84.8 \%&7.6 \%&1.9 \%\\
   $\pi\rho\,\omega$&5.5 \%&85.3 \%&7.4 \%&1.8 \%\\ \hline\hline
   $J^{P}=1/2^+$&$\bar{P}N(^2{\rm P}_{1/2})$ &$\bar{P}^\ast N(^2{\rm P}_{1/2})$ &$\bar{P}^\ast N(^4{\rm P}_{1/2})$
   & --- \\\hline
   $\pi$&50.0 \%&16.7 \%&33.3 \%&--- \\
   $\pi\rho\,\omega$&50.0 \%&16.7 \%&33.3 \%&--- \\\hline
   $J^{P}=3/2^+$&$\bar{P}N(^2{\rm P}_{3/2})$ &$\bar{P}^\ast N(^2{\rm P}_{3/2})$ &$\bar{P}^\ast N(^4{\rm P}_{3/2})$
   &$\bar{P}^\ast N(^4{\rm F}_{3/2})$  \\\hline
   $\pi$&12.5 \%&4.2 \%&83.3 \%&0.0 \%\\
   $\pi\rho\,\omega$&12.5 \%&4.2 \%&83.4 \%&0.0 \%\\ \hline
  \end{tabular}
  }
 \end{center}
\end{table}

\begin{figure}[tb]
 \centering
 \includegraphics[width=11cm,clip]{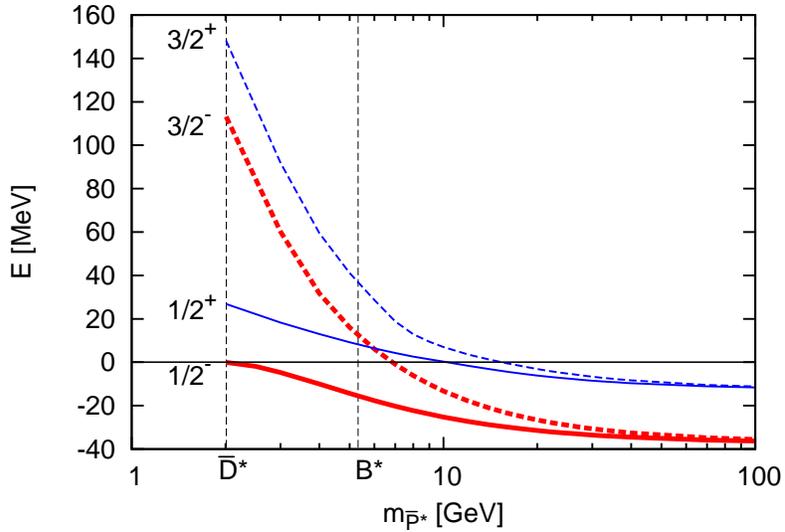}
 \caption{Energies of the $J^P=1/2^{\pm}$ and $3/2^{\pm}$ states with $I=0$ measured from the $\bar{P}N$ threshold as functions
 of heavy vector meson mass $m_{P^\ast}$, when the $\pi\rho\,\omega$
 potential is used. The solid (dashed) line shows
 the $1/2^-$ ($3/2^-$) state, and the dashed-dotted (dashed-two-dotted)
 line shows the $1/2^+$ ($3/2^+$) state.
 The horizontal axis sets as a
 logarithmic scale.}
 \label{massdif}
\end{figure}

With the finite masses of the heavy mesons as the term breaking the HQS, we calculate the energies of the $1/2^\pm$ and $3/2^\pm$ states as functions of heavy vector meson mass $m_{P^\ast}$ (Fig.~\ref{massdif}). The negative (positive) energy corresponds to a binding (resonance) energy. For the resonance energy, we plot the real parts of the complex energies extracted from the phase shift. In Fig.~\ref{massdif}, we observe that the energy difference between the states with $1/2^\pm$ and $3/2^\pm$ decreases as the heavy vector meson mass $m_{P^\ast}$ increases, and the energies finally reach $-37.4$ MeV for $J^P=1/2^-$, $3/2^-$ and $-12.8$ MeV for $J^P=1/2^+$, $3/2^+$ in the heavy quark limit.
This result is quite interesting.
Namely, two states of $1/2^{-}$ and $3/2^{-}$ ($1/2^{+}$ and $3/2^{+}$) in charm and bottom sectors stem from the same origin in the heavy quark limit.

We comment on the non-exotic $P^{(*)}N$ channel, whose quark configuration is given by $Q\bar{q}qqq$.
 The present boson-exchange potential can be applied to these systems by changing the sign of the vertices according to the G-parity. In the non-exotic channels, there can be the mixing effects with the hadronic states $P^{(*)}N$ and the compact three-quark states $Qqq$ due to the light quark-antiquark annihilation processes. Nevertheless, the emergence of the HQS doublet or singlet states in the heavy quark limit is always guaranteed as long as the HQS is taken into account. See \ref{sec:examples_nonexotic} for the discussion based on the Weinberg-Tomozawa interaction with SU(8) flavor-spin symmetry.

\section{Mass spectrum and structure of brown muck}\label{sec:spectrum}

So far we have discussed the brown muck and spin-complex as a useful object to understand the structures of multi-hadrons with a heavy quark. Because there exist internal degrees of freedom (light quarks and gluons, light hadrons and the relative angular momentum), the brown muck has not only the ground state but also several excited states.
In this section, we discuss the mass spectrum of the brown muck in relation to its internal structure in excited heavy hadrons.

\subsection{Extraction of mass of brown muck}

As discussed in Section~\ref{sec:massformula}, we identify the mass of the brown muck as $\bar{\Lambda}$ defined in Eq.~(\ref{eq:lambda0}). This is the contribution of the light components to the hadron mass in the heavy quark limit. In reality, we need to extract $\bar{\Lambda}$ from the experimental information with finite charm and bottom quark masses. This is possible by using the heavy hadron mass formula (\ref{eq:mass_formula}) together with the knowledge of the masses of the charm hadrons $\mathrm{H}_{\mathrm{c}}$ with spin $j\pm1/2$ and the bottom hadrons $\mathrm{H}_{\mathrm{b}}$ with $j\pm1/2$ which have common isospin and parity. Then, we use the following relations from Eq.~(\ref{eq:mass_formula})
\begin{align}
M_{\mathrm{H}_{\mathrm{c}}(j+1/2)} &= m_{\mathrm{c}} + \bar{\Lambda} - \frac{\lambda_{1}}{2m_{\mathrm{c}}} + 2j \frac{\lambda_{2}(\mu)}{2m_{\mathrm{c}}} + \mathcal{O}(1/m_{\mathrm{c}}^2), \label{eq:mass_charm_1} \\
M_{\mathrm{H}_{\mathrm{c}}(j-1/2)} &= m_{\mathrm{c}} + \bar{\Lambda} - \frac{\lambda_{1}}{2m_{\mathrm{c}}} - 2(j+1) \frac{\lambda_{2}(\mu)}{2m_{\mathrm{c}}} + \mathcal{O}(1/m_{\mathrm{c}}^2), \label{eq:mass_charm_2} \\         
M_{\mathrm{H}_{\mathrm{b}}(j+1/2)} &= m_{\mathrm{b}} + \bar{\Lambda} - \frac{\lambda_{1}}{2m_{\mathrm{b}}} + 2j \frac{\lambda_{2}(\mu)}{2m_{\mathrm{b}}} + \mathcal{O}(1/m_{\mathrm{b}}^2), \label{eq:mass_bottom_1} \\
M_{\mathrm{H}_{\mathrm{b}}(j-1/2)} &= m_{\mathrm{b}} + \bar{\Lambda} - \frac{\lambda_{1}}{2m_{\mathrm{b}}} - 2(j+1) \frac{\lambda_{2}(\mu)}{2m_{\mathrm{b}}} + \mathcal{O}(1/m_{\mathrm{b}}^2), \label{eq:mass_bottom_2}
\end{align}
for $j \neq 0$ (HQS doublet). 
We use $\mu = m_{\rm c}$ ($m_{\rm b}$) for charm in Eqs.~(\ref{eq:mass_charm_1}) and (\ref{eq:mass_charm_2}) (for bottom in Eqs.~(\ref{eq:mass_bottom_1}) and (\ref{eq:mass_bottom_2})).
For given $M_{\mathrm{H}_{\mathrm{c}}(j+1/2)}$, $M_{\mathrm{H}_{\mathrm{c}}(j-1/2)}$, $M_{\mathrm{H}_{\mathrm{b}}(j+1/2)}$ and $M_{\mathrm{H}_{\mathrm{b}}(j-1/2)}$, we obtain the matrix elements $\bar{\Lambda}$, $\lambda_{1}$, $\lambda_{2}(m_{\mathrm{c}})$ and $\lambda_{2}(m_{\mathrm{b}})$ up to $\mathcal{O}(1/m_{\mathrm{c}}^2)$ and $\mathcal{O}(1/m_{\mathrm{b}}^2)$. We use $m_{\mathrm{c}}=1.30$ GeV and $m_{\mathrm{b}}=4.71$ GeV following Ref.~\cite{Neubert:1993mb}. For $j=0$ (HQS singlet), we use
\begin{eqnarray}
M_{\mathrm{H}_{\mathrm{c}}(1/2)} &=& m_{\mathrm{c}} + \bar{\Lambda} - \frac{\lambda_{1}}{2m_{\mathrm{c}}} + \mathcal{O}(1/m_{\mathrm{c}}^2), \label{eq:mass_charm_0} \\
M_{\mathrm{H}_{\mathrm{b}}(1/2)} &=& m_{\mathrm{b}} + \bar{\Lambda} - \frac{\lambda_{1}}{2m_{\mathrm{b}}} + \mathcal{O}(1/m_{\mathrm{b}}^2), \label{eq:mass_bottom_0}
\end{eqnarray}
to determine $\bar{\Lambda}$ and $\lambda_{1}$ from $M_{\mathrm{H}_{\mathrm{c}}(1/2)}$ and $M_{\mathrm{H}_{\mathrm{b}}(1/2)}$. 
Those formulae should hold in any (multi-)hadrons with a heavy quark.
The information of the specific configurations of the brown muck is reflected in the values of the parameters, such as $\bar{\Lambda}$, $\lambda_{1}$, $\lambda_{2}(m_{\rm c})$ and $\lambda_{2}(m_{\rm b})$, which are defined as the matrix elements of the HQET operators with hadronic states.
We will apply this scheme to obtain $\bar{\Lambda}$, $\lambda_{1}$, $\lambda_{2}(m_{\mathrm{c}})$ and $\lambda_{2}(m_{\mathrm{b}})$ using the experimental mass spectrum of heavy baryons. We also examine the predictions of the constituent quark model and the hadronic molecule model, where the former 
represent the diquark structure of the brown muck, and the latter expresses the spin-complex structure of the brown muck.

We note that the matrix elements $\bar{\Lambda}$, $\lambda_{1}$, and $\lambda_{2}(m_{\rm Q})$ are related to the $m_{\rm Q}$ dependence of the heavy hadron mass $M_{H}(m_{\rm Q})$. For instance, $\bar{\Lambda}$ can be expressed as
\begin{align}
   \bar{\Lambda}=
   \left.
   \frac{d}{d(1/m_{\rm Q})}\left(\frac{M_{\mathrm{H}}}{m_{\rm Q}}\right)
   \right|_{m_{\rm Q}\to \infty}
   =
   \left.
   -m_{\rm Q}^{2}
   \frac{d}{dm_{\rm Q}}\left(\frac{M_{\mathrm{H}}}{m_{\rm Q}}\right)
   \right|_{m_{\rm Q}\to \infty} .
\end{align}
In the same way, $\lambda_{1}$ and $\lambda_{2}(m_{\rm Q})$ can be related to the second derivative of the masses of the spin partners
\begin{align}
   \left.
   \frac{d^{2}}{d(1/m_{\rm Q})^{2}}\left(\frac{M_{\mathrm{H}(j\pm 1/2)}}{m_{\rm Q}}\right)
   \right|_{m_{\rm Q}\to \infty}
   =
   \begin{cases}
   -\lambda_{1}+2j\lambda_{2}(\infty) \\
   -\lambda_{1}-2(j+1)\lambda_{2}(\infty)
   \end{cases} .
\end{align}
Because the heavy quark mass $m_{\rm Q}$ can be arbitrarily adjusted in the lattice QCD simulation, it is, in principle, possible to extract the matrix elements from the $m_{\rm Q}$ dependence.

\subsection{Spectroscopy of brown muck by experimental data}

In experiments, several charm and bottom baryons have been reported to exist so far \cite{Beringer:1900zz}. In the antitriplet sector of SU(3) flavor, there are $\Lambda_{\rm c}$ ($\Lambda_{\rm b}$) with $1/2^{+}$ for the ground state, and $\Lambda_{\rm c}^{\ast}$ ($\Lambda_{\rm b}^{\ast}$) with $1/2^{-}$ and $3/2^{-}$ for the excited states, for the non-strangeness ($S=0$) sector. Two $\Lambda_{\rm b}^{\ast}$ states have been observed recently in LHCb \cite{Aaij:2012da}. Although their $J^{P}$ quantum numbers are not determined yet in experimental observation, it is natural to assign $1/2^{-}$ ($3/2^{-}$) for the states with lower (higher) mass. In the strangeness $S=-1$ sector, there are $\Xi_{\rm c}$ ($\Xi_{\rm b}$) with $1/2^{+}$. 

In the sextet sector, $\Sigma_{\rm c}^{(\ast)}$ ($\Sigma_{\rm b}^{(\ast)}$) with $1/2^{+}$ and $3/2^{+}$ exist for $S=0$. The experimental data on other baryons is not sufficient to complete the flavor partners and/or the spin partners. Such states are not considered in the present discussion.

Based on the existence of the nearby spin states, we assign the ground states of $\Lambda_{\rm c}$ and $\Xi_{\rm c}$ ($\Lambda_{\rm b}$ and $\Xi_{\rm b}$) as HQS singlets, and the excited states $\Lambda_{\rm c}^{*}$ and $\Sigma_{\rm c}^{(*)}$ ($\Lambda_{\rm b}^{*}$ and $\Sigma_{\rm b}^{(*)}$) as HQS doublets. From those charm and bottom baryons, by using Eqs.~\eqref{eq:mass_charm_1}, \eqref{eq:mass_charm_2}, \eqref{eq:mass_bottom_1} and \eqref{eq:mass_bottom_2}, we obtain the matrix elements $\bar{\Lambda}$, $\lambda_{1}$, $\lambda_{2}(m_{\rm c})$ and $\lambda_{2}(m_{\rm b})$ as summarized in Table \ref{table:charm_bottom_baryons_exp}. The ground state $\Lambda_{\rm c}$ and $\Lambda_{\rm b}$ contain the brown muck with $S,I(j^{\mathcal{P}})=0,0(0^{+})$ whose mass is given as $\bar{\Lambda} = 0.88$ GeV. The excited states of the brown muck are also calculated from the mass of the excited baryons. We find that the mass of the brown muck $\bar{\Lambda}$ is of the order of 1 GeV in the heavy baryons, which is comparable with the mass of the light hadrons. The small numbers of $\lambda_{1}$ and $\lambda_{2}(m_{\rm Q})$, together with the $1/2m_{\rm Q}$ factor in the mass formula, indicate that these corrections are indeed small (about ten percents or less to $\bar{\Lambda}$), and the large amount of the heavy baryon mass originates in $\bar{\Lambda}$.

\begin{figure}[tb]
 \begin{center}
  \includegraphics[angle=-90,width=100mm]{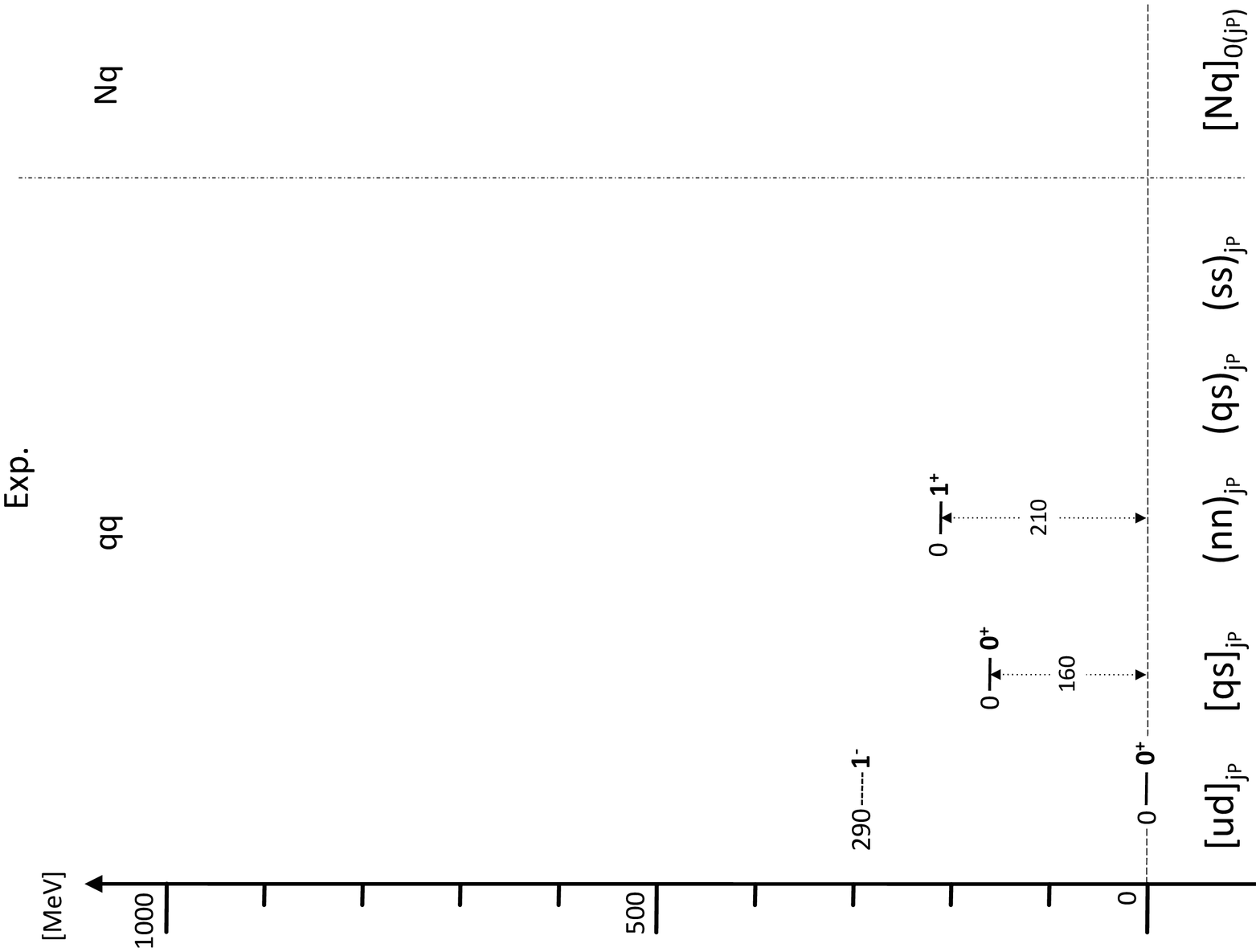}
 \end{center}
 \caption{The mass spectrum of the brown muck extrapolated from the experimental data of the charm and bottom baryons. The masses are measured from the ground state with $S,I(j^{\cal P})=0,0(0^{+})$. See also Table~\ref{table:charm_bottom_baryons_exp}.
 }
\label{fig:brown_muck_mass_exp}
\end{figure}

The mass splittings between states are important quantities in hadron mass spectroscopy.
This is also the case for the brown muck.
In Table \ref{table:charm_bottom_baryons_exp}, the excitation energies of the brown muck are denoted by $\delta \bar{\Lambda}$, which are measured from the mass of the ground state with $S,I(j^{\mathcal{P}})=0,0(0^{+})$ in $\Lambda_{\mathrm{c}}$ and $\Lambda_{\mathrm{b}}$ ($1/2^{+}$). We plot the obtained mass spectrum of the brown muck in the right panel in Fig.~\ref{fig:brown_muck_mass_exp}. The excited state with $S,I(j^{\cal P})=0,0(1^{-})$ in $\Lambda^{\ast}_{\mathrm{c}}$ and $\Lambda^{\ast}_{\mathrm{b}}$  ($1/2^{-}$, $3/2^{-}$) has an excitation energy of $\delta \bar{\Lambda} = 290$ MeV. We also observe that the brown muck with $-1,1/2(0^{+})$ in $\Xi_{\mathrm{c}}$ and $\Xi_{\mathrm{b}}$ ($1/2^{+}$)  has the excitation energy $\delta \bar{\Lambda} =160$ MeV measured from the mass of the brown muck in $\Lambda_{\rm c}$ and $\Lambda_{\rm b}$ ($1/2^{+}$). 
In the flavor sextet sector, the brown muck with $0,1(1^{+})$ in $\Sigma^{(\ast)}_{\mathrm{b}}$ and $\Sigma^{(\ast)}_{\mathrm{c}}$ ($1/2^{+}$, $3/2^{+}$) has the excitation energy $\delta \bar{\Lambda} = 210$ MeV.

\begin{table}[tb]
\caption{The matrix elements $\bar{\Lambda}$, $\lambda_{1}$, $\lambda_{2}(m_{\mathrm{c}})$ and $\lambda_{2}(m_{\mathrm{b}})$ of charm and bottom baryons observed in experiments.
The brown muck is characterized by its strangeness $S$, isospin $I$ and the total spin and parity $j^{\mathcal{P}}$. $\bar{\Lambda}$ and $\delta \bar{\Lambda}$ are given in units of GeV, and $\lambda_{1}$, $\lambda_{2}(m_{\mathrm{c}})$ and $\lambda_{2}(m_{\mathrm{b}})$ are in units of GeV$^2$.
 In the last row, we show the corresponding diquarks in the constituent quark model, which are denoted by $[q_{1}q_{2}]_{j^{\cal{P}}}$ for triplet in light flavor SU(3) symmetry and $(q_{1}q_{2})_{j^{\cal{P}}}$ for sextet ($q_{1}$, $q_{2}$=$u$, $d$, $s$; $n=u$, $d$) (see also Section~4.3).}
\begin{center}
\begingroup
\renewcommand{\arraystretch}{1.2}
\begin{tabular}{|c|c|c|c|c|c|c|}
\hline
$S,I(j^{\cal{P}})$ & baryons ($J^{P}$) &  $\bar{\Lambda}$ & $\delta \bar{\Lambda}$  & $\lambda_{1}$ & $\lambda_{2}(m_{\mathrm{c}})$, $\lambda_{2}(m_{\mathrm{b}})$ & diquark  \\
\hline
$0,0(0^{+})$ & $\Lambda_{\mathrm{c}}$, $\Lambda_{\mathrm{b}}$ ($1/2^{+}$) & 0.88 & 0 & $-0.28$ & --- & $[ud]_{0^{+}}$  \\
$0,0(1^{-})$ & $\Lambda^{\ast}_{\mathrm{c}}$, $\Lambda^{\ast}_{\mathrm{b}}$  ($1/2^{-}$, $3/2^{-}$) & 1.17 & 0.29 & $-0.39$ & 0.014, 0.012  & $[ud]_{1^{-}}$ \\
$-1,1/2(0^{+})$ & $\Xi_{\mathrm{c}}$, $\Xi_{\mathrm{b}}$ ($1/2^{+}$) & 1.04 & 0.16 & $-0.32$ & --- & $[ns]_{0^{+}}$ \\
$0,1(1^{+})$ & $\Sigma^{(\ast)}_{\mathrm{c}}$, $\Sigma^{(\ast)}_{\mathrm{b}}$ ($1/2^{+}$, $3/2^{+}$) & 1.09 & 0.21 & $-0.29$ & 0.028, 0.032 & $(nn)_{1^{+}}$ \\
\hline
\end{tabular}
\endgroup
\end{center}
\label{table:charm_bottom_baryons_exp}
\end{table}%

\subsection{Brown muck spectroscopy by theoretical studies}

In the previous subsection, we have obtained the mass spectrum of the brown muck based on the experimental information. However, the number of the observed charm and bottom baryons is not yet sufficient to perform the systematic spectroscopy of the brown muck. Especially, there lacks much information about further excited states. Theoretically, it is predicted that there exist higher excited states forming flavor partners and spin partners in the constituent quark model, as recently studied in detail by Roberts and Pervin in Ref.~\cite{Roberts:2007ni}.  In the hadronic molecule models, the existence of exotic baryons is also predicted, for instance, by the potential model as discussed in Section~\ref{sec:examples_exotic} (see also Refs.~\cite{Yasui:2009bz,Yamaguchi:2011xb,Yamaguchi:2011qw,Yamaguchi:2013ty}).
 These theoretical calculations enable us to extract the masses of the higher excited states of the brown muck 
 with different quantum numbers. Moreover, the structure of the brown muck in these predictions is related to the model space, for instance, the diquark $qq$ in the constituent quark model and the spin-complex with $Nq$ structure in the hadronic molecule model.  
 For the study of internal structure, from now on, we make use of the theoretical predictions of the heavy hadron spectrum.

\subsubsection{Constituent quark model}

\begin{figure}[p]
\rotatebox{90}{\begin{minipage}{\textheight}
  \begin{center}
   \includegraphics[angle=0,width=200mm]{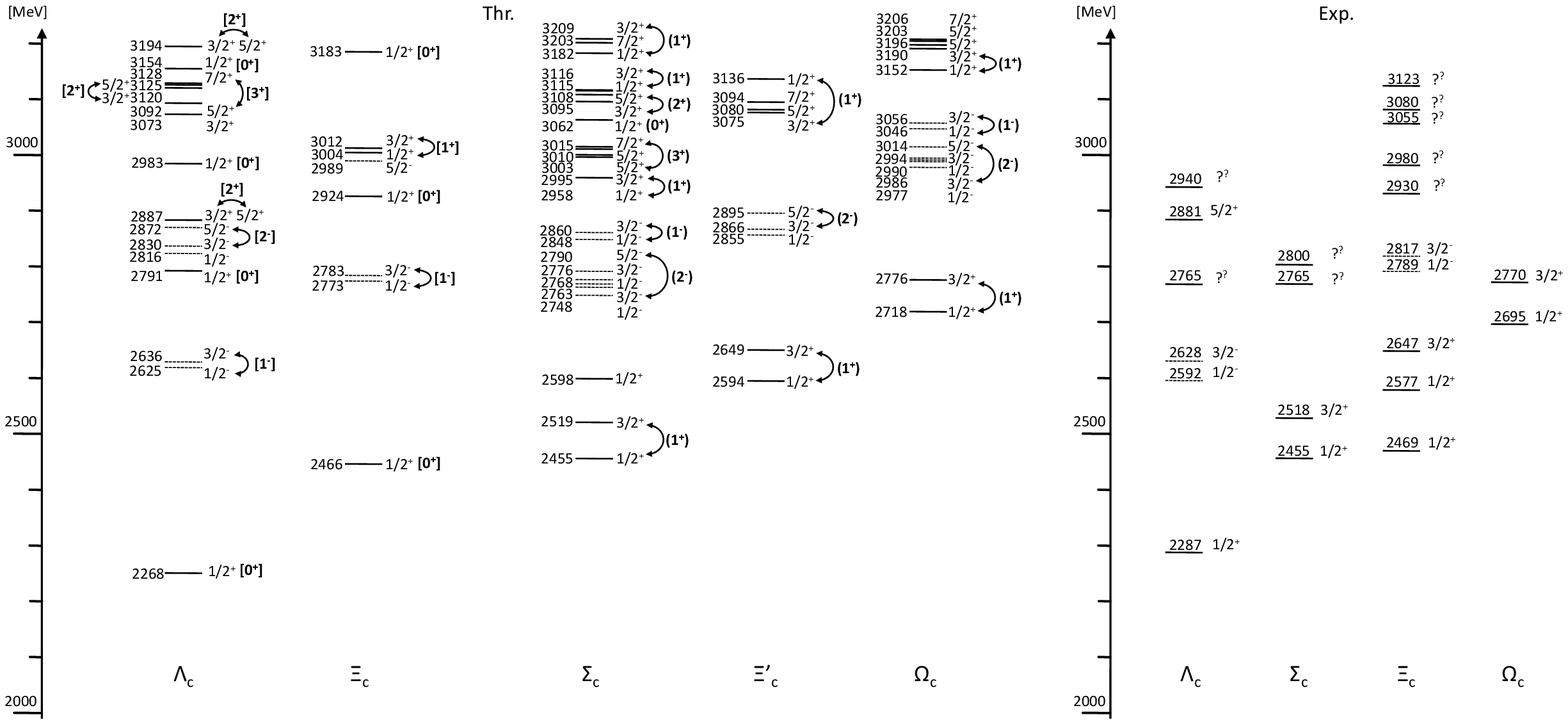}
  \end{center}
 \caption{Comparison of the mass spectrum of the charm baryons in the constituent quark model (CQM)~\cite{Roberts:2007ni} (left) and the experimental data~\cite{Beringer:1900zz} (right). The masses are given in units of MeV. $J^{P}$ with half-integer $J$ is the total spin and parity of the baryon in Ref.~\cite{Roberts:2007ni}, and $j^{\mathcal{P}}$ ($\mathcal{P}=P$) with integer $j$ is the total spin and parity of the brown muck identified in Ref.~\cite{Roberts:2007ni}. We use the square and round brackets as $[j^{\mathcal{P}}]$ and $(j^{\mathcal{P}})$ to indicate the flavor antitriplet and sextet states, respectively.}
\label{fig:charm_baryon_mass}
\end{minipage}
}
\end{figure}

\begin{figure}[p]
\rotatebox{90}{\begin{minipage}{\textheight}
  \begin{center}
   \includegraphics[angle=0,width=200mm]{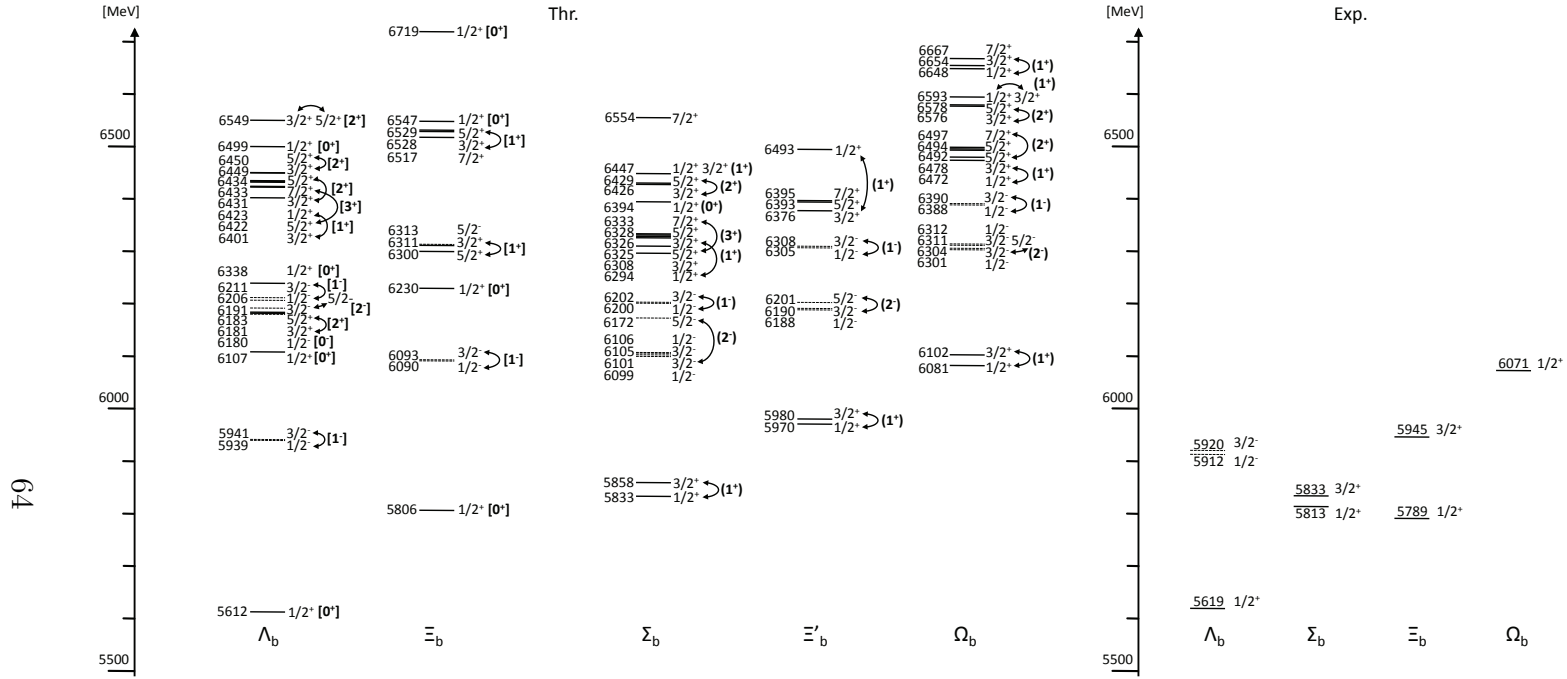}
  \end{center}
 \caption{Comparison of the mass spectrum of the bottom baryons in the constituent quark model (CQM)~\cite{Roberts:2007ni} (left) and the experimental data~\cite{Beringer:1900zz} (right) The conventions are the same as Fig.~\ref{fig:charm_baryon_mass}.}
\label{fig:bottom_baryon_mass}
\end{minipage}
}
\end{figure}

There have been in literature many discussions about the masses of heavy baryons in the quark model calculations~\cite{Copley:1979wj,Capstick:1986bm}. As one of the recent works, we consider the model by Roberts and Pervin \cite{Roberts:2007ni} where the baryons are classified in terms of the HQS. We summarize their results in Figs.~\ref{fig:charm_baryon_mass} and \ref{fig:bottom_baryon_mass} for charm and bottom sectors, respectively, in comparison with experimental data. In their analysis, some spin partners (HQS doublet/singlet) were identified, when the wave functions of the light quarks were similar to each other.

It was shown that $\Lambda_{\rm c}$ ($\Lambda_{\rm b}$) in the ground state is identified as the HQS singlet state which contains the brown muck with $S,I(j^{\mathcal{P}})=0,0(0^{+})$. Their excited states, $\Lambda_{\rm c}^{\ast}$ ($\Lambda_{\rm b}^{\ast}$) with $J^{P}=1/2^{-}$ and $3/2^{-}$, are identified as the HQS doublet states which contain the brown muck with $0,0(1^{-})$. Similarly, $\Xi_{\rm c}$ ($\Xi_{\rm b}$) with $J^{P}=1/2^{+}$ is the HQS singlet state containing the brown muck with $-1,1/2(0^{+})$. In the sextet sector, $\Sigma_{\rm c}^{(\ast)}$ ($\Sigma_{\rm b}^{(\ast)}$) with $1/2^{+}$ and $3/2^{+}$ are the HQS doublet states containing the brown muck with $0,1(1^{+})$. Those assignments are consistent with the result from the spectrum observed in experiments. In addition, many other states, which will be explored in future experiments, were predicted in Ref.~\cite{Roberts:2007ni}. Some of them are identified as the HQS doublets/singlets. 

For those excited baryons, we calculate the matrix elements $\bar{\Lambda}$, $\lambda_{1}$, $\lambda_{2}(m_{c})$ and $\lambda_{2}(m_{b})$ by applying Eqs.~\eqref{eq:mass_charm_1}, \eqref{eq:mass_charm_2}, \eqref{eq:mass_bottom_1} and \eqref{eq:mass_bottom_2}. The results are shown in Tables \ref{table:Lambda} and \ref{table:Xi} for the strangeness $S=0$ and $S=-1$ in the flavor antitriplet sector, and in Tables \ref{table:Sigma}, \ref{table:Xi'} and \ref{table:Omega} for $S=0$, $S=-1$ and $S=-2$ states in the sextet sector. In those tables, $j^{\mathcal{P}}$ is the spin and parity of the brown muck, and $J^{P}=j\pm 1/2^{P}$ ($P=\cal{P}$) is the total angular momenta and parity of the baryons containing the common brown muck ($j\ge 1$). For $j=0$, there is no spin partner. Because this model is based on the three-quark configuration, the brown muck in the predicted heavy baryons is considered to be a diquark state. We denote it by the valence quark component as $[qq]$ and $(qq)$, where the square (round) bracket stand for the antisymmetric (symmetric) combination corresponding to triplet (sextet) in light flavor SU(3) symmetry.

Let us see the details of the obtained matrix elements. We find that the value of $\bar{\Lambda}$ for the ground states $\Lambda_{\rm c}$ and $\Lambda_{\rm b}$ $(1/2^{+})$ is in good agreement with those extracted from the experimental spectrum in Table~\ref{table:charm_bottom_baryons_exp}. It is also the case for the excited state $\Lambda_{\rm c}^{*}$ and $\Lambda_{\rm b}^{*}$ $(1/2^{-},3/2^{-})$, $\Xi_{\rm c}$ and $\Xi_{\rm b}$ $(1/2^{+})$, and $\Sigma_{\rm c}^{(*)}$ and $\Sigma_{\rm b}^{(*)}$ $(1/2^{+},3/2^{+})$. This is of course a consequence of the success of the quark model prediction for the experimental data of these states in Ref.~\cite{Roberts:2007ni}. It also indicates that the brown muck of these low-lying baryons is dominated by the diquark structure.
We note that the values of $\lambda_{1}$, $\lambda_{2}(m_{\rm c})$ and $\lambda_{2}(m_{\rm b})$ together with the factors $1/2m_{\rm c}$ and $1/2m_{\rm b}$ are smaller than that of $\bar{\Lambda}$, and hence the $1/m_{\rm Q}$ expansion works well.

In addition to the observed states, the quark model predicts several excited states \cite{Roberts:2007ni}. From this information, we can construct the mass spectrum of the brown muck. One of the most important quantities in the mass spectrum is the excitation energy from the ground state. We define the excitation energy $\delta \bar{\Lambda}$ measured from the lowest energy state with a given flavor quantum numbers. In the antitriplet sector, the ground state is $j^{\mathcal{P}}=0^{+}$ for both $(S,I)=(0,0)$ and $(-1,1/2)$. In contrast, in the sextet sector, the ground state is $j^{\mathcal{P}}=1^{+}$ for all channels with $(S,I)=(0,1)$, $(-1,1/2)$ and $(-2,0)$. Moreover, the mass splitting of the ground states in the sextet is almost equal spacing ($m_{(qs)}-m_{(ud)}=124$ MeV $\simeq$ $m_{(ss)}-m_{(qs)}=116$ MeV), which indicates that the Gell-Mann--Okubo formula works well also for the brown muck. In this way, the structure of the brown muck in the constituent quark model follows the flavor SU(3) symmetry.

\begin{figure}[tb]
  \begin{center}
   \includegraphics[angle=-90,width=100mm]{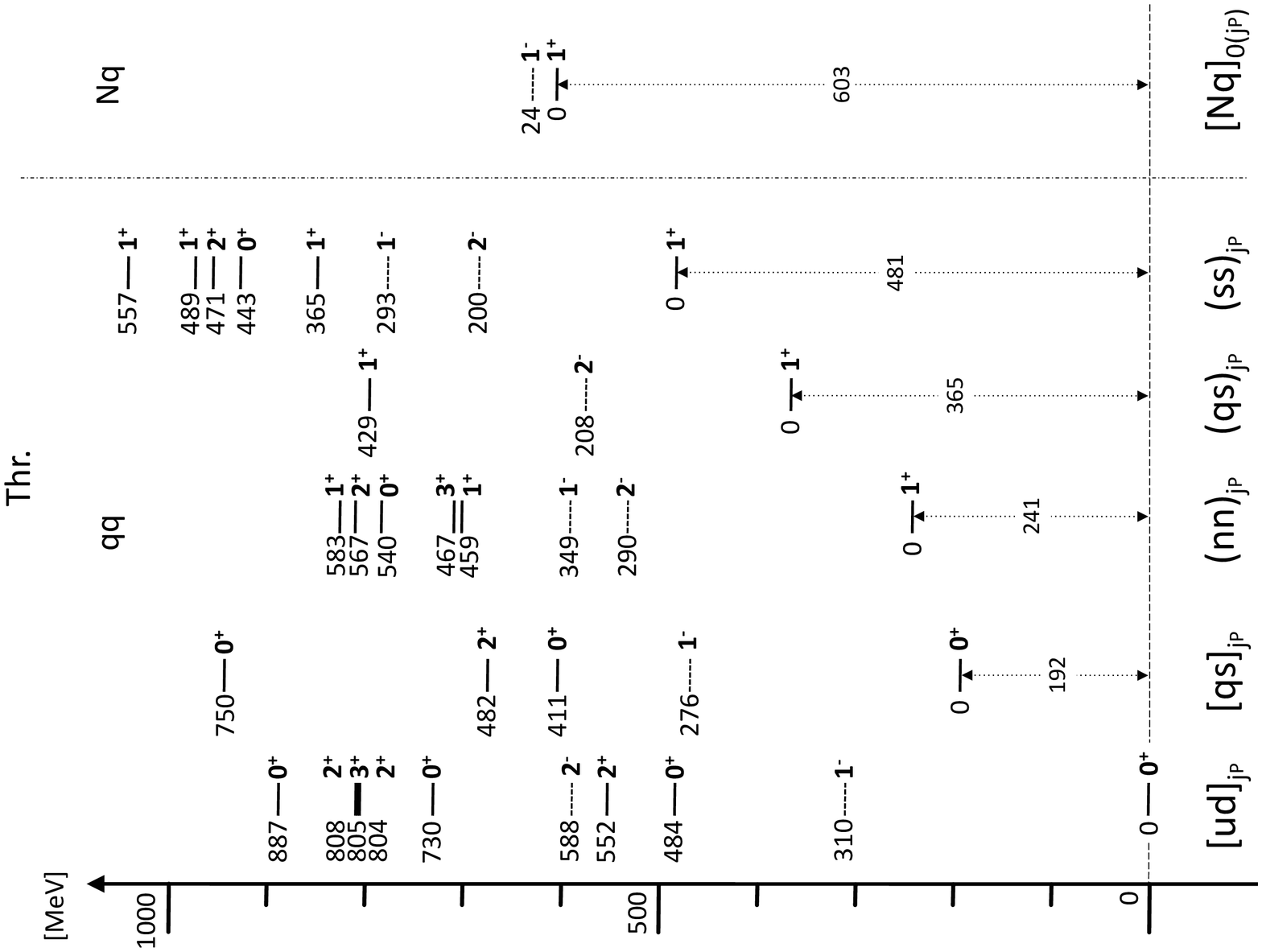}
  \end{center}
 \caption{The mass spectrum of the brown muck extrapolated from the prediction of the constituent quark model~\cite{Roberts:2007ni} and the boson exchange potential model~\cite{Yasui:2009bz,Yamaguchi:2011xb,Yamaguchi:2011qw}. The numbers in left of the bars are the excitation energy measured from the ground state of each flavor. The numbers in the dashed arrows are the mass difference of the ground states in each quantum number from the lowest $[ud]$ brown muck in $0^{+}$.
}
\label{fig:brown_muck_mass_model}
\end{figure}

From the analyses above, we obtain the mass spectrum of the brown muck, as summarized in Fig.~\ref{fig:brown_muck_mass_model}.
Some lower states are comparable with ones in Fig.~\ref{fig:brown_muck_mass_exp}, and many higher states are predicted.
In any case, the theoretical results suggest rich structures in the excited spectrum of the brown muck in the diquark picture. When the experimental measurements of the charm and bottom baryons are further performed in future, we will be able to understand more about the structure of the brown muck.

\begin{table}[p]
\caption{The matrix elements $\bar{\Lambda}$, $\lambda_{1}$, $\lambda_{2}(m_{\mathrm{c}})$ and $\lambda_{2}(m_{\mathrm{b}})$ from Eq.~(\ref{eq:mass_formula}) for $\Lambda_{\mathrm{c}}$ and $\Lambda_{\mathrm{b}}$ baryons with masses $M_{\Lambda_{\mathrm{c}}}$ and $M_{\Lambda_{\mathrm{b}}}$ obtained in the constituent quark model~\cite{Roberts:2007ni}. $j^{\mathcal{P}}$ is the total spin and parity of the brown muck $[ud]$, $J^{P}$ is the total spin and parity of the heavy baryon belonging to the HQS multiplet. $\delta \bar{\Lambda}$ is the excitation energy from the ground state (0.877 GeV for $j^{\mathcal{P}}=1/2^{+}$ in this sector). $\bar{\Lambda}$ and $\delta \bar{\Lambda}$ are given in units of GeV, and $\lambda_{1}$, $\lambda_{2}(m_{\mathrm{c}})$ and $\lambda_{2}(m_{\mathrm{b}})$ are in units of GeV$^2$.
}
\begin{center}
\begingroup
\renewcommand{\arraystretch}{1.2}
\begin{tabular}{|c|c|c|c|c|c|c|c|}
\hline
$j^{\mathcal{P}}$ & $J^{P}$ & $M_{\Lambda_{\mathrm{c}}}$, $M_{\Lambda_{\mathrm{b}}}$ &  \multirow{1}{*}{$\bar{\Lambda}$} & $\delta \bar{\Lambda}$ & $\lambda_{1}$ & $\lambda_{2}(m_{\mathrm{c}})$ & $\lambda_{2}(m_{\mathrm{b}})$ \\
\hline
 & & 2.268, 5.612 & 0.877 & 0 & $-$0.237 & --- & --- \\
\cline{3-8}
 \multirow{2}{*}{$0^{+}$} & \multirow{2}{*}{$1/2^{+}$} & 2.791, 6.107 & 1.361 & 0.484 & $-$0.338 & --- & --- \\
\cline{3-8}
 &  & 2.983, 6.338 & 1.607 & 0.730 & $-$0.198 & --- & --- \\
\cline{3-8}
 & & 3.154, 6.499 & 1.764 & 0.887 & $-$0.233 & --- &  --- \\
\hline
 & & (2.887,\,2.887), (6.181,\,6.183) & 1.429 & 0.552 & $-$0.412 & 0.0 & 0.003 \\
\cline{3-8}
 $2^{+}$ & $(3/2^{+},5/2^{+})$ & (3.120,\,3.125), (6.431,\,6.434) & 1.685 & 0.808 & $-$0.360 & 0.002 & 0.005 \\
\cline{3-8}
 & & (3.194,\,3.194), (6.449,\,6.450) & 1.681 & 0.804 & $-$0.554 & 0.0 & 0.001 \\
\hline
 $3^{+}$ & $(5/2^{+},7/2^{+})$ & (3.092,\,3.128), (6.422,\,6.433) & 1.682 & 0.805 & $-$0.339 & 0.007 & 0.007 \\
\hline
 $1^{-}$ & $(1/2^{-},3/2^{-})$ & (2.625,\,2.636), (5.930,\,5.941) & 1.187 & 0.310 & $-$0.378 & 0.005 & 0.017 \\
\hline
 $2^{-}$ & $(3/2^{-},5/2^{-})$ & (2.830,\,2.872), (6.191,\,6.206) & 1.465 & 0.588 & $-$0.234 & 0.011 & 0.014 \\
\hline
\end{tabular}
\endgroup
\end{center}
\label{table:Lambda}
\end{table}%

\begin{table}[p]
\caption{The matrix elements for $\Xi_{\mathrm{c}}$ and $\Xi_{\mathrm{b}}$ based on the result of Ref.~\cite{Roberts:2007ni}. The contained brown muck is $[ns]$ with $n=u$ or $d$. The conventions are the same as Table~\ref{table:Lambda}.}
\begin{center}
\begingroup
\renewcommand{\arraystretch}{1.2}
\begin{tabular}{|c|c|c|c|c|c|c|c|}
\hline
 $j^{\mathcal{P}}$ & $J^{P}$ & $M_{\Xi_{\mathrm{c}}}$, $M_{\Xi_{\mathrm{b}}}$ & $\bar{\Lambda}$ & $\delta\bar{\Lambda}$ & $\lambda_{1}$ & $\lambda_{2}(m_{\mathrm{c}})$ & $\lambda_{2}(m_{\mathrm{b}})$ \\
\hline
 & & 2.466, 5.806 & 1.069 & 0 & $-$0.251 & & \\
\cline{3-8}
 \multirow{2}{*}{$0^{+}$} & \multirow{2}{*}{$1/2^{+}$} & 2.924, 6.230 & 1.480 & 0.411 & $-$0.373 & --- & --- \\
\cline{3-8}
 & & 3.183, 6.547 & 1.819 & 0.750 & $-$0.165 & --- & --- \\
\cline{3-8}
 & & ------ & --- & --- & --- & --- & --- \\
\hline
 & & (3.012,\,3.004), (6.311,\,6.300) & 1.551 & 0.482 & $-$0.405 & $-$0.002 & $-$0.010 \\
\cline{3-8}
 $2^{+}$ & $(3/2^{+},5/2^{+})$ & ------ & --- & --- & --- & --- & --- \\
\cline{3-8}
 & & ------ & --- & --- & --- & --- & --- \\
\hline
 $3^{+}$ & $(5/2^{+},7/2^{+})$ & ------ & --- & --- & --- & --- & --- \\
\hline
 $1^{-}$ & $(1/2^{-},3/2^{-})$ & (2.773,\,2.783), (6.090,\,6.093) & 1.345 & 0.276 & $-$0.351 & 0.004 & 0.005 \\
\hline
 $2^{-}$ & $(3/2^{-},5/2^{-})$ & ------ & --- & --- & --- & --- & --- \\
\hline
\end{tabular}
\endgroup
\end{center}
\label{table:Xi}
\end{table}%

\begin{table}[p]
\caption{The matrix elements for $\Sigma_{\mathrm{c}}$ and $\Sigma_{\mathrm{b}}$ based on the result of Ref.~\cite{Roberts:2007ni}. The contained brown muck is $(nn)$. The conventions are the same as Table~\ref{table:Lambda}.}
\begin{center}
\begingroup
\renewcommand{\arraystretch}{1.2}
\begin{tabular}{|c|c|c|c|c|c|c|c|}
\hline
 $j^{\mathcal{P}}$ & $J^{P}$ & $M_{\Sigma_{\mathrm{c}}}$, $M_{\Sigma_{\mathrm{b}}}$ & $\bar{\Lambda}$ & $\delta \bar{\Lambda}$ & $\lambda_{1}$ & $\lambda_{2}(m_{\mathrm{c}})$ & $\lambda_{2}(m_{\mathrm{b}})$ \\
\hline
 $0^{+}$ & $1/2^{+}$ & 3.062, 6.397 & 1.658 & 0.540 & $-$0.269 & --- & --- \\
\hline
 & & (2.455,\,2.519), (5.833,\,5.858) & 1.118 & 0 & $-$0.208 & 0.027 & 0.039 \\
\cline{3-8}
 \multirow{2}{*}{$1^{+}$} & \multirow{2}{*}{$(1/2^{+},3/2^{+})$}   & (2.958,\,2.995), (6.294,\,6.326) & 1.577 & 0.459 & $-$0.278 & 0.016 & 0.050  \\
\cline{3-8}
 & & (3.115,\,3.116), (6.447,\,6.447) & 1.701 & 0.583 & $-$0.283 & 0.0 & 0.0 \\
\cline{3-8}
 & & ------ & --- & --- & --- & --- & --- \\
\hline
 $2^{+}$ & $(3/2^{+},5/2^{+})$ & (3.095,\,3.108), (6.426,\,6.429) &1.685 & 0.567 & $-$0.305 & 0.003 & 0.003 \\
\hline
 $3^{+}$ & $(5/2^{+},7/2^{+})$ & (3.003,\,3.015), (6.325,\,6.333) & 1.585 & 0.467 & $-$0.324 & 0.002 & 0.005 \\
\hline
 $1^{-}$ & $(1/2^{-},3/2^{-})$ & (2.848,\,2.860), (6.200,\,6.202) & 1.467 & 0.349 & $-$0.232 & 0.005 & 0.003 \\
\hline
 $2^{-}$ & $(3/2^{-},5/2^{-})$ & (2.763,\,2.790), (6.101,\,6.172) & 1.416 & 0.290 & $-$0.164 & 0.007 & 0.067 \\
\hline
\end{tabular}
\endgroup
\end{center}
\label{table:Sigma}
\end{table}%

\begin{table}[p]
\caption{The matrix elements for $\Xi'_{\mathrm{c}}$ and $\Xi'_{\mathrm{b}}$ based on the result of Ref.~\cite{Roberts:2007ni}. The contained brown muck is $(ns)$ with $n=u$ or $d$. The conventions are the same as Table~\ref{table:Lambda}.}
\begin{center}
\begingroup
\renewcommand{\arraystretch}{1.2}
\begin{tabular}{|c|c|c|c|c|c|c|c|}
\hline
 $j^{\mathcal{P}}$ & $J^{P}$ & $M_{\Xi'_{\mathrm{c}}}$, $M_{\Xi'_{\mathrm{b}}}$ & $\bar{\Lambda}$ & $\delta\bar{\Lambda}$ & $\lambda_{1}$ & $\lambda_{2}(m_{\mathrm{c}})$ & $\lambda_{2}(m_{\mathrm{b}})$ \\
\hline
 $0^{+}$ & $1/2^{+}$ & ------  & --- & --- & --- & --- & --- \\
\hline
& & (2.594,\,2.649), (5.970,\,5.980) & 1.242 & 0 & $-$0.230 & 0.024 & 0.016 \\
\cline{3-8}
\multirow{2}{*}{$1^{+}$} & \multirow{2}{*}{$(1/2^{+},3/2^{+})$} & (3.136,\,3.075), (6.493,\,6.376) & 1.671 & 0.429 & $-$0.324 & $-$0.026 & $-$0.184 \\
\cline{3-8}
 & & ------ & --- & --- & --- & --- & --- \\
\cline{3-8}
 & & ------ & --- & --- & --- & --- & --- \\
\hline
 $2^{+}$ & $(3/2^{+},5/2^{+})$ &  ------ & --- & --- & --- & --- & --- \\
\hline
 $3^{+}$ & $(5/2^{+},7/2^{+})$ &  ------ & --- & --- & --- & --- & --- \\
\hline
 $1^{-}$ & $(1/2^{-},3/2^{-})$ &  ------ & --- & --- & --- & --- & --- \\
\hline
 $2^{-}$ & $(3/2^{-},5/2^{-})$ & (2.866,\,2.895), (6.190,\,6.201) & 1.450 & 0.208 & $-$0.348 & 0.008 & 0.103 \\
\hline
\end{tabular}
\endgroup
\end{center}
\label{table:Xi'}
\end{table}%

\begin{table}[p]
\caption{The matrix elements for $\Omega_{\mathrm{c}}$ and $\Omega_{\mathrm{b}}$ based on the result of Ref.~\cite{Roberts:2007ni}. The contained brown muck is $(ss)$. The conventions are the same as Table~\ref{table:Lambda}.}
\begin{center}
\begingroup
\renewcommand{\arraystretch}{1.2}
\begin{tabular}{|c|c|c|c|c|c|c|c|}
\hline
 $j^{\mathcal{P}}$ & $J^{P}$ & $M_{\Omega_{\mathrm{c}}}$, $M_{\Omega_{\mathrm{b}}}$ & $\bar{\Lambda}$ & $\delta \bar{\Lambda}$ & $\lambda_{1}$ & $\lambda_{2}(m_{\mathrm{c}})$ & $\lambda_{2}(m_{\mathrm{b}})$ \\
\hline
 $0^{+}$ & $1/2^{+}$ & 3.234, 6.511 & 1.801 & 0.443 & $-$0.334 & --- & --- \\
\hline
 & & (2.718,\,2.776), (6.081,\,6.102) & 1.358 & 0 & $-$0.257 & 0.025 & 0.033 \\
\cline{3-8}
 \multirow{2}{*}{$1^{+}$}  & \multirow{2}{*}{$(1/2^{+},3/2^{+})$} & (3.152,\,3.190), (6.472,\,6.478) & 1.723 & 0.365 & $-$0.400 & 0.016 & 0.009 \\
\cline{3-8}
 & & (3.275,\,3.280), (6.593,\,6.593) & 1.847 & 0.489 & $-$0.342 & 0.002 & 0.0 \\
\cline{3-8}
 & & (3.299,\,3.321), (6.648,\,6.654) & 1.915 & 0.557 & $-$0.257 & 0.010 & 0.009 \\
\hline
 $2^{+}$ & $(3/2^{+},5/2^{+})$ & (3.262,\,3.273), (6.576,\,6.578) & 1.829 & 0.471 & $-$0.364 & 0.003 & 0.002 \\
\hline
 $3^{+}$ & $(5/2^{+},7/2^{+})$ & ------ & --- & --- & --- & --- & --- \\
\hline
 $1^{-}$ & $(1/2^{-},3/2^{-})$ & (3.046,\,3.056), (6.388,\,6.390) & 1.651 & 0.293 & $-$0.263 & 0.004 & 0.003  \\
\hline
 $2^{-}$ & $(3/2^{-},5/2^{-})$ & (2.986,\,3.014), (6.304,\,6.311) & 1.558 & 0.200 & $-$0.375 & 0.007 & 0.007 \\
\hline
\end{tabular}
\endgroup
\end{center}
\label{table:Omega}
\end{table}%

\subsubsection{Hadronic molecule model}

In Section~\ref{sec:examples_exotic}, we have discussed the spin-complex in the exotic baryon composed of $\bar{P}^{(\ast)}N$. Because this model is based on the hadronic molecule picture, we have obtained the brown muck with the spin-complex structure of $[Nq]_{0(1^{+})}$ for the HQS doublet $(1/2^{-},3/2^{-})$ and $[Nq]_{0(1^{-})}$ for $(1/2^{+},3/2^{+})$, where we explicitly denote the isospin $I$ (the strangeness is $S=0$). The mass spectrum of the $\bar{D}^{(\ast)}N$ and $B^{(\ast)}N$ states with finite charm and bottom quark masses are obtained and the results are summarized in Table~\ref{table:DbarN_BN_mass}.

In order to extract $\bar{\Lambda}$ and other matrix elements ($\lambda_{1}$, $\lambda_{2}(m_{\rm c})$ and $\lambda_{2}(m_{\rm b})$) for $\bar{P}^{(\ast)}N$, we apply the mass formulae (\ref{eq:mass_charm_1})-(\ref{eq:mass_bottom_2}) for those $\bar{D}^{(\ast)}N$ and $B^{(\ast)}N$ states. Consequently, we obtain $\bar{\Lambda}$, $\lambda_{1}$, $\lambda_{2}(m_{\mathrm{c}})$ and $\lambda_{2}(m_{\mathrm{b}})$ as summarized in Table~\ref{table:charm_bottom_baryons_exotic}. The mass of the spin-complex $[Nq]_{0(1^{+})}$ is $\bar{\Lambda}=1.476$ GeV, and the mass of $[Nq]_{0(1^{-})}$ is $\bar{\Lambda}=1.500$ GeV. Thus, there is the mass difference $\delta \bar{\Lambda} = 0.024$ GeV between $[Nq]_{0(1^{+})}$ and $[Nq]_{0(1^{-})}$.
We note again that the $1/m_{\rm Q}$ expansion works well due to the small values of $\lambda_{1}$, $\lambda_{2}(m_{\rm c})$ and $\lambda_{2}(m_{\rm b})$ together with the factors $1/2m_{\rm c}$ and $1/2m_{\rm b}$.
In Fig.~\ref{fig:brown_muck_mass_model}, we show the obtained mass spectrum of the spin-complex $[Nq]_{0(j^{\cal P})}$. Their masses are about 600 MeV heavier than the lowest energy state of the diquark $[ud]_{0^{+}}$.

The spin-complex with $[N\bar{q}]$ structure may exist as another component of the brown muck around this energy region, when $P^{(\ast)}N$ components are considered.
In fact, the meson-exchange interaction for $P^{(\ast)}N$ may be obtained from that in $\bar{P}^{(\ast)}N$ by the G-parity transformation. If this is the case, the spectrum of the brown muck with the $[ud]$ configuration will be affected by the mixing with the $[N\bar{q}]$ spin-complex, especially around 600 MeV above the ground state. It is therefore interesting to compare $P^{(\ast)}N$ and $\bar{P}^{(\ast)}N$ states and to complete the experimental measurements of the heavy baryons in this energy region, which will provide the information on the structure of the brown muck through the comparison with the spectrum of the brown muck in the diquark picture in Fig.~\ref{fig:brown_muck_mass_model}.

\begin{table}[tbp]
\caption{Masses of HQS doublets $(1/2^{-},3/2^{-})$ and $(1/2^{+},3/2^{+})$ for $\bar{\mathrm{D}}^{(\ast)}\mathrm{N}$ and $\mathrm{B}^{(\ast)}\mathrm{N}$ with the $\pi \rho\, \omega$ potential from
 Ref.~\cite{Yamaguchi:2011qw}. The values are given in GeV.}
\begin{center}
\begingroup
\renewcommand{\arraystretch}{1.2}
\begin{tabular}{|c|c|c|}
\hline
                                 & $\bar{\mathrm{D}}^{(\ast)}\mathrm{N}$ & $\mathrm{B}^{(\ast)}\mathrm{N}$ \\
\hline
$(1/2^{-},3/2^{-})$ & $(2.805, 2.920)$ & $(6.196,6.226)$ \\
$(1/2^{+},3/2^{+})$ & $(2.834,2.955)$ & $(6.225,6.251)$ \\
\hline
\end{tabular}
\endgroup
\end{center}
\label{table:DbarN_BN_mass}
\end{table}%

\begin{table}[tbp]
\caption{The matrix elements $\bar{\Lambda}$, $\lambda_{1}$, $\lambda_{2}(m_{\mathrm{c}})$ and $\lambda_{2}(m_{\mathrm{b}})$ of exotic baryons which contain the spin-complex (SC) $[Nq]_{I(j^{\mathcal{P}})}$ with isospin $I$, total spin and parity $j^{\mathcal{P}}$. The $\pi \rho\, \omega$ potential is used. $\delta{\bar{\Lambda}}$ is the difference of $\bar{\Lambda}$ from the ground state $[Nq]_{0(1^{+})}$. $\bar{\Lambda}$ and $\delta \bar{\Lambda}$ are given in units of GeV, and $\lambda_{1}$, $\lambda_{2}(m_{\mathrm{c}})$ and $\lambda_{2}(m_{\mathrm{b}})$ are in units of GeV$^2$.}
\begin{center}
\begingroup
\renewcommand{\arraystretch}{1.2}
\begin{tabular}{|c|c|c|c|c|c|c|}
\hline
SC & baryons ($J^{P}$) &  $\bar{\Lambda}$ & $\delta{\bar{\Lambda}}$ & $\lambda_{1}$  & $\lambda_{2}(m_{\mathrm{c}})$, $\lambda_{2}(m_{\mathrm{b}})$  \\
\hline
$[Nq]_{0(1^{+})}$ & $\bar{\mathrm{P}}^{(\ast)}\mathrm{N}$ ($1/2^{-}$, $3/2^{-}$) & 1.476 & 0 & $-$0.272 & 0.0500, 0.0469 \\
$[Nq]_{0(1^{-})}$ & $\bar{\mathrm{P}}^{(\ast)}\mathrm{N}$ ($1/2^{+}$, $3/2^{+}$) & 1.500 & 0.024 & $-$0.296 & 0.0526, 0.0408 \\
\hline
\end{tabular}
\endgroup
\end{center}
\label{table:charm_bottom_baryons_exotic}
\end{table}%

\section{Summary}

We have discussed the decoupling of the heavy-quark spin from the total spin of the light components in the multi-hadrons composed of a heavy hadron and surrounding light hadrons in the heavy quark limit. In those systems, the HQS plays an important role for the classification of the spectrum and the structure of the heavy multi-hadrons. We have introduced the spin-complex for the state composed of the light quarks and gluons together with the light hadrons in a heavy hadron. The spin-complex is one of the configurations of the brown muck which is defined as everything except for the heavy-quark spin in the multi-hadrons. The decoupling of the heavy-quark spin and the total spin $j$ of the spin-complex induces the HQS doublet $(j-1/2,j+1/2)$ for $j \neq 0$ with degenerate masses and the HQS singlet for $j=0$. 

We have studied the consequences of this general statement by examining the exotic baryon system made of a heavy meson and a nucleon. Various meson-baryon states are classified in terms of the spin-complex basis, by which the structure of the hadrons in the heavy quark limit is investigated. Employing a meson-exchange potential for the interaction of the heavy meson and the nucleon, we have found that the Hamiltonian of the system is block-diagonalized in the spin-complex basis and the eigenvalues become the same for the $j\pm 1/2$ sectors containing the same spin-complex. Thus, the spin degeneracy indeed occurs, when the interaction is constructed by respecting the HQS. Moreover, the wave function of the bound state exhibits the mixing ratios of the meson-baryon components obtained from the group theoretical argument. From these results, we conclude that the HQS is a useful guiding principle to study the structure of the multi-hadron systems.

We have also studied the excitations of the brown muck. The mass of the brown muck is defined as the the leading order contribution of the mass formula of the heavy hadrons in the $1/m_{\rm Q}$ expansion. This can be evaluated from the masses of the charm and bottom hadrons in the same HQS multiplet. We use the experimental spectrum of the heavy baryons and obtain the masses of the brown muck in several channels. Theoretical predictions in the constituent quark model and the hadronic molecule model are also used to calculate the excitation spectrum of the brown muck. It is found that the ground states of the brown muck are dominated by the diquark configuration for ordinary baryons with three valence quarks, while the mixing with the spin-complex component may become important in the excited states. The spectroscopy of excited baryons in future experiments will be helpful to pin down the structure of the brown muck, through the comparison with the brown muck spectrum in the quark model.

The spectroscopy of the brown muck is intimately related with the diquarks in the heavy baryons. The spin-complex will be also an interesting object in the quark-gluon plasma and the quark matter in the deconfinement phase, when there exist several color non-singlet structures. In fact, there have been discussions that light diquarks may survive around the critical temperature in the quark-gluon plasma \cite{Shuryak:2003ty,Shuryak:2004tx,Lee:2007wr,Oh:2009zj}, and diquarks may also exist in the color superconductivity \cite{Alford:2007xm,Fukushima:2010bq}. Those studies will be performed in J-PARC and GSI-FIAR as well as the BNL-RHIC and CERN-LHC and so on.

\section*{Acknowledgements} 
We thank to Prof.~M.~Oka and Prof.~D.~Jido for fruitful discussions.
This work is supported in part by Grant-in-Aid for Scientific Research on 
Priority Areas ``Elucidation of New Hadrons with a Variety of Flavors 
(E01: 21105006)" (S.~Y. and A.~H.), by Grant-in-Aid for Scientific Research from MEXT
(Grants No. 24105702, No. 24740152 (T.~H.)) and from JSPS (Grants No. 24-3518 (Y.~Y.), No. 15-5858 (S.~O.)),
and by the Grants-in-Aid for Scientific Research from JSPS (Grant No. 25247036) (S.~Y.).

\appendix 
\section{The $\pi \rho \, \omega$ exchange potential}
\label{sec:appendix1}

We consider the potential model including $\pi, \rho \,,$ and  $\omega$ exchanges to demonstrate the spin degeneracy occurs in the exotic hadrons.
Here we consider the $\bar{P}^{(\ast)}N$ systems with several $J^{P}$.

The interaction Lagrangian of heavy mesons and light vector mesons $v$ ($v = \rho \,, \omega$) respecting HQS is given at the leading order by
\begin{eqnarray}
 {\cal L}_{vHH} = -i \beta {\rm Tr}[H_b v^{\mu} (\rho_{\mu})_{ba} \bar{H}_a] 
+ i\lambda {\rm Tr} [H_b \sigma^{\mu \nu} (F_{\mu \nu}(\rho))_{ba} \bar{H}_a] .
\end{eqnarray}
The vector meson field is defined by $\rho_{\mu} = i g_V \hat{\rho}_{\mu}/\sqrt{2}$ and $\hat{\rho}_{\mu}$ is given as 
\begin{eqnarray}
 \hat{\rho}_{\mu} = 
\begin{pmatrix}
 \frac{\rho^0}{\sqrt{2}} + \frac{\omega}{\sqrt{2}} & \rho^+ \\
 \rho^- & - \frac{\rho^0}{\sqrt{2}} + \frac{\omega}{\sqrt{2}}
\end{pmatrix}_{\mu} ,
\end{eqnarray}
where $g_V \simeq 5.8 $ is the universal vector meson coupling. The vector meson field tensor is given by  $F_{\mu \nu}(\rho) = \partial_{\mu} \rho_{\nu} 
 -\partial_{\nu} \rho_{\mu} +[\rho_{\mu}, \rho_{\nu}] $.
The coupling constants are given as $\beta = 0.9$, $\lambda = 0.56$ following Ref.~\cite{Casalbuoni:1996pg}.
We use the $vNN$ vertex from 
\begin{align}
{\cal L}_{vNN} &=
g_{\omega NN}\left[ \bar{N} \gamma_{\mu} \omega^{\mu} N
 + \frac{\kappa_{\omega}}{2m_N} \bar{N} \sigma_{\mu \nu} N\partial^{\nu} \omega^{\mu} \right] \nonumber \\
 &\quad + g_{\rho NN}\left[ \bar{N} \gamma_{\mu} \vec{\tau}_{N}\cdot\vec{\rho}^{\mu} N
 + \frac{\kappa_{\rho}}{2m_N} \bar{N} \sigma_{\mu \nu}  \vec{\tau}_{N} N\cdot\partial^{\nu}\vec{\rho}^{\mu} \right]
\end{align}
where $g_{\rho NN}^2/4\pi = 0.84$, $g_{\omega NN}^2/4\pi = 20.0$, $\kappa_{\rho}=6.1$ and $\kappa_{\omega}=0.0$~\cite{Machleidt:2001} (see also Ref.~\cite{Machleidt}).

The one-boson exchange potentials derived from the vertices of the effective Lagrangians are given as
\begin{align}
 V_{1/2^-}^v &=
\begin{pmatrix}
 C_v^{\prime} & 2\sqrt{3}C_v & \sqrt{6}T_v \\
 2\sqrt{3} C_v & C_v^{\prime} -4 C_v & \sqrt{2}T_v \\
 \sqrt{6}T_v & \sqrt{2} T_v & C_v^{\prime} +2C_v +2T_v
\end{pmatrix},  \\
 V^v_{3/2^-} &=
\begin{pmatrix}
  C_v^{\prime} & -\sqrt{3}T_v & \sqrt{3}T_v & 2\sqrt{3} C_v \\
 -\sqrt{3}T_v & C_v^{\prime}+2 C_v & -2T_v & -T_v \\
 \sqrt{3}T_v & -2T_v & C_v^{\prime} +2C_{v} & T_v \\
2\sqrt{3}C_v & -T_v & T_v & C_v^{\prime} -4 C_v
\end{pmatrix}, 
\end{align}
for the $1/2^-$ and $3/2^-$ states, where $C_v^{\prime}$, $C_v$ and $T_v$ are defined as 
\begin{align}
 C_{\rho}^{\prime} &= \frac{g_V g_{\rho NN} \beta}{\sqrt{2} m_{\rho}^2} C(r;m_{\rho}) 
 \vec{\tau}_{P} \!\cdot\! \vec{\tau}_{N}, \\
 C_{\rho} &= \frac{g_V g_{\rho NN}\lambda (1+\kappa_{\rho})}{\sqrt{2} m_N} \frac{1}{3}
 C(r;m_{\rho}) \vec{\tau}_{P} \!\cdot\! \vec{\tau}_{N}, \\ 
 T_{\rho} &= \frac{g_V g_{\rho NN} \lambda (1+\kappa_{\rho})}{\sqrt{2}m_N} \frac{1}{3} 
 T(r;m_{\rho}) \vec{\tau}_{P} \!\cdot\! \vec{\tau}_{N}, \\
 C_{\omega}^{\prime} &= \frac{g_V g_{\omega NN}\beta}{\sqrt{2} m_{\omega}^2} C(r;m_{\omega}), \\
 C_{\omega} &= \frac{g_V g_{\omega NN}\lambda(1+\kappa_{\omega})}{\sqrt{2}m_N} \frac{1}{3} C(r;m_{\omega}), \\
 T_{\omega} &= \frac{g_V g_{\omega NN}\lambda(1+\kappa_{\omega})}{\sqrt{2}m_N} \frac{1}{3} T(r;m_{\omega}).
\end{align}
The total Hamiltonian for the $\bar{P}^{(\ast)}N$ states is given by combining with the kinetic term and the pion exchange potential as
\begin{eqnarray}
 H_{J^P} = K_{J^P} + V^{\pi}_{J^P} +\sum_{v=\rho, \, \omega} V^{v}_{J^P}.
\end{eqnarray}
The particle basis and the spin-complex basis are related by unitary matrix $U_{J^P}$ in Eqs.~(\ref{eq:transformation_1/2^-}) and (\ref{eq:transformation_3/2^-}).
Then, the Hamiltonians $H_{J^{P}}$ are transformed as,
\begin{align}
 H_{1/2^-}^{\mathrm{SC}} &= U^{-1}_{1/2^-} H_{1/2^-} U_{1/2^-} \nonumber \\
&=
\left(
\begin{array}{cc}
H^{\mathrm{SC}(0^{+})}_{1/2^{-}} & 0 \\
0 & H^{\mathrm{SC}(1^{+})}_{1/2^{-}}
\end{array}
\right)
\end{align}
with
\begin{align}
 H^{\mathrm{SC}(0^{+})}_{1/2^{-}} &= K_{0} \!-\! 3\bar{C} +C^{\prime}_v, \\
 H^{\mathrm{SC}(1^{+})}_{1/2^{-}} &=
\left(
\begin{array}{cc}
 K_{0} \!+\! \bar{C} +C^{\prime}_v -2\bar{T} & -2\sqrt{2} \,\bar{T} \\
 -2\sqrt{2} \,\bar{T}  & K_{2} \!+\!  \bar{C} +C^{\prime}_v \!-\! 2\,\bar{T} 
\end{array}
\right),
\end{align}
for $1/2^-$, where we define
\begin{align}
 \bar{C} &= C +2 C_v ,\quad
 \bar{T} = T - T_v .
\end{align}
In the same way, the $3/2^{-}$ channel is decomposed as
\begin{align}
 H_{3/2^-}^{\mathrm{SC}} &= U^{-1}_{3/2^-} H_{3/2^-} U_{3/2^-} \nonumber \\
 &=
 \left(
\begin{array}{cc}
 H^{\mathrm{SC}(1^{+})}_{3/2^{-}} & 0 \\
 0 &  H^{\mathrm{SC}(2^{+})}_{3/2^{-}}
\end{array}
\right),
\end{align}
with
\begin{align}
 H^{\mathrm{SC}(1^{+})}_{3/2^{-}} &=
\left(
\begin{array}{cc}
 K_{0} \!+\! \bar{C}  +C_v^{\prime} & 2\sqrt{2} \bar{T}  \\
 2\sqrt{2}\bar{T}  & K_{2} \!+\! \bar{C}  +C_v^{\prime}  \!-\! 2\,\bar{T} 
\end{array}
\right), \\
 H^{\mathrm{SC}(2^{+})}_{3/2^{-}} &=
 \left(
\begin{array}{cc}
 K_{2} \!-\! 3\,\bar{C} +C^{\prime}_v & 0 \\
 0 & K_{2} \!+\! \bar{C}  +C_v^{\prime} \!+\! 2\bar{T} 
\end{array}
\right). 
\end{align}
Thus we obtain the block-diagonal forms with the spin-complex basis in the same way with the one-pion-exchange potential.

Similarly, the block-diagonal forms for various quantum numbers are derived. The one vector meson exchange potentials are as follows:
\begin{align}
 V^v_{5/2^-} &=
\begin{pmatrix}
 C_v^{\prime} & 2\sqrt{3}C_v & -\sqrt{\frac{6}{7}}T_v & \frac{6}{\sqrt{7}}T_v \\
 2\sqrt{3}C_v & C_v^{\prime} -4C_v & -\sqrt{\frac{2}{7}} T_v & 2 \sqrt{\frac{3}{7}}T_v \\
 -\sqrt{\frac{6}{7}} T_v & -\sqrt{\frac{2}{7}}T_v & C_v^{\prime} +(2 C_v-\frac{10}{7}T_v) 
 & -\frac{4}{7}\sqrt{6}T_v \\
 \frac{6}{\sqrt{7}}T_v & 2 \sqrt{\frac{3}{7}}T_v & -\frac{4}{7}\sqrt{6}T_v 
 & C_v^{\prime} +(2 C_v+\frac{10}{7}T_v)
\end{pmatrix}, 
\\
 V^v_{7/2^-} &=
\begin{pmatrix}
 C_v^{\prime} & -3\sqrt{\frac{3}{7}}T_v & 2\sqrt{3}C_v & \sqrt{\frac{15}{7}}T_v \\
 -3\sqrt{\frac{3}{7}}T_v & C_v^{\prime} + (2 C_v+\frac{4}{7}T_v) & -\frac{3}{\sqrt{7}}T_v 
 & -\frac{6}{7}\sqrt{5}T_v \\
 2\sqrt{3}C_v & -\frac{3}{\sqrt{7}}T_v & C_v^{\prime} -4C_v & \sqrt{\frac{5}{7}}T_v \\
 \sqrt{\frac{15}{7}}T_v & -\frac{6}{7}\sqrt{5}T_v & \sqrt{\frac{5}{7}}T_v & C_v^{\prime} 
 +(2 C_v-\frac{4}{7}T_v)
\end{pmatrix},
\\
 V^v_{1/2^+} &=  \left(
\begin{array}{cccc}
 C_v^{\prime} & 2\sqrt{3} C_v & \sqrt{6}T_v  \\
 2\sqrt{3}C_v & C_v^{\prime} -4C_V & \sqrt{2} T_v \\
 \sqrt{6}T_v & \sqrt{2}T_v & C_v^{\prime} + (2C_v + 2T_v)
\end{array}
\right)  \, ,
\\
 V^v_{3/2^+} &=  \left(
\begin{array}{cccc}
 C_v^{\prime} & 2\sqrt{3}C_v & -\sqrt{\frac{3}{5}}T_v
 & 3\sqrt{\frac{3}{5}}T_v \\
 2\sqrt{3}C_v & C_v^{\prime} -4C_v & -\frac{1}{\sqrt{5}}T_v
 & \frac{3}{\sqrt{5}}T_v \\
 -\sqrt{\frac{3}{5}}T_v& -\frac{1}{\sqrt{5}}T_v&
 C_v^{\prime} + (2C_v-\frac{8}{5}T_v) & -\frac{6}{5}T_v \\
 3\sqrt{\frac{3}{5}}T_v& \frac{3}{\sqrt{5}}T_v&
 -\frac{6}{5}T_v& C_v^{\prime} + (2C_v +\frac{8}{5}T_v) \\
\end{array}
\right)  \, , \\
 V^v_{5/2^+} &=
\begin{pmatrix}
 C_v^{\prime} & -\frac{3}{5}\sqrt{10}T_v & 2\sqrt{3}C_v & 2\sqrt{\frac{3}{5}}T_v \\
 -\frac{3}{5}\sqrt{10}T_v & C_v^{\prime} +(2 C_v+\frac{2}{5}T_v) & -\sqrt{\frac{6}{5}}T_v 
 & -\frac{4}{5}\sqrt{6}T_v \\
 2\sqrt{3}C_v & -\sqrt{\frac{6}{5}}T_v & C_v^{\prime} -4C_v & \frac{2}{\sqrt{5}}T_v \\
 2\sqrt{\frac{3}{5}}T_v & -\frac{4}{5}\sqrt{6}T_v & \frac{2}{\sqrt{5}}T_v 
 & C_v^{\prime} +(2C_v-\frac{2}{5}T_v)
\end{pmatrix}, \\
 V^v_{7/2^+} &=
\begin{pmatrix}
 C_v^{\prime} & 2\sqrt{3}C_v & -T_v & \sqrt{5}T_v \\
 2\sqrt{3}C_v & C_v^{\prime} -4C_v & -\frac{1}{\sqrt{3}}T_v & \sqrt{\frac{5}{3}}T_v \\
 -T_v & -\frac{1}{\sqrt{3}}T_v & C_v^{\prime} +(2C_v-\frac{4}{3}T_v) & -\frac{2}{3}\sqrt{5}T_v \\
 \sqrt{5}T_v & \sqrt{\frac{5}{3}}T_v & -\frac{2}{3}\sqrt{5}T_v 
 & C_v^{\prime} +(2C_v+\frac{4}{3}T_v)
\end{pmatrix}\, .
\end{align}

Utilizing the unitary matrix $U_{J^P}$ in Eqs.~(\ref{eq:transformation_5/2^-}), (\ref{eq:transformation_7/2^-}), (\ref{eq:transformation_1/2^+}), (\ref{eq:transformation_3/2^+}), (\ref{eq:transformation_5/2^+}) and (\ref{eq:transformation_7/2^+}), we obtain the Hamiltonians of the $\pi\rho \, \omega$ potential in the spin-complex basis. The results for negative parity are
\begin{align}
H_{5/2^-}^{\mathrm{SC}} &= U_{5/2^-}^{-1} H_{5/2^-} U_{5/2^-}  \notag \\
 &=
\left(
\begin{array}{cc}
H_{5/2^-}^{\mathrm{SC}(2^{+})} & 0 \\
0 & H_{5/2^-}^{\mathrm{SC}(3^{+})}
\end{array}
\right),
\end{align}
with
\begin{align}
H_{5/2^-}^{\mathrm{SC}(2^{+})}
&=
\left(
\begin{array}{cc}
 K_2 -3 \bar{C} +C_v^{\prime} &0 \\
 0 & K_2+ \bar{C} + 2\bar{T}+C_v^{\prime}
\end{array}
\right), \\
H_{5/2^-}^{\mathrm{SC}(3^{+})}
&=
\left(
\begin{array}{cc}
  K_2 + \bar{C}  -\frac{4}{7}\bar{T}+C_v^{\prime} & \frac{12 \sqrt{3}}{7}\bar{T} \\
  \frac{12 \sqrt{3}}{7}\bar{T} & K_4 +\bar{C} -\frac{10}{7}\bar{T} +C_v^{\prime}
\end{array}
\right),
\end{align}
for $5/2^+$ and
\begin{align}
 H_{7/2^-}^{\mathrm{SC}} &= U_{7/2^-}^{-1}H_{7/2^-}U_{7/2^-}  \notag \\
  &=
\left(
\begin{array}{cc}
 H_{7/2^-}^{\mathrm{SC}(3^+)} & 0 \\
 0 &  H_{7/2^-}^{\mathrm{SC}(4^+)}
\end{array}
\right),
\end{align}
with
\begin{align}
 H_{7/2^-}^{\mathrm{SC}(3^+)} &=
\left(
\begin{array}{cc}
 K_2+ \bar{C}  -\frac{4}{7}\bar{T} +C_v^{\prime} & \frac{12 \sqrt{3}}{7}\bar{T} \\
 \frac{12 \sqrt{3}}{7}\bar{T} & K_4 +\bar{C}  -\frac{10}{7}\bar{T} +C_v^{\prime} 
\end{array}
\right), \\
 H_{7/2^-}^{\mathrm{SC}(4^+)} &=
\left(
\begin{array}{cc}
 K_4 -3 \bar{C}+C_v^{\prime} & 0 \\
 0 & K_4 +\bar{C} +2\bar{T} +C_v^{\prime}
\end{array}
\right), 
\end{align}
for $7/2^-$.
The results for positive parity are
\begin{align}
H_{1/2^+}^{\mathrm{SC}} &= U^{-1}_{1/2^+} H_{1/2^+} U_{1/2^+} \notag \\
 &=
\left(
\begin{array}{cc}
H_{1/2^+}^{\mathrm{SC}(0^{-})} & 0 \\
 0 & H_{1/2^+}^{\mathrm{SC}(1^{-})}
\end{array}
\right),
\end{align}
with
\begin{align}
H_{1/2^+}^{\mathrm{SC}(0^{-})} &= K_1 + \bar{C}  -4 \bar{T} +C_v^{\prime},  \\
H_{1/2^+}^{\mathrm{SC}(1^{-})} &=
\left(
\begin{array}{cc}
 K_1 -3 \bar{C} +C_v^{\prime} & 0 \\
 0 & K_1 + \bar{C} +2 \bar{T} +C_v^{\prime}
\end{array}
\right) ,
\label{eq:AppHSC1/2+}
\end{align}
for $1/2^+$,
\begin{align}
H_{3/2^+}^{\mathrm{SC}} &= U^{-1}_{3/2^+} H_{3/2^+} U_{3/2^+}  \notag \\
&=
\left(
\begin{array}{cc}
H_{3/2^+}^{\mathrm{SC}(1^{-})} & 0 \\
 0 & H_{3/2^+}^{\mathrm{SC}(2^{-})}
\end{array}
\right),
\end{align}
with
\begin{align}
H_{3/2^+}^{\mathrm{SC}(1^{-})} &=
\left(
\begin{array}{cc}
 K_1 -3 \bar{C}+C_v^{\prime} & 0  \\
 0 & K_1 + \bar{C}  +2 \bar{T} +C_v^{\prime} 
\end{array}
\right), 
\label{eq:AppHSC3/2+}
\\
H_{3/2^+}^{\mathrm{SC}(2^{-})} &=
\left(
\begin{array}{cc}
  K_1 + \bar{C} -\frac{2}{5}\bar{T} +C_v^{\prime}  & \frac{6 \sqrt{6}}{5}\bar{T} \\
  \frac{6 \sqrt{6}}{5}\bar{T} & K_3 +\bar{C}  -\frac{8}{5}\bar{T} +C_v^{\prime}
\end{array}
\right),
\end{align}
for $3/2^+$,
\begin{align}
 H_{5/2^+}^{\mathrm{SC}} &= U^{-1}_{5/2^+} H_{5/2^+} U_{5/2^+}  \notag \\
 &=
\left(
\begin{array}{cc}
 H_{5/2^+}^{\mathrm{SC}(2^{-})} & 0 \\
 0 & H_{5/2^+}^{\mathrm{SC}(3^{-})}
\end{array}
\right),
\end{align}
with
\begin{align}
 H_{5/2^+}^{\mathrm{SC}(2^{-})} &=
\left(
\begin{array}{cc}
 K_1 +\bar{C}  -\frac{2}{5}\bar{T} +C_v^{\prime} & \frac{6 \sqrt{6}}{5}\bar{T} \\
 \frac{6 \sqrt{6}}{5}\bar{T} & K_3 +\bar{C}  -\frac{8}{5}\bar{T} +C_v^{\prime}
\end{array}
\right), \\
 H_{5/2^+}^{\mathrm{SC}(3^{-})} &=
\left(
\begin{array}{cc}
 K_3 -3\bar{C} +C_v^{\prime} & 0 \\
 0 & K_3 +\bar{C}  +2 \bar{T} +C_v^{\prime}
\end{array}
\right),
\end{align}
for $5/2^+$, and
\begin{align}
 H_{7/2^+}^{\mathrm{SC}} &= U^{-1}_{7/2^+} H_{7/2^+} U_{7/2^+}  \notag \\
 &=
\left(
\begin{array}{cc}
 H_{7/2^+}^{\mathrm{SC}(3^-)} & 0 \\
 0 & H_{7/2^+}^{\mathrm{SC}(4^-)}
\end{array}
\right),
\end{align}
with 
\begin{align}
 H_{7/2^+}^{\mathrm{SC}(3^-)} &=
\left(
\begin{array}{cc}
 K_3 -3 \bar{C} +C_v^{\prime} & 0 \\
 0 & K_3 +\bar{C}  +2 \bar{T} +C_v^{\prime}
\end{array}
\right), \\
 H_{7/2^+}^{\mathrm{SC}(4^-)} &=
\left(
\begin{array}{cc}
 K_3 +\bar{C} -\frac{2}{3}\bar{T} +C_v^{\prime} & \frac{4 \sqrt{5}}{3}\bar{T} \\
 \frac{4 \sqrt{5}}{3}\bar{T} & K_5 +\bar{C}  -\frac{4}{3}\bar{T} +C_v^{\prime}
\end{array}
\right),
\end{align}
for $7/2^+$.
In these potentials, $\bar{C}$, $\bar{T}$, and $C_{v}^{\prime}$ represent the modification by the vector meson exchange potential, which appear in many components. Nevertheless, the relation (\ref{eq:identity_hamiltonian}) still holds.

\section{SU(8) Weinberg-Tomozawa interaction}\label{sec:examples_nonexotic}

Here we study the results of the SU(8) Weinberg-Tomozawa model for the $P^{(*)}N$ and $\bar{P}^{(*)}N$ channels in Refs.~\cite{GarciaRecio:2008dp,Gamermann:2010zz,GarciaRecio:2012db} from the viewpoint of the spin-complex basis. The coupled-channel $s$-wave meson-baryon scattering amplitude has been studied in the charmed baryon sector~\cite{GarciaRecio:2008dp}, in the exotic charmed baryon sector~\cite{Gamermann:2010zz}, and in the bottom sector~\cite{GarciaRecio:2012db}. In the hadronic molecule picture, the dynamically generated states in these calculations should contain the spin-complex with $[N\bar{q}]$ and $[Nq]$ configurations. Because this model encodes the HQS as a part of SU(8), it is illustrative to see how the spin symmetry emerges in the results of the charm and bottom sector.

The model describes the scattering amplitude $T_{J}$ for spin $J$ as
\begin{align}
   T_{J}(\sqrt{s}) = [1-V_{J}(\sqrt{s}) G_{J}(\sqrt{s})]^{-1} V_{J}(\sqrt{s}) ,
\end{align}
where $G_{J}(\sqrt{s})$ is the two-body loop function and $\sqrt{s}$ is the total energy of the system. The interaction kernel $V_{J}(\sqrt{s})$ is given by
\begin{align}
   V_{J,ab}(\sqrt{s}) 
   = D_{J,ab}\frac{2\sqrt{s}-M_{a}-M_{b}}{4f_{a}f_{b}}
   \sqrt{\frac{E_{a}+M_{a}}{2M_{a}}}
   \sqrt{\frac{E_{b}+M_{b}}{2M_{b}}}
   \label{eq:WTint} ,
\end{align}
where $M_{a}$, $E_{a}$, $f_{a}$ are the baryon mass, the energy of the baryon, and the meson decay constant in channel $a$, respectively, and we have suppressed the flavor indices. The coupling strength $D_{J,ab}$ is determined by the group theoretical argument, and explicit numbers are tabulated in Refs.~\cite{GarciaRecio:2008dp,Gamermann:2010zz}. 

Let us consider this model for the $P^{(*)}N$ (non-exotic) system and the $\bar{P}^{(*)}N$ (exotic) system in the heavy quark limit. In the exotic sector, the $\bar{P}^{(*)}N$ channels are the lowest energy channels, while the $P^{(*)}N$ system in the non-exotic sector has in general open channels at lower energy, such as $\pi\Sigma_{\rm Q}$ and $\pi\Lambda_{\rm Q}$. However, the transition to these open channels requires the heavy flavor exchange, which is suppressed by $1/m_{\rm Q}$ in comparison with the light flavor exchange processes. Thus, the $P^{(*)}N$ and $\bar{P}^{(*)}N$ systems can be regarded as isolated systems in the heavy quark limit. In this case, the baryon in the scattering is always the nucleon and the HQS requires $f_{P}=f_{P^{*}}\equiv f$ in the present convention, so Eq.~\eqref{eq:WTint} reduces to
\begin{align}
   V_{J,ab}(\sqrt{s}) 
   = D_{J,ab}\frac{\sqrt{s}-M_{N}}{2f^{2}}
   \frac{E_{N}+M_{N}}{2M_{N}}
   \equiv D_{J,ab}\alpha(\sqrt{s})
   \label{eq:WTintHQlimit} .
\end{align}
This shows that the dependence on the spin and channel is included in $D_{J,ab}$ exclusively, which is decoupled from the energy dependence of the interaction in $\alpha(\sqrt{s})$. In the heavy quark limit, the loop functions for $PN$ and $P^{*}N$ are identical and do not depend on the spin, namely, $G_{1/2}=\text{diag}(G,G)$ and $G_{3/2}=G$.

In the following, we concentrate on the isoscalar channel. The explicit forms of the $D_{J,ab}$ matrices for the non-exotic $P^{(*)}N$ channel are~\cite{GarciaRecio:2008dp}
\begin{align}
   D_{1/2}^{P^{(*)}N}
   =& \begin{pmatrix}
   -3 & -\sqrt{27} \\
   -\sqrt{27} & -9
   \end{pmatrix} , \\
   D_{3/2}^{P^{(*)}N}
   =&0 ,
\end{align} 
where the $J=3/2$ channel only has the $P^{*}N$ component. Corresponding couplings for the exotic $\bar{P}^{(*)}N$ sector are~\cite{Gamermann:2010zz}
\begin{align}
   D_{1/2}^{\bar{P}^{(*)}N}
   =&\begin{pmatrix}
   0 & -\sqrt{12} \\
   -\sqrt{12} & 4
   \end{pmatrix}, \\
   D_{3/2}^{\bar{P}^{(*)}N}
   =&-2 .
\end{align}
We now introduce the spin-complex basis for this system. Noting that the basis for the non-exotic channels is given by $\ket{NQ\bar{q}}$, we obtain the basis transformation matrix for the non-exotic system as 
\begin{align}
   \renewcommand{\arraystretch}{1.5}
   \begin{pmatrix}
   \ket{PN(^{2}{\mathrm S}_{1/2})}\\
   \ket{P^{*}N(^{2}{\mathrm S}_{1/2})}
   \end{pmatrix}
   =&
   U_{1/2}^{P^{(*)}N}
   \begin{pmatrix}
   \ket{[N\bar{q}]_{0^{-}}^{(0,S)}Q}_{1/2^{-}}\\
   \ket{[N\bar{q}]_{1^{-}}^{(1,S)}Q}_{1/2^{-}}
   \end{pmatrix},
   \quad
   U_{1/2}^{P^{(*)}N}
   =\begin{pmatrix}
   \frac{1}{2} & \frac{\sqrt{3}}{2} \\
   \frac{\sqrt{3}}{2} & -\frac{1}{2}
   \end{pmatrix} .
\end{align}
The transformation matrix for the exotic system is identical with the $s$-wave part of Eq.~\eqref{eq:U_1/2-}:
\begin{align}
   \renewcommand{\arraystretch}{1.5}
   \begin{pmatrix}
   \ket{\bar{P}N(^{2}{\mathrm S}_{1/2})}\\
   \ket{\bar{P}^{*}N(^{2}{\mathrm S}_{1/2})}
   \end{pmatrix}
   =&
   U_{1/2}^{\bar{P}^{(*)}N}
   \begin{pmatrix}
   \ket{[Nq]_{0^{+}}^{(0,S)}\bar{Q}}_{1/2^{-}}\\
   \ket{[Nq]_{1^{+}}^{(0,S)}\bar{Q}}_{1/2^{-}}
   \end{pmatrix},
   \quad
   U_{1/2}^{\bar{P}^{(*)}N}
   =\begin{pmatrix}
   -\frac{1}{2} & \frac{\sqrt{3}}{2} \\
   \frac{\sqrt{3}}{2} & \frac{1}{2}
   \end{pmatrix}  .
\end{align}
Using these matrices, we have the interaction kernel in the spin-complex basis as
\begin{align}
   V_{1/2}^{P^{(*)}N, {\rm SC}}
   =&(U_{1/2}^{P^{(*)}N})^{-1}
   \begin{pmatrix}
   -3\alpha & -\sqrt{27}\alpha \\
   -\sqrt{27}\alpha & -9\alpha
   \end{pmatrix}U_{1/2}^{P^{(*)}N}
   =
   \begin{pmatrix}
   -12\alpha & 0 \\
   0 & 0
   \end{pmatrix} , \\
   V_{3/2}^{P^{(*)}N, {\rm SC}}
   =& 0 ,
\end{align} 
for the non-exotic system and 
\begin{align}
   V_{1/2}^{\bar{P}^{(*)}N, {\rm SC}}
   =&(U_{1/2}^{\bar{P}^{(*)}N})^{-1}
   \begin{pmatrix}
   0 & -\sqrt{12}\alpha \\
   -\sqrt{12}\alpha & 4\alpha
   \end{pmatrix}U_{1/2}^{\bar{P}^{(*)}N}
   =
   \begin{pmatrix}
   6\alpha & 0 \\
   0 & -2\alpha
   \end{pmatrix} , \\
   V_{3/2}^{\bar{P}^{(*)}N ,{\rm SC}}
   =& -2\alpha ,
\end{align} 
for the exotic system. In the spin-complex basis, the interaction kernel is diagonalized. Because the loop function is proportional to the unit matrix, the original coupled-channel problem reduces to the product of single-channel problems in the spin-complex basis with $m_{\rm Q}\to \infty$.

Possible bound or resonance state is expressed by the pole of the scattering amplitude. The pole condition is given by $1-VG=0$. Because the positive (negative) sign of $V$ represents the repulsive (attractive) interaction, the pole conditions are summarized as
\begin{align}
   1+12\alpha G
   =&
   0 \quad (1/2^{-}, P^{(*)}N) \\
   1+2\alpha G
   =&
   0 \quad (1/2^{-}, \bar{P}^{(*)}N) \\
   1+2\alpha G
   =&
   0 \quad (3/2^{-}, \bar{P}^{(*)}N) .
\end{align} 
The existence of the bound state depends on the finite part of the loop function $G$, but it is shown that one bound state exist for the attractive interaction in the limit of large meson mass~\cite{Hyodo:2006yk,Hyodo:2006kg} under the natural renormalization scheme~\cite{Hyodo:2008xr}. The above equations indicate that the ground state is the HQS singlet in the non-exotic channel ($\ket{[N\bar{q}]_{0^{-}}^{(0,S)}Q}$), and the HQS doublet state in the exotic channel ($\ket{[Nq]_{1^{+}}^{(0,S)}\bar{Q}}_{1/2^{-}}, \ket{[Nq]_{1^{+}}^{(0,S)}\bar{Q}}_{3/2^{-}}$). 

\begin{table}[tb]
\caption{Ratios of the coupling strengths of the ground states found in Refs.~\cite{GarciaRecio:2008dp,Gamermann:2010zz,GarciaRecio:2012db} in comparison with the values in the heavy quark limit.}
\begin{center}
\begin{tabular}{|l|l|r|l|r|}
   \hline
         & \multicolumn{2}{|l|}{Non-exotic system} & \multicolumn{2}{|l|}{Exotic system} \\
   \hline
Sector & State & $|g_{PN}/g_{P^{*}N}|$ & State & $|g_{\bar{P}N}/g_{\bar{P}^{*}N}|$ \\
   \hline
$m_{\rm Q}\to\infty$ & $1/2^{-}$, singlet & $1/\sqrt{3}\sim 0.58$
                 & $1/2^{-}$, doublet & $\sqrt{3}\sim 1.73$  \\
Bottom~\cite{GarciaRecio:2012db} 
            & $5797.6$ MeV & $4.9/8.3\sim 0.59$ & & \\
Charm~\cite{GarciaRecio:2008dp,Gamermann:2010zz} 
            & $2595.4$ MeV & $3.69/5.70\sim 0.65$ 
            & $2805.0$ MeV & $1.5/1.4\sim 1.07$ \\
            \hline
\end{tabular}
\end{center}
\label{tbl:coupling}
\end{table}%

Using the wave function of the spin-complex basis, we can extract the ratio of the $PN/P^{*}N$ component as in Section~\ref{sec:fractions}. The wave function of the HQS singlet in the non-exotic sector is
\begin{align}
   \ket{0^{-}}_{1/2^{-}}
   =&
   \frac{1}{2}\ket{PN}
   +\frac{\sqrt{3}}{2}\ket{P^{*}N}
\end{align} 
and the $1/2^{-}$ part of the HQS doublet is 
\begin{align}
   \ket{1^{+}}_{1/2^{-}}
   =&
   \frac{\sqrt{3}}{2}\ket{PN}
   +\frac{1}{2}\ket{P^{*}N} .
\end{align} 
The mixing ratio of the wave function is reflected in the coupling strength $g_{a}$ obtained from the residue of the pole. The HQS implies that the ratios of the coupling strengths should be
\begin{align}
   \left|\frac{g_{PN}}{g_{P^{*}N}}\right|
   =&
   \frac{1}{\sqrt{3}}  \quad (1/2^{-}, P^{(*)}N) \\
   \left|\frac{g_{\bar{P}N}}{g_{\bar{P}^{*}N}}\right|
   =&
   \sqrt{3} \quad (1/2^{-}, \bar{P}^{(*)}N ) 
\end{align} 
in the heavy quark limit. In Table~\ref{tbl:coupling} we show the ratios of the coupling strengths of the ground states in Refs.~\cite{GarciaRecio:2008dp,Gamermann:2010zz,GarciaRecio:2012db} together with the values in the heavy quark limit. We see that the actual coupled-channel calculation with finite $m_{\rm Q}$ provides the coupling strengths similar to the values indicated by the HQS. In addition, the ratio of the coupling constants approaches the value in the heavy quark limit as the quark mass is increased from charm to bottom. In this way, the spin-complex basis provides a new insight into the calculations in the charm and bottom sectors.

\bibliographystyle{elsarticle-num}







\end{document}